\documentclass[12pt,a4paper]{article}
\pdfoutput=1
\usepackage{jheppub}
\usepackage{slashed}
\usepackage{amsmath}

\usepackage{IEEEtrantools}
\newcommand{\beq}{\begin{equation}}
\newcommand{\eeq}{\end{equation}}
\newcommand{\nn}{\nonumber}

\newcommand{\mD}{\mathcal{D}}

\newcommand{\mN}{\mathcal{N}}

\newcommand{\mS}{\mathcal{S}}

\newcommand{\p}{\partial}
\newcommand{\ov}{\overline}
\newcommand{\f}{\frac}

\newcommand{\al}{\alpha}
\newcommand{\be}{\beta}
\newcommand{\ga}{\gamma}         \newcommand{\Ga}{\Gamma}
\newcommand{\de}{\delta}        
\newcommand{\ep}{\epsilon}

\newcommand{\ze}{\zeta}
\newcommand{\et}{\eta}
\newcommand{\te}{\theta}

\newcommand{\ka}{\kappa}
\newcommand{\la}{\lambda}       

\newcommand{\rh}{\rho}

\newcommand{\si}{\sigma}         \newcommand{\Si}{\Sigma}

\newcommand{\ta}{\tau}
      
\newcommand{\ph}{\phi}          \newcommand{\Ph}{\Phi}
\newcommand{\ps}{\psi} 

\newcommand{\ch}{\chi}
        
\newcommand{\half}{\frac{1}{2}}

\newcommand{\bph}{\bar{\ph}}
\newcommand{\bPh}{\bar{\Ph}}

\newcommand{\mbp}{\mathbf{p}}
\newcommand{\bps}{\bar{\psi}}

\newcommand{\lan}{\langle}
\newcommand{\ran}{\rangle}

\newcommand{\SB}{\mathcal{S}_B}
\newcommand{\SF}{\mathcal{S}_F}
\newcommand{\Sf}{\mathcal{S}_f}

\newcommand{\TB}{\mathcal{T}_B}
\newcommand{\TF}{\mathcal{T}_F}
\newcommand{\Tf}{\mathcal{T}_f}

\preprint{TIFR/TH/15-15}

\title{Unitarity, Crossing Symmetry and Duality in the scattering of 
${\cal N}=1$ Susy Matter Chern-Simons theories}
\author[a]{Karthik Inbasekar,}
\author[b]{Sachin Jain,}
\author[a]{Subhajit Mazumdar,}
\author[a]{Shiraz Minwalla,}
\author[a]{V.Umesh,}
\author[c]{Shuichi Yokoyama}
\affiliation[a]{Department of Theoretical Physics,Tata Institute of Fundamental Research,Homi
Bhabha road, Mumbai 400005, India}
\affiliation[b]{Department of Physics, Cornell University, Ithaca, New York 14853, USA.}
\affiliation[c]{Physics Department, Technion - Israel Institute of Technology, Technion City -
Haifa 3200003}
\emailAdd{ikarthik@theory.tifr.res.in}
\emailAdd{sj339@cornell.edu}
\emailAdd{subhajitmazumdar@theory.tifr.res.in}
\emailAdd{minwalla@theory.tifr.res.in}
\emailAdd{vumesh.physics@gmail.com}
\emailAdd{shuichitoissho@gmail.com}

\abstract{We study the most general renormalizable ${\cal N}=1$ 
$U(N)$ Chern-Simons gauge theory coupled to a single (generically massive)
fundamental matter multiplet. At leading order in the $\lq$t Hooft large $N$ limit
we present computations and conjectures for the
$2 \times 2$ $S$ matrix in these theories; our results apply at all 
orders in the $\lq$t Hooft coupling and the matter self interaction. Our 
$S$ matrices are in perfect agreement with the recently conjectured strong weak coupling self
duality of this class of theories. The consistency of our results with unitarity requires a
modification of the usual rules of crossing symmetry in precisely the manner anticipated in 
\href{http://xxx.lanl.gov/abs/1404.6373}{{\tt arXiv:1404.6373}}, lending substantial support to the
conjectures of that paper. In a certain range of coupling constants our $S$ matrices have a pole  
whose mass vanishes on a self dual codimension one surface in the space of 
couplings.}

\begin{document}

\maketitle
\section{Introduction}\label{intro}

Non-Abelian $U(N)$ gauge theories in three spacetime 
dimensions are dynamically rich. At low energies parity preserving gauge self interactions 
are generically governed by the Yang-Mills action
\begin{equation}\label{ymact}
\frac{1}{g^2_{YM}} \int d^3x ~{\rm Tr} ~ F_{\mu\nu}^2\ .
\end{equation}
As $g_{YM}^2$ has the dimensions of mass, gluons are strongly coupled in the 
IR. In the absence of parity invariance the gauge field Lagrangian generically 
includes an additional Chern-Simons term and schematically takes the form 
\begin{equation}\label{ymcs}
 \frac{i\kappa}{4 \pi}~\int ~{\rm Tr} \left( AdA + \frac{2}{3} A^3 \right)
- \frac{1}{4g^2_{YM}} \int d^3x ~{\rm Tr} ~  F_{\mu\nu}^2\ .
\end{equation}
The Lagrangian \eqref{ymcs} describes a system of {\it massive} gluons; 
with mass $m \propto {\kappa g^2_{YM}}$. At energies much lower than $g^2_{YM}$ 
\eqref{ymcs} has no local degrees of freedom. The effective low energy 
dynamics is topological, and is governed by the action \eqref{ymcs} with 
the Yang-Mills term set to zero. This so called pure Chern-Simons theory was 
solved over twenty five years ago by Witten \cite{Witten:1988hf}; his beautiful and nontrivial  
exact solution has had several applications in the study of two 
dimensional conformal field theories and the mathematical study of 
knots on three manifolds.

Let us now add matter fields with standard, minimally coupled kinetic terms, 
(in any representation of the gauge group) to \eqref{ymcs}. The resulting low 
energy dynamics is particularly simple in the limit in which all matter 
masses are parametrically smaller than $g^2_{YM}$. In order to focus on this 
regime we take the limit $g^2_{YM} \to \infty$ with masses of matter 
fields held fixed. In this limit the Yang-Mills term in \eqref{ymcs} 
can be ignored and we obtain a Chern-Simons self coupled gauge theory 
minimally coupled to matter fields. While the gauge fields are non propagating, 
they mediate nonlocal interactions between matter fields. 

In order to gain 
intuition for these interactions it is useful to first consider 
the special case $N=1$, i.e. the case of an Abelian gauge theory interacting 
with a unit charge scalar field. The gauge equation of motion 
\begin{equation}\label{geom}
\kappa \varepsilon^{\mu\nu\rho} F_{\nu\rho}= 2 \pi J^\mu 
\end{equation}
ensures that each matter particle traps $\frac{1}{\kappa}$ units of flux 
(where $i\int F=2 \pi$ is defined as a single unit of flux). It follows as 
a consequence of the Aharonov-Bohm effect that exchange of two 
unit charge particles results in a phase $\frac{ \pi}{\kappa}$; in other
words the Chern-Simons interactions turns the scalars into anyons with 
anyonic phase $\pi \nu=\frac{ \pi}{\kappa}$. 

The interactions induced between matter particles by the exchange of 
non-abelian Chern-Simons gauge bosons are similar with one additional twist. 
In close analogy with the discussion of the previous paragraph, the 
exchange of two scalar matter quanta in representations $R_1$ and $R_2$ 
of $U(N)$  results in the phase $ \frac{\pi T_{R_1}.T_{R_2}}{\kappa}$ where 
$T_{R}$ is the generator of $U(N)$ in the representation $R$. The new element 
in the non-abelian theory is that the phase obtained upon interchanging 
two particles is an operator (in $U(N)$ representation space) rather than a 
number. The eigenvalues of this operator are given by 
\begin{equation}\label{evo}
\nu_R'= \frac{c_2(R_1)+ c_2(R_2) - c_2(R')}{2 \kappa}
\end{equation}
where $c_2(R)$ is the quadratic Casimir of the representation $R$ and 
$R'$ runs over the finite set of representations that appear in the Clebsh-Gordon decomposition of
the tensor product of $R_1$ and $R_2$. In other words
the interactions mediated by non-abelian Chern-Simons coupled gauge fields 
turns matter particles into non-abelian anyons. 

In some ways anyons are qualitatively different  from either bosons or 
fermions. For example anyons (with fixed anyonic phases) are never free:  
there is no limit in which the multi particle anyonic Hilbert 
space can be regarded as a `Fock space' of a single particle state space. 
Thus while matter Chern-Simons theories are regular relativistic quantum field 
theories from a formal viewpoint, it seems possible that they will display 
dynamical features never before encountered in the study of quantum field 
theories. This possibility provides one motivation for the  
intensive study of these theories.

Over the last few years matter Chern-Simons theories have been intensively 
studied in two different contexts. The ${\cal N}=6$ supersymmetric ABJ and ABJM theories
\cite{Aharony:2008gk,Aharony:2008ug} have been exhaustively studied from the viewpoint of the 
AdS/CFT correspondence \cite{Maldacena:1997re,Aharony:1999ti}. Several other supersymmetric
Chern-Simons theories with ${\cal N} \geq 2$ supersymmetry have also been intensively studied, 
sometimes motivated by brane constructions in string theory. The technique 
of supersymmetric localization has been used to perform exact computations 
of several supersymmetric quantities
\cite{Drukker:2008zx,Chen:2008bp,Bhattacharya:2008bja,Kim:2009wb,Kapustin:2009kz,
Marino:2011nm} (indices, supersymmetric Wilson loops, three sphere partition
functions). These studies have led, in particular, to the conjecture and detailed check for `Seiberg
like' Giveon-Kutasov dualities between Chern-Simons matter theories
with $\mN\geq 2$ supersymmetry \cite{Benini:2011mf,Giveon:2008zn}. Most of these impressive 
studies have, however, focused on observables \footnote{These observables 
include partition functions, indices, Wilson lines and correlation functions 
of local gauge invariant operators. Note that gauge invariant operators do 
not pick up anyonic phases when they go around each other precisely because 
they are gauge singlets.} that are not directly sensitive to the anyonic 
nature of of the underlying excitations and have exhibited no qualitative 
surprises.

Qualitative surprises arising from the effectively anyonic nature of the 
matter particles seem most likely to arise in observables built out 
of the matter fields themselves rather than gauge invariant composites 
of these fields. There exists a well defined gauge invariant observable of 
this sort; the $S$ matrix of the matter fields. While this quantity has been 
somewhat studied for highly supersymmetric Chern-Simons theories, the results 
currently available (see e.g.
\cite{Agarwal:2008pu,Bargheer:2012cp,Bianchi:2011fc,Chen:2011vv,Bianchi:2011dg,Bianchi:2014iia,
Bianchi:2012cq}) have all be obtained in perturbation
theory.  Methods based on supersymmetry have not yet proved powerful enough
to obtain results for $S$ matrices at all orders in the coupling constant, even 
for the maximally supersymmetric ABJ theory. For a very special class of 
matter Chern-Simons theories, however, it has recently been demonstrated that 
large $N$ techniques are powerful enough to compute $S$ matrices at all orders 
in a `t Hooft coupling constant, as we now pause to review. 

Consider large $N$ Chern-Simons coupled to a finite number of matter 
fields in the fundamental representation of $U(N)$.
\footnote{These theories were initially studied because of 
their conjectured dual description in terms of Vasiliev equations of 
 higher spin gravity.} It was realized in \cite{Giombi:2011kc} that the usual large 
$N$ techniques are roughly as effective in these theories as in vector 
models even in the absence of supersymmetry (see
\cite{Aharony:2011jz,Maldacena:2011jn,Maldacena:2012sf,Chang:2012kt,Aharony:2012nh,Jain:2012qi,
Yokoyama:2012fa,GurAri:2012is,Aharony:2012ns,Jain:2013py,Takimi:2013zca,Jain:2013gza,
Yokoyama:2013pxa,Bardeen:2014paa,Jain:2014nza,Bardeen:2014qua,Gurucharan:2014cva,
Dandekar:2014era,Frishman:2014cma,Moshe:2014bja,Bedhotiya:2015uga,Gur-Ari:2015pca} for related
works). In particular large $N$ techniques have recently been used in \cite{Jain:2014nza} to compute
the $ 2 \rightarrow 2$ $S$ matrices of the matter particles in purely bosonic/fermionic fundamental
matter theories coupled to a Chern-Simons gauge field. 

Before reviewing the results of \cite{Jain:2014nza} let us pause to work out 
the effective anyonic phases for two particle systems of quanta in the 
fundamental/ antifundamental representations at large $N$. 
\footnote{The application of large $N$ techniques to these theories has led 
to conjectures for strong weak coupling dualities between classes of these 
theories. The simplest such duality relates a Chern-Simons theory coupled to a 
single fundamental bosonic multiplet to another Chern-Simons theory 
coupled to a single fermionic multiplet. This duality was first clearly 
conjectured in \cite{Aharony:2012nh}, building on the results of 
\cite{Maldacena:2011jn,Maldacena:2012sf}, and following up on an earlier suggestion in
\cite{Giombi:2011kc}. The discovery of a three dimensional 
Bose-Fermi duality was the first major qualitative surprise in the study of 
Chern-Simons matter theories, and is intimately connected with the effectively
anyonic nature of the matter excitations, as explained, for instance, 
in  \cite{Jain:2014nza}.}  Following \cite{Jain:2014nza} 
we refer to any matter quantum that transforms in the (anti)fundamental 
of $U(N)$ a(n) (anti)particle. A two particle system can couple into 
two representations $R'$ (see \eqref{evo}); the symmetric representation 
(two boxes in the first row
of the Young Tableaux) and the antisymmetric representation (two boxes in the
first column of the Young Tableaux). It is easily verified that the 
anyonic phase $\nu_{R'}$ (see \eqref{evo}) is of order $\frac{1}{N}$ 
(and so negligible in the large $N$ limit) for both choices of $R'$. 
On the other hand a particle - antiparticle system can couple into 
$R'$ which is either the adjoint of the singlet. $\nu_{R'}$ once again 
vanishes in the large $N$ limit when $R'$ is the adjoint. However when $R'$ is the singlet
representation it turns out that $\nu_{sing}=\frac{N}{\kappa}= \lambda$ and so is of order unity in
the large $N$ limit. In summary two particle systems are always non anyonic in the 
large $N$ limit of these special theories. Particle - antiparticle 
systems are also non anyonic in the adjoint channel. However they are 
effectively anyonic - with an interesting finite anyonic phase- in the singlet
channel. See \cite{Jain:2014nza} for more details. This preparation makes 
clear that  qualitative surprises related to anyonic 
physics in the two quantum scattering in these theories might occur only in 
particle - antiparticle scattering in the singlet sector. 

The authors of \cite{Jain:2014nza} used large $N$ techniques to explicitly 
evaluate the $S$ matrices in all three non-anyonic channels in the theories 
they studied (see below for more details of this process). They also used 
a mix of consistency checks and physical arguments involving crossing 
symmetry to conjecture a formula for the particle - antiparticle $S$ matrix 
in the singlet channel. The conjecture of  \cite{Jain:2014nza} for the 
$S$ matrix in the singlet channel has two unexpected novelties related to the 
anyonic nature of the two particle state
\begin{itemize}
\item{1.} The singlet $S$ matrix in both the bosonic and fermion theories has 
a contact term localized on forward scattering. In particular the $S$ matrix 
is not an analytic function of momenta. 
\item{2.} The analytic part of the singlet $S$ matrix is given by the 
analytic continuation of the $S$ matrix in any of the other three 
channels  $\times$ $\frac{\sin \pi \lambda}{\pi \lambda}$. In other words 
the usual rules of crossing symmetry to the anyonic channel are 
modified by a factor determined by the anyonic phase.
\end{itemize}

The modification of the usual rules of analyticity and crossing symmetry 
in the anyonic channel of $2 \times 2$ scattering was a major surprise
 of the analysis of \cite{Jain:2014nza}. The authors of \cite{Jain:2014nza} 
offered physical explanations - involving the anyonic nature of 
scattering in the singlet channel for both these unusual features of the 
$S$ matrix. The simple (though non rigorous) explanations proposed in 
\cite{Jain:2014nza} are universal in nature; they should apply equally well 
to all large $N$ Chern-Simons theories coupled to fundamental matter, and not just the particular
theories studied in \cite{Jain:2014nza}. This fact suggests a simple 
strategy for testing the conjectures of \cite{Jain:2014nza} which we 
employ in this paper. We simply redo the $S$ matrix computations 
of \cite{Jain:2014nza} in a different class of Chern-Simons theory coupled 
to fundamental matter and check that the conjectures of 
\cite{Jain:2014nza} - unmodified in all details - indeed continue to yield 
sensible results (i.e. results that pass all necessary consistency checks) 
in the new system. We now describe the system we study and the nature 
of our results in much more detail. 

The theories we study are the most general power counting renormalizable 
${\cal N}=1$ $U(N)$ gauge theories coupled to a single fundamental multiplet 
(see \eqref{action} below). In order to study scattering in these theories 
we imitate the strategy of \cite{Jain:2014nza}. 
The authors of \cite{Jain:2014nza}
worked in lightcone gauge; in this paper we work in a supersymmetric 
generalization of lightcone gauge \eqref{lcg}. In this gauge (which preserves
manifest offshell supersymmetry) the gauge self interaction term vanishes. 
This fact - together with planarity at large $N$ -   allows us to find a 
manifestly supersymmetric Schwinger-Dyson equation for the exact propagator 
of the matter supermultiplet. This equation turns out to be easy to solve; 
the solution gives simple exact expression for the all orders propagator
for the matter supermultiplet (see subsection \S\ref{prop}). 

With the exact propagator in hand, we then proceed to write down an exact 
Schwinger-Dyson equation for the offshell four point function of the 
matter supermultiplet. The resultant integral equation is quite complicated; 
as in \cite{Jain:2014nza} we have been able to solve this equation only 
in a restricted kinematic range ($q_\pm=0$ in the notation of fig
\ref{gaugeintSD}). In this kinematic regime, however, we have been able 
to find a completely explicit (if somewhat complicated) solution of 
the resulting equation(see subsection \S\ref{inteqn}-\S\ref{offshellsoln}). 

In order to evaluate the $S$ matrices we then proceed to take the onshell limit 
of our explicit offshell results. As explained in detail in \cite{Jain:2014nza},
the 3 vector $q^\mu$ has the interpretation of momentum transfer for 
both channels of particle- particle scattering and also for particle 
antiparticle scattering in the adjoint channel. In these channels the 
fact that we know the offshell four point amplitudes 
only when $q_\pm=0$ forces us to study scattering in a particular Lorentz 
frame; any frame in which momentum transfer happens along the spatial $q^3$ 
direction. In any such frame we obtain explicit results for all $2 \times 2$
scattering matrices in these three channels. The results are then
covariantized to formulae that apply to any frame. Following this method 
we have obtained explicit results for the 
$S$ matrices in these three channels. Our results are presented in detail in 
subsections \S\ref{onshell} - \S\ref{onssN2}. As we explain in detail below, 
our explicit results have exactly 
the same interplay with the proposed strong weak coupling self duality 
of the set of ${\cal N}=1$ Chern-Simons fundamental matter theories 
(see subsection \ref{dualitycon}) as that described in \cite{Jain:2014nza}; 
duality maps particle - particle $S$ matrices in the symmetric and antisymmetric
channels to each other, while it maps the particle - antiparticle $S$ matrix 
in the adjoint channel to itself. 

As in \cite{Jain:2014nza} our explicit offshell results do not 
permit a direct computation of the $S$ matrix for particle - antiparticle scattering
in the singlet channel. This is because the three vector $q^\mu$ 
is the center of mass momentum for this scattering 
process and so must be timelike, which is impossible if $q^\pm=0$. 
Our explicit results for the $S$ matrices in the other channels, together 
with the conjectured modified crossing symmetry rules of 
\cite{Jain:2014nza}, however, yield a conjectured formula for the 
$S$ matrix in this channel.

 In section \ref{unitarityx} we subject our 
conjecture for the particle - antiparticle $S$ matrix 
to a very stringent consistency check; we verify that 
it obeys the nonlinear unitarity equation \eqref{uni1}\footnote{At large $N$ 
this equation may be shown to  close on $2 \times 2$ scattering.}. 
From the purely algebraic point of view the fact that our complicated S 
matrices are unitary appears to be a minor miracle- one that certainly 
fails very badly for the $S$ matrix obtained using the usual rules of crossing 
symmetry. We view this result as very strong evidence for the correctness 
of our formula, and indirectly for the modified crossing symmetry 
rules of \cite{Jain:2014nza}. 

Our proposed formula for particle - antiparticle scattering in the singlet 
channel has an interesting analytic structure. As a function of $s$ (at
fixed $t$) our $S$ matrix has the expected two particle cut starting at 
$s=4 m^2$. In a certain range of interaction parameters it also has poles 
at smaller (though always positive) values of $s$. These poles represent 
bound states; when they exist these bound states must be absolutely stable 
even at large but finite $N$, simply because they are the lightest 
singlet sector states (baring the vacuum) in the theory; recall that 
our theory has no gluons. Quite remarkably it turns out that 
the mass of this bound state supermultiplet vanishes 
at $w=w_c(\lambda)$ where $w$ is the  superpotential interaction 
parameter of our theory (see \eqref{action}) and $w_c(\lambda)$ is the 
simple function listed in \eqref{wcr}. In other words a one parameter 
tuning of the superpotential is sufficient to produce massless bound
states in a theory of massive `quarks'; we find this result quite 
remarkable. Scaling $w$ to $w_c$ permits a parametric separation 
between the mass of this bound state and all other states in the theory. 
In this limit there must exist a decoupled QFT description of the 
dynamics of these light states even at large but finite $N$; it seems 
likely to us that this dynamics is governed by a ${\cal N}=1$ Wilson-Fisher fixed point. We leave
the detailed investigation of this 
suggestion to future work. 

The $S$ matrices computed and conjectured in this paper turn out to simplify 
dramatically at $w=1$, at which point the system \eqref{action} turns out 
to enjoy an enhanced  ${\cal N}=2$ supersymmetry. In the three non-anyonic 
channels our $S$ matrix reduces simply to its tree level counterpart at $w=1$. 
It follows, in other words, that the $S$ matrix is not renormalized as 
a function of $\lambda$ in these channels. This result illustrates the 
conflict between naive crossing symmetry and unitarity in a simple setting. 
Naive crossing symmetry would yield a singlet channel $S$ matrix that is 
also tree level exact. However tree level $S$ matrices by themselves can never
obey the unitarity equations (they do not have the singularities 
needed to satisfy the Cutkosky's rules obtained by gluing them together). 
The resolution to this paradox appears simply to be that the naive crossing
symmetry rules are wrong in the current context. Applying the 
conjectured crossing symmetry rules of \cite{Jain:2014nza} we find a 
singlet channel $S$ matrix that continues to be very simple, but is not 
tree level exact, and in fact satisfies the unitarity equation. 

In this paper we have limited our attention to the study of ${\cal N}=1$ 
theories with a single fundamental matter multiplet. Were we to extend 
our analysis to theories with two multiplets we would encounter, in 
particular, the ${\cal N}=3$ theory. Extending to the study of 
a theory with four multiplets (and allowing for the the gauging of 
a $U(1)$ global symmetry)  would allow us to study the 
${\cal N}=6$ $U(N) \times U(1)$ ABJ theory. We believe it would not be 
difficult to adapt the methods of this paper to find explicit all orders 
results for the $S$ matrices of all these theories at leading order in 
large $N$. We expect to find scattering matrices that are 
unitary precisely because they transform under crossing symmetry in the unusual manner conjectured in 
\cite{Jain:2014nza}. It would be particularly interesting to find 
explicit results for the ${\cal N}=6$ theory in order to  
facilitate a detailed comparison with the perturbative computations 
of $S$ matrices in ABJM theory
\cite{Agarwal:2008pu,Bargheer:2012cp,Bianchi:2011fc,Chen:2011vv,Bianchi:2011dg,Bianchi:2014iia,
Bianchi:2012cq}, which appear to report results 
that are crossing symmetric but (at least naively) conflict with unitarity.

\section{Review of Background Material}

\subsection{Renormalizable \texorpdfstring{${\cal N} =1$}{N=1} theories with a single fundamental 
multiplet} \label{rnt}

In this paper we study $2 \times 2$ scattering in the most general 
renormalizable ${\cal N}=1$ supersymmetric $U(N)$ Chern-Simons theory coupled 
to a single fundamental matter multiplet. Our theory is defined in 
superspace by the Euclidean action
\cite{Avdeev:1992jt,Ivanov:1991fn}
\begin{align}\label{action}
 \mS_{\mN=1}=-\int d^3x d^2\te\biggl[ \f{\ka}{2\pi}Tr\biggl(&-\f{1}{4}D_\al \Ga^{\be} D_\be
\Ga^{\al}-\f{1}{6} D^\al \Ga^\be
\{\Ga_\al,\Ga_\be\}-\f{1}{24} \{\Ga^\al,\Ga^\be\}\{\Ga_\al,\Ga_\be\}\biggr) \nn\\
&-\f{1}{2}(D^\al\bar{\Ph}+i \bar{\Ph} \Ga^\al)(D_\al\Ph-i \Ga_\al \Ph)+m_0
\bar{\Ph}\Ph+\f{\pi w}{\ka} (\ov{\Ph}\Ph)^2\biggr] \ .
\end{align}

The integration in \eqref{action} is over the three Euclidean spatial 
coordinates and the two anticommuting spinorial coordinates $\theta^\alpha$ 
(the $SO(3)$ spinorial indices $\alpha$ range over two allowed values 
$\pm$). The fields $\Phi$ and $\Gamma^\alpha$ in \eqref{action} are, 
respectively, complex and real superfields \footnote{ See Appendix \S \ref{superspace} for our
conventions for superspace}. They may be expanded in components as 
\begin{align}\label{superfieldexpansions}
&\Ph = \ph+\te\ps-\te^2 F \ , \nn \\
&\bar{\Ph} =\bar{\ph}+\te\bar{\ps}-\te^2 \bar{F} \ ,\nn\\
&\Ga^\al=\ch^\al-\te^\al B+i \te^\be A_\be^{ \ \al}-\te^2(2\la^\al-i\p^{\al\be}\ch_\be) \ , 
\end{align}
where $\Gamma_\alpha$ is an $N \times N$ matrix in color space, while $\Phi$ is 
an $N$ dimensional column. 

The superderivative $D_\alpha$ in \eqref{action} is defined by
\begin{align}\label{superderivativepos}
 D_\al=\f{\p}{\p\te^\al}+ i \te^\be \p_{\al\be} \ , D^\al=C^{\al\be}D_\be \ ,
\end{align}
where $C^{\al\be}$ is the charge conjugation matrix. See 
Appendix \S \ref{gamma} for notations and conventions. 

The theories 
\eqref{action} are characterized by one dimensionless coupling constant
$w$, a dimensionful mass scale $m_0$, and two integers $N$ (the rank of the 
gauge group $U(N)$) and $\kappa$, the level of the Chern-Simons theory. 
\footnote{The precise definition of $\kappa$ is defined as follows. 
Let $k$ denote the level of the WZW theory related to Chern-Simons theory after
all fermions have been integrated out. $\kappa$ is the related to $k$ by 
$\kappa= k+ {\rm sgn }(k) N$
}
In the large $N$ limit of interest to us in this paper, the $\lq$t Hooft coupling 
$\lambda= \frac{N}{\kappa}$ is a second effectively continuous dimensionless
parameter.

The action \eqref{action} enjoys invariance under the super gauge transformations 
\begin{align}\label{gaugetransformations}
&\de\Ph=i  K \Ph \ ,\nn\\
&\de\bar{\Ph}= - i \bar{\Ph} K \ , \nn\\
&\de\Ga_\al=D_\al K+ [\Ga_\al, K] \ ,
\end{align}
where $K$ is a real superfield (it is an $N \times N$ matrix in color space).

\eqref{action} is manifestly invariant under the two supersymmetry 
transformations generated by the supercharges $Q_\alpha$
\beq\label{supercharge}
Q_\al= i (\f{\p}{\p\te^\al}-i \te^\be \p_{\be\al})
\eeq
that act on $\Phi$ and $\Gamma_\al$ as 
\begin{align}\label{stran}
&\de_\al\Phi= Q_\al \Phi \ , \nn  \\
&\de_\al \Ga_\be= Q_\al \Ga_\be\ .
\end{align}
The differential operators $Q_\alpha$ and $D_\alpha$ obey the algebra
\begin{align}\label{algebra}
& \{Q_\al,Q_\be\}=2 i\p_{\al\be} \ , \nn\\
&\{D_\al,D_{\be}\}=2 i\p_{\al\be} \ , \nn\\
& \{Q_\al,D_\be\}=0 \ .
\end{align}

At the special value $w=1$, the action \eqref{action} actually has enhanced 
supersymmetry; it is  ${\cal N}=2$ (four supercharges) supersymmetric. 
\footnote{This may be confirmed, for instance, by checking that 
\eqref{compact} at $w=1$ 
is identical to the $\mN=2$ superspace Chern-Simons action 
coupled to a single chiral multiplet in the fundamental representation with 
no superpotential (see Eq 2.3 of \cite{Gaiotto:2007qi}) expanded in components 
in Wess-Zumino gauge.}  

The physical content of the theory \eqref{action} is most transparent when 
the Lagrangian is expanded out in component fields in the so called 
 Wess-Zumino gauge -  defined by the requirement
\beq
B=0 \ , \ch=0\ .
\eeq
Imposing this gauge, integrating over $\theta$ and eliminating auxiliary 
fields we obtain the component field action  \footnote{Our trace conventions are
\beq
Tr(T^a T^b)=\f{1}{2}\de^{ab}\ , \ \sum_a (T^a)_i^{\ j} (T^a)_k^{\ l}=\f{1}{2} \de_i^{\ l}
\de_k^{\ j }\ .
\eeq
The gauge covariant derivatives in \eqref{compact} are
\begin{align}
& \mD^\mu\bph=\p^\mu\bph+i  \bph A^\mu\ , \ \mD_\mu\ph=\p_\mu\ph-i A_\mu \ph\ ,\nn\\
&\slashed{\mD}\bps=\ga^\mu(\p_\mu \bps+i \bps A_\mu )\ , \ \slashed{\mD}\ps=\ga^\mu(\p_\mu \ps-i
A_\mu \ps ).
\end{align}
}
\begin{align}\label{compact}
 \mS_{\mN=1}=\int d^3x \biggl(&-\f{\ka}{2\pi} \ep^{\mu\nu\rh}\text{Tr}\big(A_\mu\p_\nu A_\rh
-\f{2i}{3} A_\mu A_\nu A_\rh \big) + \mD^\mu \bph\mD_\mu \ph+m_0^2 \bph \ph- \bps
(i\slashed{\mD}+m_0) \ps \nn\\
&+\f{4\pi wm_0}{\ka} (\bph\ph)^2+ \f{4\pi^2 w^2}{\ka^2} (\bph\ph)^3-\f{2\pi}{\ka}(1+w)
(\bph\ph)(\bps\ps)-\f{2\pi w}{\ka} (\bps\ph)(\bph\ps)\nn\\
& +\f{\pi}{\ka}(1-w) \big((\bph\ps)(\bph\ps)+(\bps\ph)(\bps\ph)\big)\biggr)
\end{align}
displaying that \eqref{action} is the action for one fundamental boson and
one fundamental fermion coupled to a Chern-Simons gauge field. Supersymmetry
sets the masses of the bosonic and fermionic fields equal, and imposes 
several relations between a priori independent coupling constants.

\subsection{Conjectured Duality}\label{dualitycon}

It has been conjectured \cite{Jain:2013gza} that the theory 
\eqref{action} enjoys a strong weak coupling self duality. The theory \eqref{action} with 
$\lq$t Hooft coupling $\lambda$ and self coupling parameter $w$ is conjectured 
to be dual to the theory with $\lq$t Hooft coupling $\lambda'$ and self coupling 
$w'$ where
\beq\label{dtransf}
\la'=\la -\text{Sgn}(\la)\ , \ w'= \f{3-w}{1+w} \, ~~~ m_0'= \f{-2
m_0}{1+w}\ .
\eeq
As we will explain below, the pole mass for the matter multiplet in this 
theory is given by 
\begin{equation}\label{exactmass}
m=\f{2 m_0}{2+(-1+w) \la \ \text{Sgn(m)}} \ .
\end{equation}
It is easily verified that under duality
\begin{equation}\label{mop}
m'=-m\ .
\end{equation}

The concrete prior evidence for this duality is the perfect matching of 
$S^2$ partition functions of the two theories. This match works provided \cite{Jain:2013gza} 
\begin{equation}\label{cond}
\lambda m(m_0, w) \geq 0\ ,
\end{equation}
Through this paper we will assume 
that \eqref{cond} is obeyed. Note that the condition \eqref{cond} is preserved by duality (i.e. a
theory and its conjectured dual either both obey or both violate \eqref{cond}). 

Note that $w=1$ is a fixed point for the duality map \eqref{dtransf}; this 
was necessary on physical grounds (recall that our theory has enhanced
${\cal N}=2$ supersymmetry only at $w=1$). In the special case $w=1$ and 
$m_0=0$, the duality conjectured in this subsection reduces to
the previously studied duality \cite{Benini:2011mf}  (a variation on Giveon- 
Kutasov duality \cite{Giveon:2008zn}). Over the last few years this 
supersymmetric duality has been subjected to (and has successfully passed)
several checks performed with the aid of supersymmetric localization, including 
the matching of three sphere partition function, superconformal indices 
and Wilson loops on both sides of the
duality \cite{Drukker:2008zx,Chen:2008bp,Bhattacharya:2008bja,Kim:2009wb,Kapustin:2009kz,
Marino:2011nm, Aharony:2012ns} .

\subsection{Properties of free solutions of the Dirac equation}

In subsequent subsections we will investigate the constraints
imposed supersymmetry on the $S$ matrices of the the theory \eqref{action}. 
Our analysis will make heavy use of the properties of the free solutions 
to Dirac's equations, which we review in this subsection.

Let $u_\alpha$ and $v_\alpha$ are positive and negative energy solutions to 
Dirac's equations with mass $m$. Let  
$p^\mu=( \sqrt{m^2+ {\mbp}^2}, {\mbp})$. Then $u_\alpha$ and $v_\alpha$ obey 
\begin{align}\label{diracuv}
 ({\slashed p} - m) u(p) = 0 \ ,\\\nn
 ({\slashed p} + m) v(p) = 0 \ .
\end{align}
We choose to normalize these spinors so that  
\begin{equation}\label{normalz}
 \begin{split}
   {\bar u}({\mathbf p})\cdot u({\mathbf p}) &= -2m ~~~~~~~~~~~~~~~ {\bar v}({\mathbf p})\cdot
v({\mathbf p}) = 2m \\
   u({\mathbf p}) u^*({\mathbf p}) &= -({\slashed p} + m)C ~~~~~~~ v({\mathbf p}) v^*({\mathbf p})=
-({\slashed p} - m)C\ .
 \end{split}  
\end{equation}
$C$ in \eqref{normalz} is the charge conjugation matrix defined to obey
 the equation 
\begin{equation}\label{cc}
C \gamma^\mu C^{-1} = - (\gamma^\mu)^T \ .
\end{equation}
Throughout this paper we use $\gamma$ matrices that obey 
the algebra\footnote{We use the mostly plus convention for $\eta_{\mu\nu}$, the corresponding
Euclidean algebra obeys $\{\ga^\mu,\ga^\nu\}=-2\de^{\mu\nu}$. See Appendix \S \ref{gamma} for
explicit representations of the $\ga$ matrices and charge conjugation matrix $C$.}
\begin{equation}\label{gm}
\{\ga^\mu,\ga^\nu\}=-2\eta^{\mu\nu}\ .
\end{equation}
We also choose all three $\gamma^\mu$ matrices to be purely imaginary \footnote
{This 
is possible in 3 dimensions; recall the unconventional choice of sign in 
\eqref{gm}.}
and to obey 
\begin{equation}\label{gcc}
(\gamma^\mu)^\dagger= -\eta^{\mu\mu} \gamma^\mu   ~~~{\rm no ~~sum}
\end{equation}
with these conventions it is easily verified that $C=\gamma^0$ obeys 
\eqref{cc} and so we choose  
$$C= \gamma^0\ .$$

Using the conventions spelt out above, it is easily verified that 
$u(\mbp)$ and $v^*(\mbp)$ obey the same equation (i.e. complex conjugation flips the 
two equations in \eqref{diracuv}), and have the same normalization. It follows
that it is possible to pick the (as yet arbitrary) phases of $u(\mbp)$ and 
$v(\mbp)$ to ensure that 
\begin{equation}\label{starprop}
u_\alpha(\mbp)= - v_\alpha^*(\mbp), ~~~v_\alpha(\mbp)=-u_\alpha^*(\mbp)
\end{equation}
\footnote{Note that ${\bar u}^\alpha = u^{*\alpha}=C^{\alpha\beta}u^*_\beta$ and not
$(u^\alpha)^*$. Thus, $(u^{*\alpha})^* = -u^\alpha$, where we have used 
the fact that $C=\gamma^0$ is imaginary. Similarly 
$(u^\alpha)^* = -u^{*\alpha}$. Likewise for $v$. Care should be taken while 
complex conjugating dot products of spinors, for instance 
$(v^*({\mathbf p}_i)v^*({\mathbf p}_j))^*=-(v({\mathbf p}_i)v({\mathbf p}_j))$, 
$(u({\mathbf p}_i)u({\mathbf p}_j))^*=-(u^*({\mathbf p}_i)u^*({\mathbf p}_j))$, 
and so on. }. We will adopt the choice \eqref{starprop} throughout our paper.

Notice that the replacement $m \to -m$ interchanges the equations for 
$u$ and $v$. It follows that $u(m) \propto v(-m)$. Atleast with the choice of 
phase that we adopt in this paper (see below) we find
\begin{equation}\label{uvm}
u(m, p)= - v(-m, p), ~~~v(m, p)=-u(-m, p)\ .
\end{equation}

To proceed further it is useful to make a particular choice of $\gamma$ matrices
and to adopt a particular choice of phase for $u$. We choose the $\gamma^\mu$ 
matrices listed in \S \ref{gamma} and take $u({\mbp})$ and $v({\mbp})$ 
to be given by
\begin{align}\label{uvsol}
& u(\mbp)=\begin{pmatrix}
         -\sqrt{p^0-p^1}\\
          \f{p^3+im}{\sqrt{p^0-p^1}}
        \end{pmatrix} \ , \
\bar{u}(\mbp)=\begin{pmatrix}
               \f{i p^3 +m }{\sqrt{p^0-p^1}} & i \sqrt{p^0-p^1}
              \end{pmatrix}\ , \nn\\
& v(\mbp)=\begin{pmatrix}
         \sqrt{p^0-p^1}\\
          \f{-p^3+im}{\sqrt{p^0-p^1}}
        \end{pmatrix} \ , \
\bar{v}(\mbp)=\begin{pmatrix}
               \f{-i p^3 +m }{\sqrt{p^0-p^1}} & -i \sqrt{p^0-p^1}
              \end{pmatrix}\ ,
\end{align}
where 
$$ p^0=+\sqrt{m^2+{\mbp}^2}\ .$$
Notice that the arguments of the square roots in \eqref{uvsol} are always 
positive; the square roots in \eqref{uvsol} are defined to be positive 
(i.e. $\sqrt{x^2}=|x|$). It is easily verified that the solutions 
\eqref{uvsol} respect \eqref{uvm} as promised. 

In the rest of this section we discuss an analytic rotation of the spinors 
to complex (and in particular negative) values of the $p^\mu$ (and in particular $p^0$). 
This formal construction will prove useful in the study of the 
transformation properties of the $S$ matrix under crossing symmetry.

Let us define $$\sqrt{a e^{i \al}}= |\sqrt{a}|e^{i \frac{ \al}{2}}.$$
Clearly our function is single valued only on a double cover of 
the complex plane. In other words our square root function is well defined 
if $\alpha$ is specified modulo $4 \pi$, but is not well defined if 
$\alpha$ is specified modulo $2 \pi$. We define
\begin{equation}\label{anrot} \begin{split} 
u(\mbp, \alpha)&= u(e^{i \alpha} p^\mu)=\begin{pmatrix}
         -e^{i \frac{\alpha}{2}} \sqrt{p^0-p^1}\\
          \frac{p^3 e^{ i \frac{\alpha}{2}} +im e^{-i \frac{\alpha}{2}}}{\sqrt{p^0-p^1}}
        \end{pmatrix} \ , \  \\
 v(\mbp, \alpha)&=v(e^{i \alpha} p^\mu)=- \begin{pmatrix}
         -e^{i \frac{\alpha}{2}}\sqrt{p^0-p^1}\\
           \frac{p^3 e^{ i \frac{\alpha}{2}} -im e^{-i \frac{\alpha}{2}}}{\sqrt{p^0-p^1}}
        \end{pmatrix} \ , \ \\
u^*(\mbp, \alpha)&= \begin{pmatrix}
         -e^{-i \frac{\alpha}{2}} \sqrt{p^0-p^1}\\
          \frac{p^3 e^{ -i \frac{\alpha}{2}} -im e^{i \frac{\alpha}{2}}}{\sqrt{p^0-p^1}}
        \end{pmatrix} \ , \  \\
 v^*(\mbp, \alpha)&=- \begin{pmatrix}
         -e^{-i \frac{\alpha}{2}}\sqrt{p^0-p^1}\\
           \frac{p^3 e^{ -i \frac{\alpha}{2}} +im e^{+i \frac{\alpha}{2}}}{\sqrt{p^0-p^1}}
        \end{pmatrix} \ , \
\end{split}
\end{equation} 
with $\alpha \in [0, 4 \pi)$. 
It follows immediately from these definitions that 
\begin{eqnarray}\label{uvrel}
u(\mbp, \alpha+\pi) =& -i v(\mbp, \alpha)\ , \ v(\mbp,\al+\pi)&= -i u(\mbp,\al)\ , \nn\\
u(\mbp,\al-\pi) = &i v(\mbp,\al)\ , \ v(\mbp,\al-\pi)&=i u(\mbp,\al)\ ,\nn\\
u^*(\mbp, \alpha) =& -v(\mbp, -\alpha)\ , \ v^*(\mbp, \alpha)&=-u(\mbp, -\alpha)\ .
\end{eqnarray}

Notice, in particular, that the choice $\alpha=\pi$ and 
$\alpha=-\pi$ both amount to the replacement of $p^\mu$ with - $p^\mu$. 
Note also that the complex conjugation of $u(p, \alpha)$ is equal to the 
function $u^*(p)$ with $p$ rotated by $-\alpha$.

\subsection{Constraints of supersymmetry on scattering}\label{susyscat}

In this paper we will study $2 \times 2$ scattering of particles in an 
${\cal N}=1$ supersymmetric field theory. In this subsection we set up our
conventions and notations and 
explore the constraints of supersymmetry on scattering amplitudes. 

Let us consider the scattering process 
\begin{equation}
1+2 \rightarrow 3+4
\end{equation}
where $1, 2$ represent initial state particles and $3,4$ are final state 
particles. Let the $i^{th}$ particle be associated with the superfield 
$\Phi_i$. As a scattering amplitude represents the transition between 
free incoming and free outgoing onshell particles, the initial and final 
states of $\Phi_i$ are effectively subject to the free equation of motion
\begin{equation}\label{feom}
\left( D^2 + m_i \right) \Phi_i=0 \ 
\end{equation}
where $D^2=\f{1}{2}D^\al D_\al$. The general solution to this free 
equation of motion is 
\begin{align} \label{pohos}
\Phi(x,\te)= \int \f{d^2p}{\sqrt{2 p^0}(2\pi)^2 }
\bigg[\bigg(& a(\mathbf{p}) (1+m\te^2)+\te^\al
u_\al(\mathbf{p}) \al(\mathbf{p}) \bigg)e^{i p.x}\nn\\
&+\bigg(a^{c\dag}(\mathbf{p})(1+m\te^2)+\te^\al
v_\al(\mathbf{p})\al^{c\dag}(\mathbf{p})\bigg) e^{-i p.x}\bigg]
\end{align}
where $a/ a^\dagger$ are annihilation/creation operator for the bosonic 
particles  and $\alpha/\alpha^\dagger$ are annihilation/creation operators for 
the fermionic particles respectively
\footnote{Similarly $a^c/ a^{c\dagger}$ and $\alpha^c/\alpha^{c\dagger}$ are the
annihilation/creation operators for the bosonic and fermionic anti-particles respectively.}. The
bosonic and fermionic 
oscillators obey the commutation relations 
\begin{equation}\label{bcr}
[a(\mathbf{p}), a^\dagger (\mathbf{p}')]= (2 \pi)^2 \delta^2( \mathbf{p}-\mathbf{p'}),
~~~[a(\mathbf{p}), a^\dagger(\mathbf{p'})]= 
(2 \pi)^2 \delta^2 \left(\mathbf{p}-\mathbf{p'} \right) \ .
\end{equation}
($a^c$ and $\alpha^c$ obey analogous commutation relations). 

The action of the supersymmetry operator on a free onshell superfield 
is simple 
\begin{eqnarray}\label{osactfs}
\begin{split}
[Q_\alpha, \Phi_i] &=\\ Q_\alpha \Phi_i &=  i \int \f{d^2p}{(2\pi)^2
\sqrt{2p^0}}\bigg[\bigg(u_\alpha({\mathbf p})(1+m\te^2)\alpha({\mathbf
p})+\theta^\beta(-u_\beta({\mathbf p}) u^*_\alpha({\mathbf p})) a({\mathbf p})\bigg)e^{i
p.x}\nn\\
&+\bigg(v_\alpha({\mathbf p})(1+m\te^2)\alpha^{c\dagger}({\mathbf p})+\theta^\beta(v_\beta({\mathbf
p}) v^*_\alpha({\mathbf p})) a^{c\dagger}({\mathbf p})\bigg) e^{-i p.x}\bigg]\ .
\end{split}
\end{eqnarray}
In other words, the action of the supersymmetry generator on onshell 
superfields is given by 
\begin{equation} \begin{split}
-i Q_\alpha & = u_\alpha({\mathbf
p_i})\left(a \partial_\alpha + a^c \partial_{\alpha^c} \right)
+u_\alpha^* ({\mathbf p}_i) \left( 
-\alpha \partial_a + \alpha^c \partial_{a^c}  \right)\\
&+ v_\alpha({\mathbf p}_i) \left(a^\dagger \partial_\alpha^\dagger 
+(a^c)^\dagger \partial_{(\alpha^c)^\dagger} \right) 
+v_\alpha^*({\mathbf p}_i) \left(\alpha^\dagger \partial_a^\dagger 
+(\alpha^c)^\dagger \partial_{(a^c)^\dagger} \right)\ .\\
\end{split}
\end{equation}

The explicit action of $Q_\alpha$ on onshell superfields may be repackaged as 
follows. Let us define a superfield of 
annihilation operators, and another superfield for creation operators: 
\begin{equation}\label{casfield}\begin{split}
A_i({\mathbf p})= a_i({\mathbf p}) + \alpha_i({\mathbf p})\theta_i\ , \\
A^\dagger_i({\mathbf p}) = a^\dagger_i({\mathbf p}) + \theta_i \alpha_i^\dagger({\mathbf p})\ .\\
\end{split}
\end{equation}
Here $\theta_i$ is a new formal superspace parameter ($\theta_i$
has nothing to do with the $\theta_\alpha$ that appear in the superfield 
action \eqref{action} ). It follows from \eqref{osactfs} and 
\eqref{casfield} that 
\begin{eqnarray}\label{Qactdef} \begin{split}
[Q_\alpha, A_i({\mathbf p}_i, \theta_i)]&= Q^1_\alpha
A_i({\mathbf p}_i, \theta_i)\\
[Q_\alpha, A^\dagger_i({\mathbf p}_i, \theta_i)]&=  Q^2_\alpha
A^\dagger_i({\mathbf p}_i,
\theta_i)
\end{split}
\end{eqnarray}
where 
\begin{equation}\label{dedef} \begin{split} 
 Q^1_\be &=i\left(-u_\be({\mathbf p})\overrightarrow{\f{\partial}{\partial\theta}} + u_\be^*({\mathbf p})\theta\right)\\
 Q^2_\be&=i\left(v_\be({\mathbf p})
\overrightarrow{\f{\partial}{\partial\theta}} +
v_\be^*({\mathbf p})\theta\right)\ .
\end{split}
\end{equation}

We are interested in the $S$ matrix 
\begin{align}\label{sma} 
S({\mathbf p}_1, \theta_1, {\mathbf p}_2, \theta_2, {\mathbf p}_3, \theta_3, 
{\mathbf p}_4, \theta_4) \sqrt{(2p_1^0) (2p_2^0)(2p_3^0)(2p_4^0)} =\nn\\
\langle 0 | A_4( {\mathbf p}_4, \theta_4) A_3( {\mathbf p}_3, \theta_3) U A_2^\dagger({\mathbf p}_2,
\theta_2) A_1^\dagger({\mathbf p}_1, \theta_1) | 0 \rangle 
\end{align}
where $U$ is an evolution operator (the RHS denotes the transition amplitude 
from the in state with particles 1 and 2 to the out state with particles 3 
and 4). 

The condition that the $S$ matrix defined in \eqref{sma} is invariant under 
supersymmetry follows from the action of supersymmetries on oscillators 
given in \eqref{osactfs}. The resultant equation for the $S$ matrix may be 
written in terms of the operators defined in \eqref{dedef} as 
\begin{align}\label{susyinv}
\biggl(\overrightarrow{Q}^1_{\alpha}(\mbp_1,\te_1)
 &+ \overrightarrow{Q}^1_{\alpha}(\mbp_2,\te_2)\nn\\ 
&+ \overrightarrow{Q}^2_{\alpha}(\mbp_3, \te_3)
+ \overrightarrow{Q}^2_{\alpha}(\mbp_4, \te_4) \biggr) 
S({\mathbf p}_1, \theta_1, {\mathbf p}_2, \theta_2, {\mathbf p}_3, \theta_3, 
{\mathbf p}_4, \theta_4) = 0\ .
\end{align}

We have explicitly solved  \eqref{susyinv}; the solution\footnote{The superspace $S$ matrix
\eqref{SUSYSmatrix} encodes different processes allowed by supersymmetry in the theory. 
In particular, the presence of grassmann parameters indicates fermionic in $(\te_1,\te_2)$
and fermionic out $(\te_3,\te_4$) states. The absence of grassmann parameter indicates a bosonic
in/out state. Thus, the no $\te$ term $\SB$ encodes the $2 \to 2$ $S$ matrix for a purely bosonic
process, while the four $\te$ term $\SF$ encodes the $2\to 2$ $S$ matrix of a purely fermionic
process. Note in particular that $S$ matrices corresponding to all other $2 \to 2$ processes that
involve both bosons and fermions are completely determined in terms of the $S$ matrices $\SB$ and
$\SF$ together with \eqref{cdef} and \eqref{cstdef}. } is given by 
\begin{multline}\label{SUSYSmatrix}
S({\mathbf p}_1, \theta_1, {\mathbf p}_2, \theta_2, {\mathbf p}_3, \theta_3, 
{\mathbf p}_4, \theta_4) =
\SB + \SF ~\theta_1\theta_2\theta_3\theta_4 + \left(\half C_{12}\SB-\half
C^*_{34}\SF\right)~\theta_1\theta_2 \\+ \left(\half C_{13}\SB - \half
C^*_{24}\SF\right)~\theta_1\theta_3 + \left(\half C_{14}\SB + \half
C^*_{23}\SF\right)~\theta_1\theta_4 + \left(\half C_{23}\SB + \half
C^*_{14}\SF\right)~\theta_2\theta_3 \\+ \left(\half C_{24}\SB - \half
C^*_{13}\SF\right)~\theta_2\theta_4 + \left(\half C_{34}\SB - \half
C^*_{12}\SF\right)~\theta_3\theta_4
\end{multline}
where
\begin{align}\label{cdef}
\half C_{12}&= -\frac{1}{4m}v^*({\mathbf p}_1)v^*({\mathbf p}_2) & \half C_{23}&= -\frac{1}{4m}v^*({\mathbf p}_2)u^*({\mathbf p}_3) &\nn\\\half C_{13}&= -\frac{1}{4m}v^*({\mathbf p}_1)u^*({\mathbf p}_3) &
\half C_{24}&= -\frac{1}{4m}v^*({\mathbf p}_2)u^*({\mathbf p}_4) &\nn\\\half C_{14}&= -\frac{1}{4m}v^*({\mathbf p}_1)u^*({\mathbf p}_4) &
\half C_{34}&= -\frac{1}{4m}u^*({\mathbf p}_3)u^*({\mathbf p}_4)
\end{align}
and
\begin{align}\label{cstdef}
\half C^*_{12}&= \frac{1}{4m}v({\mathbf p}_1)v({\mathbf p}_2) & \half C^*_{23}&= \frac{1}{4m}v({\mathbf p}_2)u({\mathbf p}_3) &\nn\\\half C^*_{13}&= \frac{1}{4m}v({\mathbf p}_1)u({\mathbf p}_3) &
\half C^*_{24}&= \frac{1}{4m}v({\mathbf p}_2)u({\mathbf p}_4) &\nn\\\half C^*_{14}&= \frac{1}{4m}v({\mathbf p}_1)u({\mathbf p}_4) &
\half C^*_{34}&= \frac{1}{4m}u({\mathbf p}_3)u({\mathbf p}_4)
\end{align}
Note that the general solution to \eqref{susyinv} is given in terms of 
two arbitrary functions $\SB$ and $\SF$ of the four momenta; 
\eqref{susyinv} determines the remaining six functions in the general expansion of the $S$ matrix in
terms of these two functions. See Appendix 
\ref{manN1ex} for a check of these relations from another viewpoint 
(involving offshell supersymmetry of the effective action, see section \S\ref{off4pt})

Although we are principally interested in ${\cal N}=1$ supersymmetric theories
in this paper, we will sometimes study the special limit $w=1$ in which 
\eqref{action} enjoys an enhanced ${\cal N}=2$ supersymmetry. In this 
case the additional supersymmetry further constrains the $S$ matrix. 
In Appendix \ref{manN2} we demonstrate that the additional supersymmetry 
determines $\SB$ in terms of $\SF$. In the ${\cal N}=2$ case, in other words, 
all components of the $S$ matrix are determined by supersymmetry in terms of 
the four boson scattering matrix. 

\subsection{Supersymmetry and dual supersymmetry} \label{dsusy}

The strong weak coupling duality we study in this paper is conjectured to be
a Bose-Fermi duality. In other words
\begin{equation}\label{bfint}
a^D=\alpha, ~~~\alpha^D= a
\end{equation}
together with a similar exchange of bosons and fermions for creation 
operators (the superscript $D$ stands for `dual'). 
Suppose we define
\begin{equation}\label{casfieldd}\begin{split}
A^D_i({\mathbf p})= a_i^D({\mathbf p}) + \alpha_i^D({\mathbf p})\theta_i \ ,\\
(A^D)^\dagger_i({\mathbf p}) = (a^D)^\dagger_i({\mathbf p}) 
+ \theta_i (\alpha_i^D)^\dagger({\mathbf p})\ .\\
\end{split}
\end{equation}
The dual supersymmetries must act in the same way on $A^D$ and 
$(A^D)^\dagger$ as ordinary supersymmetries act on $A$ and $A^D$. In other
words the action of dual supersymmetries on $A^D$ and $(A^D)^\dagger$ is given 
by 
\begin{eqnarray}\label{Qactdefd} \begin{split}
[Q^D_\alpha, A^D_i({\mathbf p}_i, \theta_i)]&= (Q^D)^1_\alpha
A^D_i({\mathbf p}_i, \theta_i)\ ,\\
[Q^D_\alpha, (A^D)^\dagger_i({\mathbf p}_i, \theta_i)]&=  (Q^D)^2_\alpha
(A^D)^\dagger_i({\mathbf p}_i,
\theta_i)\ ,
\end{split}
\end{eqnarray}
where 
\begin{equation}\label{dedefd} \begin{split} 
 (Q^D)^1_\be &=i\left(-u_\be({\mathbf p}, -m)\overrightarrow{\f{\partial}{\partial\theta}} -
v_\be({\mathbf p}, -m)\theta\right)\ ,\\
 (Q^D)^2_\be&=i\left(v_\be({\mathbf p}, -m)
\overrightarrow{\f{\partial}{\partial\theta}} -
u_\be({\mathbf p}, -m)\theta\right)\ .
\end{split}
\end{equation}
The spinors in \eqref{dedefd} are all evaluated at $-m$ as duality flips the
sign of the pole mass. 

The action of the dual supersymmetries on $A$ and $A^\dagger$ is obtained from 
\eqref{dedefd} upon performing the interchange 
$ \theta  \leftrightarrow \partial_\theta$ (this accounts for the interchange
of bosons and fermions). Using also \eqref{uvm} we find that 
\begin{eqnarray}\label{Qactdefdu} \begin{split}
[Q^D_\alpha, A_i({\mathbf p}_i, \theta_i)]&= -Q^1_\alpha
A^D_i({\mathbf p}_i, \theta_i)\ ,\\
[Q^D_\alpha, A^\dagger_i({\mathbf p}_i, \theta_i)]&=  Q^2_\alpha
(A^D)^\dagger_i({\mathbf p}_i,
\theta_i)\ .
\end{split}
\end{eqnarray}
It follows, in particular, that an $S$ matrix invariant under the 
usual supersymmetries is automatically invariant under dual supersymmetries. 
In other words onshell supersymmetry `commutes' with duality.

\subsection{Naive crossing symmetry and supersymmetry}\label{naivec}

Let us define the analytically rotated supersymmetry operators \footnote{Note that the notation
$u_\be^*(\mbp,-\al)$ means that the analytically rotated function of $u^*$ in \eqref{anrot} is
evaluated at the phase $-\al$.}
\begin{equation}\label{dedefc} \begin{split} 
 Q^1_\be(\mbp, \al,\te) &=i\left(-u_\be({\mathbf p},
\al)\overrightarrow{\f{\partial}{\partial\theta}} + u_\be^*({\mathbf p}, -\al)\theta\right)\ ,\\
 Q^2_\be(\mbp, \al,\te) &=i\left(v_\be({\mathbf p},
\al)\overrightarrow{\f{\partial}{\partial\theta}} +
v_\be^*({\mathbf p}, -\al)\theta\right)\ .
\end{split}
\end{equation}
It is easily verified from these definitions that 
\begin{equation}\label{qotr}
Q^2_\alpha(\mbp, 0,-i\te)=Q^1_\alpha(\mbp, \pi, \te) \ .
\end{equation}

Using \eqref{qotr} the equation \eqref{susyinv} may equivalently be written as
\begin{align}\label{susyinvt}
\biggl(\overrightarrow{Q}^1_{\alpha}(\mbp_1, 0, \te_1)
 &+ \overrightarrow{Q}^1_{\alpha}(\mbp_2,0, \te_2)\nn\\
 &+ \overrightarrow{Q}^1_{\alpha}(\mbp_3, \pi, \te_3)
+ \overrightarrow{Q}^1_{\alpha}(\mbp_4, \pi, \te_4) \biggr) 
S({\mathbf p}_1, \theta_1, {\mathbf p}_2, \theta_2, {\mathbf p}_3, - i\theta_3, 
{\mathbf p}_4, - i \theta_4)=0
\end{align}
with $p_1 + p_2 = p_3 + p_4.$

The constraints of supersymmetry on the $S$ matrix are consistent with (naive) 
crossing symmetry. In order to make this manifest, we define a `master' function $S_M$
$$S_M(\mbp_1, \phi_1, \theta_1, \mbp_2, \phi_2, \theta_2,
\mbp_3, \phi_3, \theta_3, \mbp_4, \phi_4, \theta_4)\ .$$
The master function $S_M$ is defined so that 
\begin{equation}\label{relms}
S({\mathbf p}_1, \theta_1, {\mathbf p}_2, \theta_2, {\mathbf p}_3, -i \theta_3, 
{\mathbf p}_4, -i \theta_4) = S_M(\mbp_1, 0, \theta_1, \mbp_2, 0, \theta_2,
\mbp_3, \pi, \theta_3, \mbp_4, \pi, \theta_4) 
\end{equation}
In other words $S_M$ is $S$ with the replacement $-i \theta_3 \rightarrow \theta_3$, $-i\theta_4 \rightarrow \theta_4$, analytically rotated to general values 
of the phase $\phi_1$, $\phi_2$, $\phi_3$ and $\phi_4$. It follows from  
\eqref{susyinvt} that the master equation $S_M$ obeys the completely 
symmetrical supersymmetry equation

\begin{align}\label{susyinvsym}
\biggl(\overrightarrow{Q}^1_{\alpha}(\mbp_1, \phi_1, \te_1)
 & + \overrightarrow{Q}^1_{\alpha}(\mbp_2,\phi_2,
\te_2)+\overrightarrow{Q}^1_{\alpha}(\mbp_3,\phi_3, \te_3)\nn\\
&+ \overrightarrow{Q}^1_{\alpha}(\mbp_4, \phi_4, \te_4) \biggr) 
S_M({\mathbf p}_1, \phi_1, \theta_1, {\mathbf p}_2, \phi_2, \theta_2, {\mathbf p}_3, \phi_3,
\theta_3,{\mathbf p}_4, \phi_4, \theta_4)=0
\end{align}

The function $S_M$ encodes the scattering matrices in all channels. In order 
to extract the $S$ matrix for $ p_i + p_j \rightarrow p_k+p_m$ with 
$p_i+p_j =p_k+p_m$ (with (i, j, k, m) being any permutation of (1, 2, 3, 4)) 
we simply evaluate the function $S_M$ with $\phi_i$ and $\phi_j$ set to zero, 
$\phi_k$ and $\phi_m$ set to $\pi$, $\theta_i$ and $\theta_j$ left unchanged and
$\theta_k$ and $\theta_m$ replaced by $i \theta_k$ and $i \theta_m$. 
The fact that the master equation obeys an equation that is symmetrical in 
the labels $1, 2, 3, 4$ is the statement of (naive) crossing symmetry.

The solution to the differential equation \eqref{susyinvsym} is 
\begin{align}
 S_M({\mathbf p}_1, \phi_1, \theta_1, &{\mathbf p}_2, \phi_2, \theta_2, {\mathbf p}_3, \phi_3,
\theta_3,{\mathbf p}_4, \phi_4, \theta_4)= \tilde{\SB} + \tilde{\SF}\te_1\te_2\te_3\te_4 \nn\\
&+  \f{\tilde{\SB}}{4}\sum_{i,j=1}^4D_{ij}(\mbp_i,\phi_i,\mbp_j,\phi_j) \te_i\te_j
-\f{\tilde{\SF}}{8}\sum_{i,j,k,l=1}^4\ep^{ijkl}\tilde{D}_{ij}(\mbp_i,\phi_i,\mbp_j,\phi_j)
\te_k\te_l\label{SUSYmaster}
\end{align}
where
\begin{align}
\f{1}{2} D_{ij}(\mbp_i,\phi_i,\mbp_j,\phi_j)&=-\frac{1}{4m}u^*({\mathbf p}_i,-\phi_i)u^*({\mathbf
p}_j,-\phi_j)\ , \nn\\
\f{1}{2}\tilde{D}_{ij}(\mbp_i,\phi_i,\mbp_j,\phi_j)&= \frac{1}{4m}u({\mathbf p}_i,\phi_i)u({\mathbf
p}_j,\phi_j)\ .\label{cstdefmaster}
\end{align}
In the above equations `$*$' means complex conjugation and the spinor indices are contracted from
NW-SE as usual. To summarize, $S_M$ obeys the supersymmetric ward identity 
and is completely solved in terms of two analytic functions 
$\tilde{\SB}(\mbp_1,\mbp_2,\mbp_3,\mbp_4)$ and $\tilde{\SF}(\mbp_1,\mbp_2,\mbp_3,\mbp_4)$ of 
the momenta.

As we have explained under \eqref{susyinvsym}, the $S$ matrix
corresponding to scattering processes in any given channel can be simply 
extracted out of $S_M$. For example, let $S$ denote the 
the $S$ matrix in the channel with
$p_1,p_2$ as in-states and $p_3,p_4$ as out-states. Then 
\begin{equation}\label{sdeo}
S(\mbp_1,\te_1,\mbp_2,\te_2,\mbp_3,\te_3,\mbp_4,\te_4)=S_M({\mathbf p}_1, \pi, i\theta_1,
{\mathbf p}_2, \pi, i\theta_2, {\mathbf p}_3, 0,
\theta_3,{\mathbf p}_4, 0, \theta_4)\ .
\end{equation}
It is easily verified that \eqref{cstdefmaster} together with \eqref{uvrel} 
imply \eqref{cdef}.

Notice that \eqref{sdeo} maps ${\tilde \SB}$ to $\SB$ while ${\tilde \SF}$ 
is mapped to $-\SF$ \footnote{Of course ${\tilde \SB}$ and ${\tilde \SF}$
are evaluated at $\phi_1=\phi_2=\pi$ while $\SB$ and $\SF$ are evaluated 
at $\phi_1=\phi_2=0$; roughly speaking this amounts to the replacement 
$p_1^\mu \rightarrow -p_1^\mu$, $p_2^\mu \rightarrow -p_2^\mu$.}. The minus 
sign in the continuation of $\SF$ has an interesting explanation. The 
four fermion amplitude $\SF$ has a phase ambiguity. This ambiguity 
follows from the fact that $\SF$ is the overlap of initial and final 
fermions states. These initial and final states are written in terms of 
the spinors $u_\alpha$ and $v_\alpha$, which are defined as appropriately 
normalized solutions of the Dirac equation are inherently ambiguous upto a 
phase. It is easily verified that the quantity
$$\left(u^*(\mbp_1,-\phi_1) u(\mbp_3,\ph_3) \right)\left(u^*(\mbp_2,-\phi_2)
u(\mbp_4,\ph_4)\right)$$ 
has the same phase ambiguity as $\SF$. If we define an auxiliary quantity 
$\tilde{\Sf}$ by the equation
\beq\label{Fdefmas}
\tilde{\SF}=-\f{1}{4m^2} \left(u^*(\mbp_1,-\phi_1) u(\mbp_3,\ph_3) \right)\left(u^*(\mbp_2,-\phi_2)
u(\mbp_4,\ph_4)\right){\tilde \Sf}
\eeq
and $\Sf$ by 
\beq\label{Fdef2}
\SF = -\frac{1}{4m^2}\left(u^*(\mbp_1)u(\mbp_3)\right) \left( u^*(\mbp_2) u(\mbp_4) \right)\Sf
\eeq
then the phases of $\Sf$ and $\tilde{\Sf}$ are unambiguous and so 
potentially physical. As the quantity  
$$\left(u^*(\mbp_1,-\phi_1) u(\mbp_3,\ph_3) \right)\left(u^*(\mbp_2,-\phi_2)
u(\mbp_4,\ph_4)\right)$$
picks up a minus sign under the phase rotation that takes us from $S_M$ to $S$. 
It follows that $\tilde{\Sf}$ rotates to $\Sf$ with no minus sign.

\subsection{Properties of the convolution operator}

Like any matrices, $S$ matrices can be multiplied. The multiplication rule 
for two $S$ matrices, $S_1$ and $S_2$, expressed as functions in onshell 
superspace is given by 
\begin{align}\label{sost} 
S_1 \star S_2
\equiv 
\int d\Gamma S_1({\mathbf p}_1, \theta_1, {\mathbf p}_2, \theta_2, {\mathbf k}_3, \phi_1, 
{\mathbf k}_4, \phi_2)\exp(\phi_1\phi_3+\phi_2\phi_4)2k_1^0(2\pi)^2\delta^{(2)}({\mathbf
k}_3-{\mathbf k}_1)\nn\\2k_2^0(2\pi)^2\delta^{(2)}({\mathbf k}_4-{\mathbf k}_2)S_2 
({\mathbf k}_1,
\phi_3, {\mathbf k}_2, \phi_4, {\mathbf p}_3, \theta_3, 
{\mathbf p}_4, \theta_4)
\end{align}
where the measure $d\Gamma$ is
\beq\label{mesdef}
 d\Gamma = \frac{d^2 k_3}{2k_3^0(2\pi)^2}\frac{d^2 k_4}{2k_4^0(2\pi)^2}\frac{d^2
k_1}{2k_1^0(2\pi)^2}\frac{d^2 k_2}{2k_2^0(2\pi)^2}d\phi_1 d\phi_3 d\phi_2 d\phi_4 \ .
\eeq

It is easily verified that the onshell superfield $I$
\begin{align} \label{identop}
I({\mathbf p}_1, \theta_1, {\mathbf p}_2, \theta_2, {\mathbf p}_3, \te_3, 
{\mathbf p}_4, \te_4) & = \exp(\theta_1\theta_3 +
\theta_2\theta_4) I(\mbp_1,\mbp_2,\mbp_3,\mbp_4)\nn\\
I(\mbp_1,\mbp_2,\mbp_3,\mbp_4) & =2p_3^0(2\pi)^2\delta^{(2)}({\mathbf p}_1-{\mathbf
p}_3) 2p_4^0(2\pi)^2\delta^{(2)}({\mathbf p}_2-{\mathbf p}_4)
\end{align}
is the identity operator under this multiplication rule, i.e. 
\begin{equation}
S \star I = I \star S= S
\end{equation}
for any $S$. It may be verified that $I$ defined in \eqref{identop} 
obeys \eqref{susyinv} and so is supersymmetric. 

In Appendix \S \ref{sp} we demonstrate that if $S_1$ and $S_2$ are onshell 
superfields that obey \eqref{susyinv}, then $S_1 \star S_2$ also obeys \eqref{susyinv}. 
In other words the product of two supersymmetric $S$ matrices is also supersymmetric. 

The onshell superfield corresponding to $S^\dagger$ is given in terms of 
the onshell superfield corresponding to $S$ by the equation 
\begin{equation}\label{sdag}
S^\dagger({\mathbf p}_1, \theta_1, {\mathbf p}_2, \theta_2, {\mathbf p}_3, \te_3, 
{\mathbf p}_4, \te_4)= S^*(
{\mathbf p}_3, \theta_3, {\mathbf p}_4, \theta_4, {\mathbf p}_1, \te_1, 
{\mathbf p}_2, \te_2)\ .
\end{equation}

The equation satisfied by $S^\dagger$ can be obtained by complex conjugating and interchanging the
momenta in the supersymmetry invariance condition for $S$ (see \eqref{susyinvS1}). It follows from
the anti-hermiticity of $Q$ that
\beq\label{susyinvSdag}
\left(Q^*_{u({\mathbf p}_1)} + Q^*_{u({\mathbf p}_2)}+Q_{u({\mathbf p}_3)} + Q_{u({\mathbf
p}_4)}\right)S^*({\mathbf p}_3, \theta_3, 
{\mathbf p}_4, \theta_4,{\mathbf p}_1, \theta_3,{\mathbf p}_2, \theta_4)=0
\eeq
which implies  $[Q,S^\dagger]=0$. Thus $S^\dagger$ is supersymmetric if and only if $S$ is
supersymmetric.

\subsection{Unitarity of Scattering}

The unitarity condition 
\beq\label{uni1}
 S S^\dagger = \mathbb{I}
\eeq
may be rewritten in the language of onshell superfields as 
\begin{equation}\label{ossf}
(S \star S^\dagger -I)=0\ .
\end{equation}\footnote{As explained in \cite{Jain:2014nza}, the unitarity equation for 
$2 \times 2$ does not receive contributions from $2 \times n$ scattering
in the large $N$ limits studied in the current paper as well.}

It follows from the general results of the previous subsection that the LHS of 
\eqref{ossf} is supersymmetric, i.e it obeys \eqref{susyinv}. Recall that 
any onshell superfield that obeys \eqref{susyinv} must take the form 
\eqref{SUSYSmatrix} where $\SB$ and $\SF$ are the zero theta and 4 theta terms
in the expansion of the corresponding object. In particular, in order to 
verify that the LHS of \eqref{ossf} vanishes, it is sufficient to verify that 
its zero and 4 theta components vanish. 

Inserting the explicit solutions for $S$ and $S^\dagger$, one finds 
that the no-theta term of \eqref{ossf} is proportional to 
(we have used that $k_3\cdot k_4=p_3\cdot p_4$ onshell)
\begin{IEEEeqnarray}{l}\label{unicond1}
\int\frac{d^2 k_3}{2k_3^0(2\pi)^2}\frac{d^2 k_4}{2k_4^0(2\pi)^2}\left[\SB({\mathbf p}_1,{\mathbf
p}_2,{\mathbf k}_3,{\mathbf k}_4)\SB^*({\mathbf p}_3,{\mathbf p}_4,{\mathbf k}_3,{\mathbf
k}_4)\right.\nn\\\left.-\f{1}{16m^2}\left(2(p_3\cdot p_4 + m^2)\SB({\mathbf p}_1, {\mathbf p}_2,
{\mathbf k}_3, {\mathbf k}_4)\SB^*({\mathbf p}_3, {\mathbf p}_4, {\mathbf k}_3,
{\mathbf k}_4)\right.\right.\nn\\\left.\left.+u^*({\mathbf k}_3)u^*({\mathbf k}_4)~ v^*({\mathbf
p}_3)v^*({\mathbf p}_4)\SB({\mathbf p}_1, {\mathbf p}_2, {\mathbf k}_3,{\mathbf k}_4)\SF^*({\mathbf
p}_3, {\mathbf p}_4, {\mathbf k}_3,{\mathbf k}_4)\right.\right.\nn\\\left.\left.+v({\mathbf
p}_1)v({\mathbf p}_2)~ u({\mathbf k}_3)u({\mathbf k}_4)\SF({\mathbf p}_1, {\mathbf p}_2, {\mathbf
k}_3,{\mathbf k}_4)\SB^*({\mathbf p}_3, {\mathbf p}_4, {\mathbf k}_3,{\mathbf
k}_4)\right.\right.\nn\\\left.\left.+v({\mathbf p}_1)v({\mathbf p}_2)~ v^*({\mathbf
p}_3)v^*({\mathbf p}_4)\SF({\mathbf p}_1, {\mathbf p}_2, {\mathbf k}_3,{\mathbf k}_4)\SF^*({\mathbf
p}_3, {\mathbf p}_4, {\mathbf k}_3,{\mathbf k}_4)\right)\right]\nn\\ =
2p_3^0(2\pi)^2\delta^{(2)}({\mathbf p}_1-{\mathbf p}_3)2p_4^0(2\pi)^2\delta^{(2)}({\mathbf
p}_2-{\mathbf p}_4)\ .
\end{IEEEeqnarray}
The four theta term in \eqref{ossf} is proportional to 
\begin{IEEEeqnarray}{l}\label{unicond2}
\int\frac{d^2 k_3}{2k_3^0(2\pi)^2}\frac{d^2 k_4}{2k_4^0(2\pi)^2}\left[-\SF({\mathbf p}_1,{\mathbf
p}_2,{\mathbf k}_3,{\mathbf k}_4)\SF^*({\mathbf p}_3,{\mathbf p}_4,{\mathbf k}_3,{\mathbf
k}_4)\right.\nn\\\left.+\f{1}{16m^2}\left(2(p_3\cdot p_4 + m^2)\SF({\mathbf p}_1, {\mathbf p}_2,
{\mathbf k}_3,
{\mathbf k}_4)\SF^*({\mathbf p}_3, {\mathbf p}_4, {\mathbf k}_3,
{\mathbf k}_4)\right.\right.\nn\\\left.\left.+u({\mathbf k}_3)u({\mathbf k}_4)~ v({\mathbf
p}_3)v({\mathbf p}_4)\SF({\mathbf p}_1, {\mathbf p}_2, {\mathbf k}_3,{\mathbf k}_4)\SB^*({\mathbf
p}_3, {\mathbf p}_4, {\mathbf k}_3,{\mathbf k}_4)\right.\right.\nn\\\left.\left.+v^*({\mathbf
p}_1)v^*({\mathbf p}_2)~ u^*({\mathbf k}_3)u^*({\mathbf k}_4)\SB({\mathbf p}_1, {\mathbf p}_2,
{\mathbf k}_3,{\mathbf k}_4)\SF^*({\mathbf p}_3, {\mathbf p}_4, {\mathbf k}_3,{\mathbf
k}_4)\right.\right.\nn\\\left.\left.+v^*({\mathbf p}_1)v^*({\mathbf p}_2)~ v({\mathbf
p}_3)v({\mathbf p}_4)\SB({\mathbf p}_1, {\mathbf p}_2, {\mathbf k}_3,{\mathbf k}_4)\SB^*({\mathbf
p}_3, {\mathbf p}_4, {\mathbf k}_3,{\mathbf k}_4)\right)\right]\nn\\ =
-2p_3^0(2\pi)^2\delta^{(2)}({\mathbf p}_1-{\mathbf p}_3)2p_4^0(2\pi)^2\delta^{(2)}({\mathbf
p}_2-{\mathbf p}_4)\ .
\end{IEEEeqnarray}
The equations \eqref{unicond1} and \eqref{unicond2} are necessary and sufficient to ensure
unitarity. 

\eqref{unicond1} and \eqref{unicond2} may be thought of as constraints 
imposed by unitarity on the four boson scattering matrix $\SB$ and the four 
fermion scattering matrix $\SF$. These conditions are written in terms of the 
onshell spinors $u$ and $v$ (rather than the momenta of the scattering 
particles for a reason we now pause to review. Recall that the Dirac equation
and normalization conditions define $u_\alpha$ and $v_\alpha$ only upto an 
undetermined phase (which could be a function of momentum). An 
expression built out of $u$'s and $v$'s can be written unambiguously in terms 
of onshell momenta if and only if all undetermined phases cancel out. 
The phases of terms involving $\SF$ in \eqref{unicond1} and \eqref{unicond2} 
do not cancel. This might at first appear to be a paradox; surely the unitarity
(or lack) of an $S$ matrix cannot depend on the unphysical choice of an 
arbitrary phase. The resolution to this `paradox' is simple; the function 
$\SF$ is itself not phase invariant, but transforms under phase transformations
like $\left(u({\mathbf p}_1)u({\mathbf p}_2)\right)\left(v({\mathbf p}_3)v({\mathbf
p}_4)\right)$.
It is thus useful to define 
\beq\label{Fnewdef}
 \SF = \frac{1}{4m^2}\left(u({\mathbf p}_1)u({\mathbf p}_2)\right)\left(v({\mathbf p}_3)v({\mathbf
p}_4)\right)\Sf\ .
\eeq
The utility of this definition is that $\Sf$ does not suffer from a phase 
ambiguity. Rewritten in terms of $\SB$ and $\Sf$, the unitarity equations 
may be written entirely in terms of participating momenta (with no spinors) \footnote{See
\S\ref{unitdet} for a derivation of this result.}. 
In terms of the quantity 
\beq\label{Ydef}
Y({\mathbf p}_3,{\mathbf
p}_4) = \f{2(p_3\cdot p_4 + m^2)}{16m^2}
\eeq
and 
$$d\Gamma^\prime = \frac{d^2
k_3}{2k_3^0(2\pi)^2}\frac{d^2 k_4}{2k_4^0(2\pi)^2}$$

\begin{IEEEeqnarray}{l}\label{newunicond5}
\int d\Gamma^\prime\bigg[\SB({\mathbf p}_1,{\mathbf p}_2,{\mathbf k}_3,{\mathbf k}_4)\SB^*({\mathbf
p}_3,{\mathbf p}_4,{\mathbf k}_3,{\mathbf k}_4)\bigg.\nn\\-Y({\mathbf p}_3,{\mathbf
p}_4)\bigg(\SB({\mathbf p}_1,{\mathbf p}_2,{\mathbf k}_3,{\mathbf k}_4) + 4Y({\mathbf p}_1,{\mathbf
p}_2) \Sf({\mathbf p}_1,{\mathbf p}_2,{\mathbf k}_3,{\mathbf
k}_4)\bigg)\bigg.\bigg.\nn\\\bigg(\SB^*({\mathbf p}_3,{\mathbf p}_4,{\mathbf k}_3,{\mathbf k}_4) +
4Y({\mathbf p}_3,{\mathbf p}_4) \Sf^*({\mathbf p}_3,{\mathbf p}_4,{\mathbf k}_3,{\mathbf
k}_4)\bigg)\bigg] = 2p_3^0(2\pi)^2\delta^{(2)}({\mathbf p}_1-{\mathbf
p}_3)2p_4^0(2\pi)^2\delta^{(2)}({\mathbf p}_2-{\mathbf p}_4)\nn\\
\end{IEEEeqnarray}
and
\begin{IEEEeqnarray}{l}\label{newunicond6}
\int d\Gamma^\prime\bigg[-16Y^2({\mathbf p}_3,{\mathbf p}_4)\Sf({\mathbf p}_1,{\mathbf p}_2,{\mathbf
k}_3,{\mathbf k}_4)\Sf^*({\mathbf p}_3,{\mathbf p}_4,{\mathbf k}_3,{\mathbf
k}_4)\bigg.\nn\\+Y({\mathbf p}_3,{\mathbf p}_4)\bigg(\SB({\mathbf p}_1,{\mathbf p}_2,{\mathbf
k}_3,{\mathbf k}_4) + 4Y({\mathbf p}_1,{\mathbf p}_2) \Sf({\mathbf p}_1,{\mathbf p}_2,{\mathbf
k}_3,{\mathbf k}_4)\bigg)\bigg.\bigg.\nn\\\bigg(\SB^*({\mathbf p}_3,{\mathbf p}_4,{\mathbf
k}_3,{\mathbf k}_4) + 4Y({\mathbf p}_3,{\mathbf p}_4) \Sf^*({\mathbf p}_3,{\mathbf p}_4,{\mathbf
k}_3,{\mathbf k}_4)\bigg)\bigg] \nn= -2p_3^0(2\pi)^2\delta^{(2)}({\mathbf p}_1-{\mathbf
p}_3)2p_4^0(2\pi)^2\delta^{(2)}({\mathbf p}_2-{\mathbf p}_4).\nn\\
\end{IEEEeqnarray}

The equations \eqref{newunicond5} and \eqref{newunicond6} followed from \eqref{uni1}. It is useful
to rephrase the above equations in terms of the ``$T$ matrix'' that represents the actual
interacting
part of the ``$S$ matrix''. Using the definition of the Identity operator \eqref{identop} we can
write a superfield expansion to define the ``$T$ matrix'' as  
\begin{align}\label{tdef}
S({\mathbf p}_1, \theta_1, {\mathbf p}_2, \theta_2, {\mathbf k}_3, \te_3, 
{\mathbf k}_4, \te_4)=&I({\mathbf p}_1, \theta_1, {\mathbf p}_2, \theta_2, {\mathbf k}_3, \te_3, 
{\mathbf k}_4, \te_4)\nn\\
&+ i (2\pi)^3\de^3(p_1+p_2-p_3-p_4) T({\mathbf p}_1, \theta_1, {\mathbf p}_2, \theta_2, {\mathbf
k}_3, \te_3, 
{\mathbf k}_4, \te_4)\ .
\end{align}
The identity operator is defined in \eqref{identop} is a supersymmetry invariant. It follows
that the `` $T$ matrix'' is also invariant under supersymmetry. In other words the ``$T$ matrix''
obeys
\eqref{susyinv} and has a superfield expansion \footnote{The matrices $\TB$ and $\TF$
correspond to the $T$ matrices of the four boson and four fermion scattering respectively.}
\begin{multline}\label{SUSYTmatrix}
T({\mathbf p}_1, \theta_1, {\mathbf p}_2, \theta_2, {\mathbf p}_3, \theta_3, 
{\mathbf p}_4, \theta_4) =
\TB + \TF ~\theta_1\theta_2\theta_3\theta_4 + \left(\half C_{12}\TB-\half
C^*_{34}\TF\right)~\theta_1\theta_2 \\+ \left(\half C_{13}\TB - \half
C^*_{24}\TF\right)~\theta_1\theta_3 + \left(\half C_{14}\TB + \half
C^*_{23}\TF\right)~\theta_1\theta_4 + \left(\half C_{23}\TB + \half
C^*_{14}\TF\right)~\theta_2\theta_3 \\+ \left(\half C_{24}\TB - \half
C^*_{13}\TF\right)~\theta_2\theta_4 + \left(\half C_{34}\TB - \half
C^*_{12}\TF\right)~\theta_3\theta_4
\end{multline}
where 
\beq\label{FTnewdef}
 \TF = \frac{1}{4m^2}\left(u({\mathbf p}_1)u({\mathbf p}_2)\right)\left(v({\mathbf
p}_3)v({\mathbf p}_4)\right)\Tf\ .
\eeq
and the coefficients $C_{ij}$ are given as before in \eqref{cdef} and \eqref{cstdef}.

It follows from \eqref{tdef} that
\begin{align}\label{f1f2T}
\SB(\mbp_1,\mbp_2,\mbp_3,\mbp_4)&= I(\mbp_1,\mbp_2,\mbp_3,\mbp_4)+ i (2\pi)^3\de^3(p_1+p_2-p_3-p_4)
\TB(\mbp_1,\mbp_2,\mbp_3,\mbp_4)\ ,\nn\\
\Sf(\mbp_1,\mbp_2,\mbp_3,\mbp_4)&=I(\mbp_1,\mbp_2,\mbp_3,\mbp_4)+ i (2\pi)^3\de^3(p_1+p_2-p_3-p_4)
\Tf(\mbp_1,\mbp_2,\mbp_3,\mbp_4)\ .
\end{align}

Substituting the definitions \eqref{f1f2T} into \eqref{newunicond5} and
\eqref{newunicond6} the unitarity conditions can be rewritten as 
\begin{IEEEeqnarray}{l}\label{unicondf1T}
\int d\tilde{\Gamma}\bigg[\TB({\mathbf p}_1,{\mathbf p}_2,{\mathbf k}_3,{\mathbf
k}_4)\TB^*({\mathbf
p}_3,{\mathbf p}_4,{\mathbf k}_3,{\mathbf k}_4)\bigg.\nn\\-Y({\mathbf p}_3,{\mathbf
p}_4)\bigg(\TB({\mathbf p}_1,{\mathbf p}_2,{\mathbf k}_3,{\mathbf k}_4) + 4Y({\mathbf
p}_1,{\mathbf p}_2) \Tf({\mathbf p}_1,{\mathbf p}_2,{\mathbf k}_3,{\mathbf
k}_4)\bigg)\bigg.\bigg.\nn\\\bigg(\TB^*({\mathbf p}_3,{\mathbf p}_4,{\mathbf k}_3,{\mathbf k}_4)
+4Y({\mathbf p}_3,{\mathbf p}_4) \Tf^*({\mathbf p}_3,{\mathbf p}_4,{\mathbf k}_3,{\mathbf
k}_4)\bigg)\bigg] = i(\TB(\mbp_1,\mbp_2,\mbp_3,\mbp_4)-\TB^*(\mbp_3,\mbp_4,\mbp_1,\mbp_2))\nn\\
\end{IEEEeqnarray}
and
\begin{IEEEeqnarray}{l}\label{unicondf2T}
\int d\tilde{\Gamma}\bigg[-16Y^2({\mathbf p}_3,{\mathbf p}_4)\Tf({\mathbf p}_1,{\mathbf
p}_2,{\mathbf k}_3,{\mathbf k}_4)\Tf^*({\mathbf p}_3,{\mathbf p}_4,{\mathbf k}_3,{\mathbf
k}_4)\bigg.\nn\\+Y({\mathbf p}_3,{\mathbf p}_4)\bigg(\TB({\mathbf p}_1,{\mathbf p}_2,{\mathbf
k}_3,{\mathbf k}_4) + 4Y({\mathbf p}_1,{\mathbf p}_2) \Tf({\mathbf p}_1,{\mathbf p}_2,{\mathbf
k}_3,{\mathbf k}_4)\bigg)\bigg.\bigg.\nn\\\bigg(\TB^*({\mathbf p}_3,{\mathbf p}_4,{\mathbf
k}_3,{\mathbf k}_4) + 4Y({\mathbf p}_3,{\mathbf p}_4) \Tf^*({\mathbf p}_3,{\mathbf p}_4,{\mathbf
k}_3,{\mathbf k}_4)\bigg)\bigg]\nn\\
=4 iY(\mbp_3,\mbp_4)\left(\Tf^*(\mbp_3,\mbp_4,\mbp_1,\mbp_2)-\Tf(\mbp_1,\mbp_2,\mbp_3,
\mbp_4)\right)\nn\\
\end{IEEEeqnarray}
where 
$$d\tilde{\Gamma} = (2\pi)^3 \de^3(p_1+p_2-k_3-k_4)\frac{d^2
k_3}{2k_3^0(2\pi)^2}\frac{d^2 k_4}{2k_4^0(2\pi)^2}\ .$$
The equations \eqref{unicondf1T} and \eqref{unicondf2T} can be put in a more user friendly form by
going to the center of mass frame with the definition
\begin{align}\label{com}
&p_1 = \left(\sqrt{p^2+m^2},p,0\right), \ p_2= \left(\sqrt{p^2+m^2},-p,0\right) \nn\\
&p_3 = \left(\sqrt{p^2+m^2},p \cos(\te), p\sin(\te)\right),\ p_4= \left(\sqrt{p^2+m^2},-p
\cos(\te), -p\sin(\te)\right)
\end{align}
where $\te$ is the scattering angle between $p_1$ and $p_3$. In terms of the Mandelstam variables
\begin{align}
& s=-(p_1+p_2)^2 \ , t=-(p_1-p_3)^2 , \ u= (p_1-p_4)^2, \  s+t+u=4m^2\ ,\nn\\
& s= 4(p^2+m^2)\ , t=  -2p^2 (1-\cos(\te))\ , u=-2p^2 (1+\cos(\te))\ .
\end{align}
Using the definitions we see that \eqref{Ydef} becomes
\beq\label{Yy}
Y= \f{2 (p_3\cdot p_4 +m^2)}{16 m^2}= \f{-s+4m^2}{16 m^2}=Y(s)\ .
\eeq
Then \eqref{unicondf1T} and \eqref{unicondf2T} can be put in the form (See for instance eq 2.58-eq
2.59 of \cite{Jain:2014nza})
\begin{align}
\f{1}{8\pi\sqrt{s}}\int d\te\bigg(- Y(s) (\TB(s,\te)+4 Y(s) \Tf(s,\te))
(\TB^*(s,-(\al-\te))+4 Y(s) \Tf^*(s,-(\al-\te)))\nn\\
+ \TB(s,\te) \TB^*(s,-(\al-\te)) \bigg)= i (\TB^*(s,-\al)-\TB(s,\al))\label{umesheqB12}
\end{align}
\begin{align}
\f{1}{8\pi\sqrt{s}}\int d\te\bigg(Y(s) (\TB(s,\te)+4 Y(s) \Tf(s,\te))
(\TB^*(s,-(\al-\te))+4 Y(s) \Tf^*(s,-(\al-\te)))\nn\\ -16 Y(s)^2 \Tf(s,\te)
\Tf^*(s,-(\al-\te))
\bigg)= i
4Y(s)(-\Tf(s,\al)+\Tf^*(s,-\al))\label{umesheqF12}
\end{align}
In a later section \S\ref{unitarityx} we will use the simplified equations \eqref{umesheqB12} and
\eqref{umesheqF12} for the unitarity analysis.

\section{Exact computation of the all orders \texorpdfstring{$S$}{S} matrix}\label{excom}

In this section we will present results and conjectures for the the
$2 \times 2$ $S$ matrix of the general ${\cal N}=1$ theory 
\eqref{actionlcgauge1} at all orders in the t'Hooft coupling. In \S 
\ref{Frules} we recall the action for our theory and determine the 
bare propagators for the scalar and vector superfields. At leading order 
in the $\frac{1}{N}$ the vector superfield propagator is exact (it is not 
renormalized). However the propagator of the scalar superfield does receive 
corrections. In \S \ref{prop}, we determine the all orders propagator 
for the superfield $\Phi$ by solving the relevant 
Schwinger-Dyson equation. We will then turn to the determination
of the exact offshell four point function of the superfield $\Phi$. 
As in \cite{Jain:2014nza}, we demonstrate that this four point function is the 
solution to a linear integral equation which we explicitly write down in 
\S \ref{inteqn}. In a particular kinematic regime we present an exact solution
to this integral equation in \S \ref{offshellsoln}. In order to obtain the 
$S$ matrix, in \S  \ref{onshell} we take the onshell limit of this answer. The 
kinematic restriction on our offshell result turns out to be inconsistent 
with the onshell limit in one of the four channels of scattering (particle - 
antiparticle scattering in the singlet channel) and so we do not have 
an explicit computation of the $S$ matrix in this channel. In the other 
three channels, however, we are able to extract the full $S$ matrix (with no kinematic restriction)
albeit in a particular Lorentz frame.  In \S \ref{onshell} we present the unique covariant
expressions for the $S$ matrix consistent with our results. In \S \ref{dsm} we report our result
that the covariant $S$ matrix reported in \S\ref{onshell} is duality invariant. We present
explicit exact results for the $S$ matrices in the T and U channels of
scattering in \S\ref{ons}. In \S\ref{onss} we present the explicit conjecture for the $S$ matrix in
the singlet (S) channel. In \S\ref{onssN2} we report the explicit $S$ matrices for the $\mN=2$
theory.
\subsection{Supersymmetric Light Cone Gauge} \label{lcg}

We study the general ${\cal N}=1$ theory \eqref{action}. 
Wess-Zumino gauge, employed in subsection \S \ref{rnt} to display the 
physical content of our theory, is inconvenient for actual computations 
as it breaks manifest supersymmetry. In other words if $\Gamma_\alpha$ is 
chosen to lie in Wess-Zumino gauge, it is in general not the case that 
$ Q_\beta \Gamma_\alpha$ also respects this gauge condition. 
In all calculations presented in this paper we will work instead 
in `supersymmetric light cone gauge' \footnote{We would like to thank 
S. Ananth and W. Siegel for helpful correspondence on this subject.}
\begin{align}\label{lightconegauge}
 \Ga_-=0
\end{align}
As $\Gamma_-$ transforms homogeneously under supersymmetry (see
\eqref{stran}) it is obvious that this gauge choice is supersymmetric.
It is also easily verified that all gauge self interactions in 
\eqref{action} vanish in our lightcone gauge and the action \eqref{action} 
simplifies to 
\begin{align}\label{actionlcgauge1}
S_{tree}=-\int d^3 x d^2\te \bigg[-\f{\ka}{8\pi}
Tr(\Ga^-i\p_{--}\Ga^-)-\f{1}{2}D^\al\bar{\Ph}D_\al\Ph-\f{i}{2}\Ga^-(\bar{\Ph}
D_-\Ph-D_-\bar{\Ph}\Ph)\nn\\+m_0\bar{\Ph}\Ph+\f{\pi w}{\ka}(\bar{\Ph}\Ph)^2\bigg] \ .
\end{align}
Note in particular that \eqref{actionlcgauge1} is quadratic in $\Gamma_+$.

The condition \eqref{lightconegauge} implies, in particular,  that 
the component gauge fields in $\Gamma_\alpha$ obey
$$A_-=A_1+i A_2=0$$
(see Appendix \S \ref{susylcgauge} for more details and further discussion 
about this gauge). In other words the gauge \eqref{lightconegauge} is a 
supersymmetric generalization of ordinary lightcone gauge.

\subsection{Action and bare propagators}\label{Frules}

The bare scalar propagator that follows from \eqref{actionlcgauge1} is 
\beq\label{superpropagator1}
\lan\bar{\Ph}(\te_1,p)\Ph(\te_2,-p')\ran= \f{D^2_{\te_1,p}-m_0}{p^2+m_0^2}\de^2(\te_1-\te_2)
(2\pi)^3 \de^3(p-p') \ .
\eeq
where $m_0$ is the bare mass.
\begin{figure}[h]
\begin{center}
\includegraphics{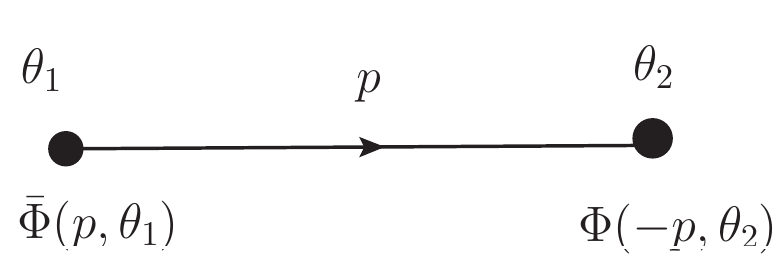}
\caption{\label{superpropfig1}Scalar superfield propagator} 
\end{center}
\end{figure}
We have chosen the convention for the momentum flow direction to be from $\bar{\Ph}$ to
$\Ph$ (see fig \ref{superpropfig1}). Our sign conventions are such that the momenta leaving a vertex
have a positive sign. The notation $D^2_{\te_1,p}$ means that the operator depends on
$\te_1$ and the momentum $p$, the explicit form for $D^2$ and some useful formulae are listed in
\S \ref{superspace}.  The gauge superfield propagator in momentum space is
\beq\label{supergaugepropagator}
\lan\Ga^-(\te_1,p)\Ga^-(\te_2,-p')\ran= -\f{8\pi}{\ka}\f{\de^2(\te_1-\te_2)}{p_{--}}(2\pi)^3
\de^3(p-p')\ 
\eeq
where $p_{--}=-(p_1+ip_2)=-p_-$. 
\begin{figure}[h]
\begin{center}
\includegraphics[width=6cm,height=2cm]{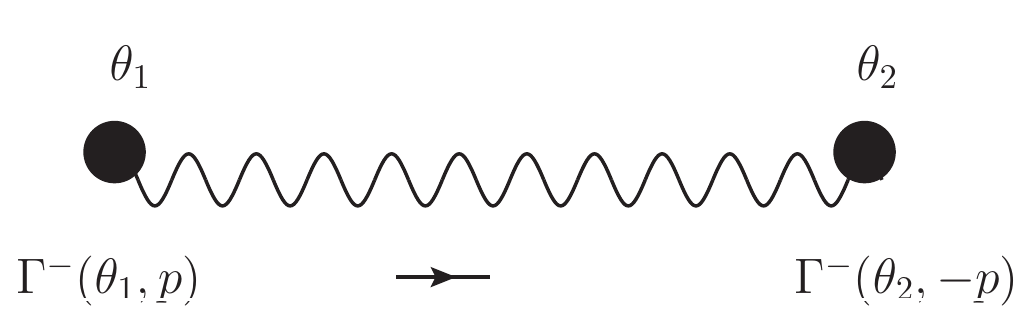}
\caption{\label{photonprop}Gauge superfield propagator, the arrow indicates direction of momentum
flow}
\end{center}
\end{figure}
Inserting the expansion \eqref{superfieldexpansions} into the LHS of 
\eqref{supergaugepropagator} and matching powers of $\theta$, we find in 
particular that 
\beq
\lan A_+(p)A_3(-p')\ran= \f{4\pi i}{\ka}\f{1}{p_-}(2\pi)^3 \de^3(p-p') \ ,\  \lan
A_3(p)A_+(-p')\ran= -\f{4\pi i}{\ka}\f{1}{p_-} (2\pi)^3 \de^3(p-p')\ , \  
\eeq
is in perfect agreement with the propagator of the gauge field in regular 
(non-supersymmetric) lightcone gauge (see Appendix A ,Eq A.7 of \cite{Giombi:2011kc})

\subsection{The all orders matter propagator} \label{prop} 

\subsubsection{Constraints from supersymmetry}

The exact propagator of the matter superfield $\Phi$ 
 enjoys invariance under supersymmetry 
transformations which implies that 
\beq \label{susyprop}
(Q_{\te_1,p}+Q_{\te_2,-p})\lan\bar{\Ph}(\te_1,p)\Ph(\te_2,-p)\ran=0 
\eeq
where the supergenerators $Q_{\te_1,p}$ were defined in \eqref{supercharge}. 
This constraint is easily solved. Let the exact scalar propagator take the form 
\begin{equation}\label{esp}
\langle \bPh(p, \theta_1) \Ph(-p', \theta_2) \rangle = (2 \pi)^3 
\delta^3(p-p') P(\theta_1, \theta_2, p)\ .
\end{equation}
The condition \eqref{susyprop} implies that the function $P$ obeys the equation
\beq \label{cps}
\bigg[\f{\p}{\p\te_1^\al}+\f{\p}{\p\te_2^\al}-p_{\al\be}(\te_1^\be-\te_2^\be)\bigg]P(\te_1,\te_2,
p)=0 \ .
\eeq
The most general solution to \eqref{cps} is 
\beq \label{fp}
C_1(p^\mu) \exp(-\te_1^\al p_{\al\be}\te_2^\be) +C_2(p^\mu)\de^2(\te_1-\te_2) \ 
\eeq
or equivalently
\begin{align} \label{pform}
P(\te_1,\te_2,p)&= \exp(-\te_1^\al p_{\al\be}\te_2^\be)\left(C_1(p^\mu) + C_2(p^\mu)
\de^2(\te_1-\te_2)\right) \ 
\end{align}
\footnote{The equivalence of \eqref{pform} and \eqref{fp} follows from 
the observation that $\theta^a A_{ab} \theta^b$ vanishes if $A_{ab}$ is symmetric
in $a$ and $b$.} 
where $C_1(p^\mu)$ is an arbitrary function of $p^\mu$ of dimension $m^{-2}$, while 
$C_2(p^\mu)$ is another function of $p^\mu$ of dimension $m^{-1}$.

It is easily verified using the formulae \eqref{expformula} that the bare propagator
\eqref{superpropagator1} can be recast in the form \eqref{pform} with 
\beq
C_1=\f{1}{p^2+m_0^2} \ , \ C_2=\f{m_0}{p^2+m_0^2} \ . 
\eeq

In a similar manner supersymmetry constrains the terms quadratic
in $\Phi$ and ${\bar \Phi}$ in the quantum effective action. In momentum 
space the most general supersymmetric quadratic effective action takes the 
form 
\begin{align} \label{susqe} 
S &= -\int \f{d^3 p}{(2 \pi)^3}
d^2 \te  \bPh(p, \theta) \left( A(p) D^2 + B(p) \right)  \Phi(-p, \theta)\\
&=-\int \f{d^3p}{(2\pi)^3} d^2 \theta_1 d^2 \theta_2
 \bPh(p, \theta_1) \exp(-\te_1^\al p_{\al\be}\te_2^\be) (A(p) + B(p) \delta^2(\theta_1 -\theta_2) )
\Phi(-p, \theta_2)
\end{align}
\footnote{In going from the first line to the second line of \eqref{susqe} 
we have integrated by parts and used the identity \eqref{expformula}. See Appendix \S
\ref{superspace} for the expressions of superderivatives and operator $D^2$ in momentum space. } 
The tree level quadratic action of our theory is clearly of the form 
\eqref{susqe} with $A(p)=1$ and $B(p)=m_0$.

\subsubsection{All orders two point function}\label{selfie2}

Let the exact 1PI quadratic effective action take the form 
\begin{equation}\label{quadee}
S_2= \int \f{d^3p}{(2\pi)^3} d^2\theta_1 d^2 \theta_2
\bPh(-p, \te_1)\left(\exp(-\te_1^\al p_{\al\be}\te_2^\be)
 + m_0 \de^2(\te_1 -\te_2) + \Sigma(p, \te_1, \te_2)\right) \Phi(p, \te_2)  \ .
\end{equation}
It follows from \eqref{susqe} that the supersymmetric self energy 
$\Sigma$ is of the form
\begin{equation}\label{fse}
\Sigma(p, \theta_1, \theta_2) = C(p)\exp(-\te_1^\al p_{\al\be}\te_2^\be)+
D(p) \de^2(\te_1-\te_2)
\end{equation} 
where $C(p)$ and $D(p)$ are as yet unknown functions of momenta. 

Imitating the steps described in section 3 of \cite{Giombi:2011kc}, 
the self energy $\Sigma$ 
\begin{figure}[h]
\begin{center}
\includegraphics[width=14.5cm,height=4cm]{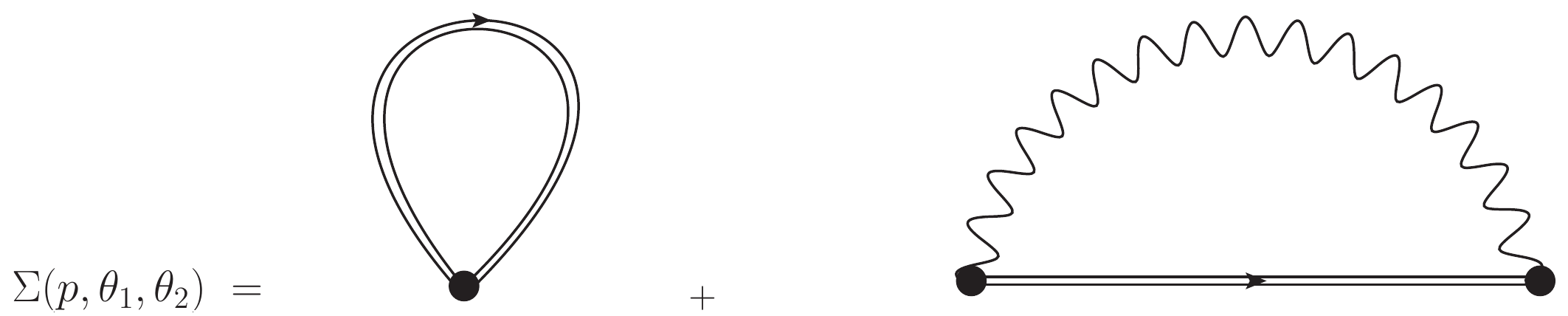}
  \caption{\label{sigmafig}Integral equation for self energy}
\end{center}
\end{figure}
defined in \eqref{quadee} may be shown to obey the 
integral equation \footnote{We work at leading order in the large $N$ limit}
\begin{align}\label{inte}
\Sigma(p, \theta_1, \theta_2) =&  \  2\pi\la w \int \f{d^3r}{(2\pi)^3}
\de^2(\te_1-\te_2)P(r, \theta_1,\theta_2)\nn\\
&- 2\pi\la \int \f{d^3
r}{(2\pi)^3}D^{\te_2,-p}_-D^{\te_1,p}_-\left(\f{\de^2(\te_1-\te_2)}{(p-r)_{--}}P(r,
\theta_1,\theta_2)\right)\nn\\
&+2\pi\la
\int\f{d^3r}{(2\pi)^3}\f{\de^2(\te_1-\te_2)}{(p-r)_{--}}D_-^{\te_1,r}D_-^{\te_2,-r}P(r,\te_1,\te_2)
\end{align}
where $P(p, \theta_1, \theta_2)$ is the exact superfield propagator. 
\footnote{The first line in the RHS of \eqref{inte} comes from the quartic interaction in Fig
\ref{sigmafig} while the second and third lines in \eqref{inte} comes from the gaugesuperfield
interaction in Fig \ref{sigmafig} . Note that each vertex in the diagram corresponding to the
gaugesuperfield interaction in Fig \ref{sigmafig} contains one factor of $D$, resulting in the two 
powers of $D$ in the second and third line of \eqref{inte}.} 
Note that the propagator $P$ depends on $\Sigma$ (in fact $P$ is obtained by 
inverting quadratic term in effective action \eqref{quadee}). In 
other words $\Sigma$ appears both on the LHS and RHS of \eqref{inte};
we need to solve this equation to determine $\Sigma$. 

Using the equations \eqref{usefulformulae}, the second and third lines on the 
RHS if \eqref{inte} may be considerably simplified (see Appendix \S\ref{selfieApp}) and we 
find
\begin{align}\label{intea}
\Sigma(p, \theta_1, \theta_2) =&  \  2\pi\la w \int \f{d^3r}{(2\pi)^3}
\de^2(\te_1-\te_2)P(r, \theta_1,\theta_2)\nn\\
&- 2\pi\la \int \f{d^3
r}{(2\pi)^3}\f{p_{--}}{(p-r)_{--}}\de^2(\te_1-\te_2)P(r,\theta_1,\theta_2)\nn\\
&+2\pi\la
\int\f{d^3r}{(2\pi)^3}\f{r_{--}}{(p-r)_{--}}\de^2(\te_1-\te_2)P(r,\te_1,\te_2)
\end{align}
Combining the second and third lines on the RHS of \eqref{intea} we see that
the factors of $p_{--}$ and $r_{--}$ cancel perfectly between the numerator 
and denominator, and \eqref{intea} simplifies to 
\begin{align}\label{inteb}
\Sigma(p, \theta_1, \theta_2) =&  \  2\pi\la (w-1) \int \f{d^3r}{(2\pi)^3}
\de^2(\te_1-\te_2)P(r, \theta_1,\theta_2) \ .
\end{align}

Notice that the RHS of \eqref{inteb} is independent of $p$, so it follows that 
$$\Sigma(p, \theta_1, \theta_2)= (m-m_0)\delta^2(\theta_1-\theta_2)$$
for some as yet undetermined constant $m$. It follows that the exact propagator
$P$ takes the form of the tree level propagator with $m_0$ replaced by $m$
i.e. 
\begin{equation}\label{ep}
P(p, \theta_1, \theta_2)
= \frac{D^2- m}{ p^2 + m^2} \de^2(\te_1- \te_2) \ .
\end{equation}

Plugging \eqref{ep} into \eqref{inteb} and simplifying we find the equation
\begin{align} \label{deb}
m-m_0&= 2\pi \la (w-1) \int \f{d^3r}{(2\pi)^3} \f{1}{r^2+m^2} \ .
\end{align}
The integral on the RHS diverges. Regulating this divergence using dimensional
regularization, we find that \eqref{deb} reduces to 
\begin{equation}\label{am}
m-m_0= \f{\la|m|}{2}(1-w) 
\end{equation}
and so  
\begin{equation}\label{exactmassn}
m=\f{2 m_0}{2+(-1+w) \la \ \text{Sgn(m)}} \ .
\end{equation}

Let us summarize. The {\it exact} 1PI quadratic effective 
action for the $\Phi$ superfield has the same form as the tree level 
effective action but with the bare mass $m_0$ replaced by the exact 
mass $m$ given in \eqref{exactmassn}. \footnote{Note that propagator for the fermion in the
superfield $\Phi$ is the usual propagator for a relativistic fermion of mass $m$. Recall, 
of course, that the propagator of $\Phi$ is not gauge invariant, and so 
its form depends on the gauge used in the computation. If we had carried 
out all computations in Wess-Zumino gauge (which breaks offshell supersymmetry)
we would have found the much more complicated expression for the fermion 
propagator reported in section 2.1 of  \cite{Jain:2013gza}. Note however that 
the gauge invariant physical pole mass $m$ of \eqref{exactmassn} agrees 
perfectly with the pole mass (reported in eq 1.6 of \cite{Jain:2013gza}) of 
the complicated propagator of \cite{Jain:2013gza}. The agreement of 
gauge invariant quantities in these rather different computations constitutes 
a nontrivial consistency check of the computations presented in this subsection.}  As explained in
\S \ref{dualitycon} the exact mass
\eqref{exactmassn} is duality invariant.

Note also that the $\mN=2$ point, $w=1$ there
is no renormalization of the mass, and the bare propagator is exact and the 
bare mass (which equals the pole mass) is itself duality invariant.

\subsection{Constraints from supersymmetry on the offshell four point function}\label{off4pt}
Much as with the two point function, the offshell four point function of matter superfields is
constrained by the supersymmetric Ward identities. Let us define 
\begin{align}\label{4ptamp}
\lan
\bar{\Ph}((p+q + \frac{l}{4} ),\te_1)\Ph(-p + \frac{l}{4},\te_2)\Ph(-(k+q) + \frac{l}{4}
,\te_3)\bar{\Ph}(k +
\frac{l}{4},\te_4)
 \ran \nn\\
= (2 \pi)^3 \delta(l)
V(\te_1,\te_2,\te_3,\te_4,p,q,k) .
\end{align}
It follows from the invariance under supersymmetry that 
\beq\label{4ptsusy1}
(Q_{\te_1,p+q}+Q_{\te_2,-p}+Q_{\te_3,-k-q}+Q_{\te_4,k})V(\te_1,\te_2,\te_3,\te_4,p,q,k)=0 \ .
\eeq
The general solution to \eqref{4ptsusy1} is easily obtained (see Appendix
\S \ref{susy4pt}). Defining 
\begin{align}\label{Xdef}
& X=\sum_{i=1}^4 \te_i \ ,\nn\\
& X_{12}=\te_1-\te_2 \ , \nn \\
& X_{13}=\te_1-\te_3 \ , \nn \\
& X_{43}=\te_4-\te_3 \ .
\end{align}
we find 
\beq\label{4ptsusyinvariantgen}
V=\exp\bigg(\f{1}{4}X.(p.X_{12}+q.X_{13}+k.X_{43})\bigg)F(X_{12},X_{13},X_{43},p,q,k) \ .
\eeq
where $F$ is an unconstrained function of its arguments. In other words 
supersymmetry fixes the transformation of $V$ under a uniform shift of all 
$\theta$ parameters
$\te_i\rightarrow \te_i+ \gamma$. 
(for $i =1 \ldots 4$ where $\gamma$ is a constant Grassman parameter). 
The undetermined function $F$ is a function of shift 
invariant combinations of the four $\te_i$. 

Let us now turn to the structure of the exact 1PI effective action for scalar 
superfields in our theory. The most general effective action consistent with 
global $U(N)$ invariance and supersymmetry takes the form 
\begin{equation}\label{eef} \begin{split}
S_4&= \frac{1}{2} \int \frac{d^3 p}{(2 \pi)^3}  \frac{d^3 k}{(2 \pi)^3} 
\frac{d^3 q}{(2 \pi)^3} d^2 \theta_1 d^2 \theta_2 d^2 \theta_3 d^2 \theta_4
\\ 
&\left(V(\te_1,\te_2,\te_3,\te_4,p,q,k)
\Phi_m(-(p+q), \theta_1) {\bar \Phi}^m(p, \theta_2) {\bar \Phi}^n(k+q, \theta_3)\Phi_n(-k,
\theta_4) 
 \right) \ .
\\
\end{split}
\end{equation}
It follows from the definition \eqref{eef} that the function $V$ may be 
taken to be invariant under the $Z_2$ symmetry 
\begin{align}\label{map}
& p\rightarrow k+q, k\rightarrow p+q ,q\rightarrow -q \ ,\nn\\
& \te_1\rightarrow \te_4 , \te_2\rightarrow \te_3, \te_3\rightarrow \te_2, \te_4\rightarrow \te_1 \
.
\end{align}

As in the case of two point functions, it is easily demonstrated that the 
invariance of this action under supersymmetry constraints the 
coefficient function $V$ that appears in \eqref{eef} to obey the equation 
\eqref{4ptsusy1}. As we have already explained above, the most general solution
to this equation is given in equation \eqref{4ptsusyinvariantgen} for 
a general shift invariant function $F$.

\subsection{An integral equation for the offshell four point function}
\label{inteqn}

\begin{figure}[h]
\begin{center}
\includegraphics[width=17cm,height=8cm]{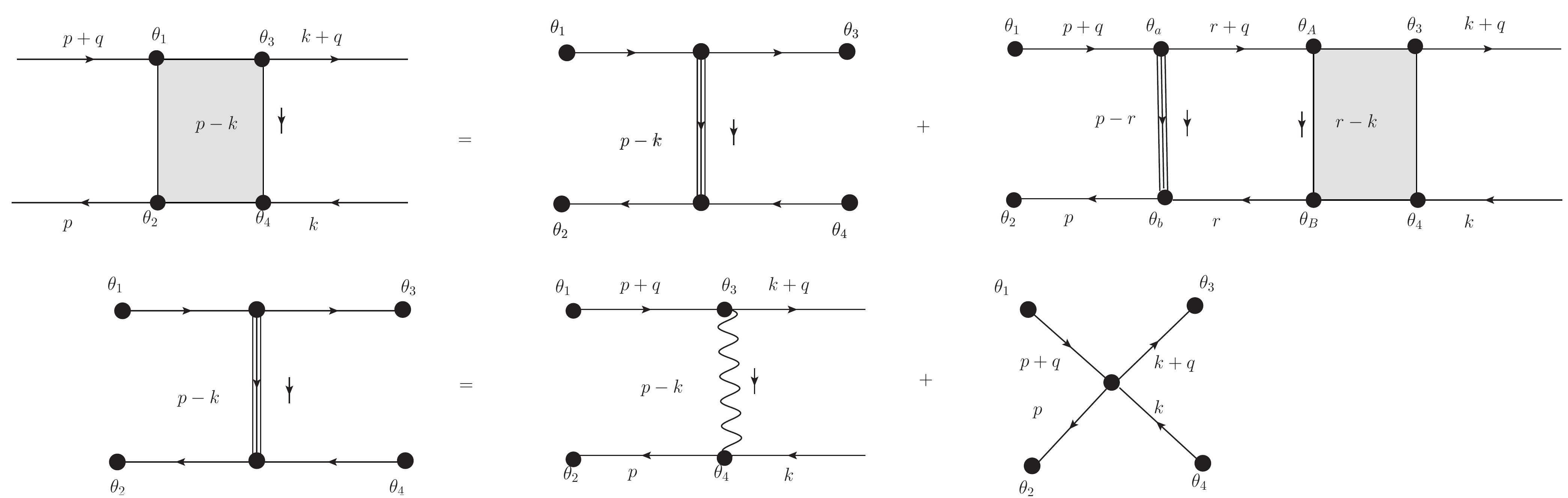}
\caption{\label{gaugeintSD}The diagrams in the first line pictorially represents the Schwinger-Dyson
equation for offshell four point function (see \eqref{int4pt}). The second line represents the tree
level contributions from the gauge superfield interaction and the quartic interactions.}
\end{center}
\end{figure}

The coefficient function $V$ of the quartic term of the exact 
IPI effective action may be shown to obey the integral equation (see Fig \ref{gaugeintSD} for a
diagrammatic representation of this equation) 
\begin{align}\label{int4pt}
V(\te_1,\te_2,\te_3,\te_4,p,q,k) =\ & V_0(\te_1,\te_2,\te_3,\te_4,p,q,k)\nn\\
&+ \int \f{d^3r}{(2\pi)^3} d^2\te_a d^2\te_b
d^2\te_A d^2\te_B \bigg(NV_0(\te_1,\te_2,\te_a,\te_b,p,q,r)\nn\\
& \qquad  P(r+q,\te_a,\te_A) P(r,\te_B,\te_b)V(\te_A,\te_B,\te_3,\te_4,r,q,k)\bigg) \ 
\end{align}

In \eqref{int4pt} $V_0$ is the tree level contribution to $V$. $V_0$ receives 
contributions from the two diagrams depicted in Fig \ref{gaugeintSD}. The explicit 
evaluation of $V_0$ is a straightforward exercise and we find (see Appendix 
\S\ref{4pttree} for details) 
\begin{align} \label{V0}
V_0(\te_1,\te_2,\te_3,\te_4,p,q,k)=&\exp\bigg(\f{1}{4}X.(p.X_{12}+q.X_{13}+k.X_{43}
)\bigg)\nn\\
&\ \ \biggl(-\f{i\pi w}{\ka}X_{12}^-X_{12}^+X_{13}^-X_{13}^+X_{43}^-X_{43}^+\nn\\
&\ \ \ \ -\f{4\pi i}{\ka(p-k)_{--}}X_{12}^+X_{13}^+X_{43}^+ (X_{12}^-+X_{34}^-)\biggr)\ .
\end{align}
In the above, the first term in the bracket is the delta function from the quartic interaction,
the second term is from the tree diagram due to the gauge superfield exchange computed in \S
\ref{4pttree}. 

We now turn to the evaluation of the coefficient $V$ in the exact 1PI effective
action. There are $2^6$ linearly independent functions of the six independent 
shift invariant Grassman variables  $X_{12}^\pm$, $X_{13}^\pm$ and $X_{43}^\pm$.
Consequently the most general $V$ consistent with supersymmetry is 
parameterized by 64 unknown functions of the three independent momenta. 
$V$ (and so $F$) is necessarily an even function of these variables.  
It follows that the most general function $F$ can be parameterized in terms 
of 32 bosonic functions of $p, k$ and $q$. In principle one could 
insert the most general supersymmetric $F$ into the integral equation 
\eqref{int4pt} and equate equal powers of $\theta_i$ on the two sides 
of  \eqref{int4pt} to obtain 32 coupled integral equations for the 32 
unknown complex valued functions. One could, then, attempt 
to solve this system of equations. This procedure would obviously be very 
complicated and difficult to implement in practice. Focusing on the special 
kinematics $q^\pm=0$ we were able to shortcircuit this laborious process, in 
a manner we now describe.

After a little playing around we were able to demonstrate that $V$ of the 
form \footnote{The variables $X, X_{ij}$ are defined in terms of $\te_i$ in \eqref{Xdef}.} 
\begin{align}\label{totform}
V(\te_1,\te_2,\te_3,\te_4,p,q,k)=\exp\bigg(\f{1}{4}X&.(p.X_{12}+q.X_{13}+k.X_{43})\bigg)
F(X_{12},X_{13},X_{43},p,q,k)\nn\\
F(X_{12},X_{13},X_{43},p,q,k)=\begin{split} X_{12}^+ X_{43}^+ \bigg(& A(p,k,q) X_{12}^-
X_{43}^-X_{13}^+X_{13}^-+B(p,k,q) X_{12}^- X_{43}^-\nn \\
&+ C(p,k,q)  X_{12}^- X_{13}^+ + D(p,k,q) X_{13}^+ X_{43}^-\bigg)\end{split}\ , \\
\end{align}
is closed under the multiplication rule induced by the RHS of \eqref{int4pt} (see Appendix \S
\ref{associativity}). Plugging in the general form of $V$ \eqref{totform} in the integral
equation \eqref{int4pt} and performing the grassmann integration, \eqref{int4pt} turns into to the following integral equations for the coefficient functions 
$A$, $B$, $C$ and $D$:
\begin{align}\label{integraleqmassiveA}
& A(p,k,q)+\f{ 2\pi i w }{\ka} \nn\\
& + i \pi \la \int \f{d^3r_E}{(2\pi)^3} \f{2
A(q_3p_-+2(q_3-im)r_-)+(q_3 r_-+2 i m p_-)(2 B q_3 +C k_-)-D r_-(q_3 p_-+2 i m r_-)
}{(r^2+m^2)((r+q)^2+m^2)(p-r)_-}\nn\\
&-i\pi\la w \int\f{d^3r_E}{(2\pi)^3} \f{4 i A m+2 B q_3^2+C q_3 k_-+2 D (q_3+i
m) r_-}{(r^2+m^2)((r+q)^2+m^2)}=0
\end{align}
\begin{align}\label{integraleqmassiveB}
 B(p,k,q) &+ i \pi \la \int \f{d^3r_E}{(2\pi)^3}\f{2 A(p+r)_- +4 B
(q_3 r_-+im (p-r)_- )-C k_-(p+r)_- - D r_-(p-3 r)_- }{(r^2+m^2)((r+q)^2+m^2)(p-r)_-}\nn\\
& -i\pi\la w \int\f{d^3r_E}{(2\pi)^3} \f{2A +4 i m B-C k_--D r_-}{(r^2+m^2)((r+q)^2+m^2)}=0
\end{align}
\begin{align}\label{integraleqmassiveC}
 C(p,k,q)-\f{4 \pi i}{\ka (p-k)_-}&+i\pi \la \int \f{d^3r_E}{(2\pi)^3}
\f{2 C \big(q_3(p+3 r)_-+2 im (p-r)_-\big)}{(r^2+m^2)((r+q)^2+m^2)(p-r)_-}\nn\\
&-i\pi\la w \int \f{d^3r_E}{(2\pi)^3} \f{2C (q_3+2 i m)}{(r^2+m^2)((r+q)^2+m^2)}=0
\end{align}
\begin{align}\label{integraleqmassiveD}
&D(p,k,q)-\f{4 \pi i}{\ka(p-k)_-}\nn\\
&+i\pi \la \int
\f{d^3r_E}{(2\pi)^3}\f{-A(4 q_3-8 im)+(q_3-2 i m)(4 B q_3 +2 Ck_-)+2 D (3 q_3+2 i m)r_-
}{(r^2+m^2)((r+q)^2+m^2)(p-r)_-}=0\ .
\end{align}

We will sometimes find it useful to view the four integral equations above 
as a single integral equation for a four dimensional column vector $E$
whose components are the functions $A$, $B$, $C$, $D$, i.e. 
\begin{equation}\label{edef} 
E(p,k,q)= \left(\begin{array}{c} A(p,k,q) \\ B(p,k,q) \\ C(p,k,q) \\ D(p,k,q) \end{array}\right)\ .
\end{equation}

The integral equations take the schematic form 
\begin{equation}\label{inteqe}
E= R + I E
\end{equation}
where $R$ is a 4 column of functions and $I$ is a matrix of integral 
operators acting on $E$. The integral equation \eqref{inteqe}
may be converted into a differential equation by differentiating both 
sides of \eqref{inteqe} w.r.t $p_+$. Using \eqref{deroz} and performing
all $d^3 r$ integrals (using  \eqref{intr3} for the integral over $r_3$) 
we obtain the differential equations
\begin{align}\label{ppeq}
 \p_{p_+}E(p,k,q)= S(p,k,q) +H(p,k_-,q) E(p,k,q) \ ,
\end{align}
where 
\beq\label{Spq}
S(p,k,q)= -\f{8 i\pi^2}{\ka} \de^2((p-k)_-,(p-k)_+)\left(\begin{array}{c} 0 \\ 0 \\
1 \\ 1 \end{array}\right)
\eeq

\beq\label{Hpq}
H(p,k_-,q_3)=\f{1}{a(p_s,q_3)}\left(
\begin{array}{cccc}
  (6 q_3-4 i m)p_- & 2 q_3 (2 i m+q_3) p_- & (2 im+q_3)k_- p_-  & -(2 i m+q_3)p_-^2  \\
 4 p_- & 4 q_3 p_-  & -2 k_- p_- & 2 p_-^2 \\
 0 & 0 & 8q_3 p_-  & 0 \\
 8 i m-4 q_3 & 4 q_3 (q_3-2 i m) & 2 (q_3-2 i m)k_-  & (4 i m+6 q_3) p_- \\
\end{array}
\right)
\eeq
and
\beq
a(p_s,q_3)=\f{\sqrt{m^2+p_s^2} \left(4 m^2+q_3^2+4 p_s^2\right)}{2\pi}\ .
\eeq

As we have explained above, the exact vertex $V$ enjoys invariance under the $\mathbb{Z}_2$
transformation \eqref{map}. In terms of the functions $A,B,C,D$, the $\mathbb{Z}_2$ action is given
by
\beq\label{map2}
E(p,k,q)=T E(k,p,-q)\ ,
\eeq
where 
\begin{align}
T=\left(\begin{array}{cccc} 
1&0&0&0\\
0&1&0&0\\
0&0&0&-1\\
0&0&-1&0\end{array}\right)\ .
\end{align}
The differential equations \eqref{ppeq} do not manifestly respect
the invariance \eqref{map2}. In fact in Appendix \S \ref{consistencycheck} we have demonstrated
that 
the differential equations \eqref{ppeq} admit solutions that enjoy the invariance \eqref{map2} if
and only if the following consistency condition is obeyed: 
\beq\label{consistencycondition}
[H(p,k_-,q),T H(k,p_-,-q)T]=0 \ .
\eeq
In the same Appendix we have also explicitly verified that this integrability condition is in fact
obeyed; this is a consistency check on \eqref{ppeq} and indirectly on the underlying integral
equations.

\subsection{Explicit solution for the offshell four point function}\label{offshellsoln}

In this subsection, we solve the system of integral equations for the unknown functions $A,
B, C, D$ presented in the previous subsection. We propose the ansatz
\begin{align}\label{momentumansatz}
& A(p,k,q)=A_1(p_s,k_s,q_3)+\f{A_2(p_s,k_s,q_3)k_-}{(p-k)_-}\ , \nn\\
& B(p,k,q)=B_1(p_s,k_s,q_3)+\f{B_2(p_s,k_s,q_3)k_-}{(p-k)_-}\ , \nn\\
& C(p,k,q)=-\f{C_2(p_s,k_s,q_3)-C_1(p_s,k_s,q_3)k_+p_-}{(p-k)_-}\ , \nn\\
& D(p,k,q)=-\f{D_2(p_s,k_s,q_3)-D_1(p_s,k_s,q_3)k_-p_+}{(p-k)_-} \ .
\end{align}
Our ansatz \eqref{momentumansatz} \footnote{We were able to arrive at this ansatz by first
explicitly computing the one loop answer and observing the functional forms. Moreover, in previous
work a very similar ansatz was already used to solve the integral equations for the fermions
(see Appendix F of \cite{Jain:2014nza}).} fixes  the solution 
in terms of 8 unknown functions of $p_s, k_s$ and $q_3$. 

Plugging the ansatz \eqref{momentumansatz} into the integral equations
\eqref{integraleqmassiveA}-\eqref{integraleqmassiveD}, one can do the angle and $r_3$ integrals (using the formulae \eqref{intangle} and \eqref{intr3} 
respectively) leaving only the $r_s$ integral to be performed. Differentiating
this expression w.r.t. to $p_s$ turns out to kill the $r_s$ integral yielding
differential equations in $p_s$ for the eight equations above. 
\footnote{Another way to obtain these differential equations is to 
plug the ansatz \eqref{momentumansatz} directly 
into the differential equations \eqref{ppeq}.}
The resulting differential equations turn out to be exactly solvable. 
Assuming that the solution respects the symmetry \eqref{map2}, it turns out 
to be given in terms of two unknown functions of $k_s$ and $q_3$. 
These can be thought of as the
integration constants that are not fixed by the symmetry requirement \eqref{map2}. Plugging the
solutions back into the integral equations we were able to determine 
these two integration  functions of $k_s$ and
$q_3$ completely. We now report our results.

The solutions for $A $ and $B$ are
\begin{align}\label{solutionsmassiveintAB}
A_1(p_s,k_s,q_3)=&e^{-2 i \la  \tan^{-1}\f{2\sqrt{m^2+p_s^2}}{q_3}}\biggl( G_1(k_s,q_3)\nn\\
& +\f{2 \pi  (w-1) (2 m-i q_3) e^{2 i \la
\big(\tan^{-1}\f{2\sqrt{k_s^2+m^2}}{q_3}+\tan^{-1}\f{2 \sqrt{m^2+p_s^2}}{q_3}\big)}}{\ka 
(e^{\f{i \pi \la q_3}{|q_3|}} (q_3 (w+3)-2 i m (w-1))+i (w-1) (2 m+i q_3) e^{2 i
\la\tan^{-1}\f{2 |m|}{q_3}})}\biggr)\ , \nn\\
A_2(p_s,k_s,q_3)=&e^{-2 i\la  \tan^{-1}\left(\f{2 \sqrt{m^2+p_s^2}}{q_3}\right)}G_2(k_s,q_3)\ ,
\nn
\end{align}
\begin{align}
B_1(p_s,k_s,q_3)=& \f{2 \pi A_1(p_s,k_s,q_3)}{q_3}\nn\\
&+\f{2 \pi}{b_1 b_2}\biggl(-i (w-1)^2 (4
m^2+q_3^2) e^{i \la  \big(\f{\pi  q_3}{| q_3| }-2
\tan ^{-1}\f{2 \sqrt{m^2+p_s^2}}{q_3}+4 \tan ^{-1}\f{2
|m|}{q_3}\big)} \nn\\
& +i (w-1)^2 (-4 m^2+8 i m q_3+3 q_3^2) e^{i \la  \big(\f{\pi
 q_3}{| q_3| }+2 \tan ^{-1}\f{2
\sqrt{k_s^2+m^2}}{q_3}\big)} \nn\\
& -8 i q_3^2 (w+1) e^{i \la\big(\f{\pi  q_3}{| q_3|
}+2 (\tan ^{-1}\f{2 \sqrt{k_s^2+m^2}}{q_3}-\tan ^{-1}\f{2
\sqrt{m^2+p_s^2}}{q_3}+\tan ^{-1}\f{2
|m|}{q_3})\big)}\nn\\
& +(w-1) (q_3+2 i m) (2 m (w-1)+i q_3 (w+3)) + e^{2 i \la 
(\f{\pi  q_3}{| q_3| }-\tan ^{-1}\f{2
\sqrt{m^2+p_s^2}}{q_3}+\tan ^{-1}\f{2| m|}{q_3})}\nn\\
&+ (w-1) (2 m-3 i q_3) (q_3 (w+3)+2 i m (w-1)) +e^{2 i \la 
\big(\tan^{-1}\f{2 \sqrt{k_s^2+m^2}}{q_3}+\tan ^{-1}\f{2
|m|}{q_3}\big)}\biggr)\ ,\nn
\end{align}
\begin{align}
B_2(p_s,k_s,q_3)=& \f{A_2(p_s,k_s,q_3)}{q_3}\ , \nn \\
G_1(k_s,q_3)= & -\f{2\pi}{\ka}\f{1}{g_1}\biggl(-8 i q_3^2 (w+1) e^{i \la \big(\f{\pi
q_3}{|q_3|}+2(\tan^{-1}\f{2 \sqrt{k_s^2+m^2}}{q_3}+\tan^{-1}\f{2|m|}{q_3})\big)}\nn\\
&+i (w-1)^2 (q_3-2 im)^2 e^{i \la \big(\f{\pi q_3}{|q_3| }+4 \tan^{-1}\f{2| m|}{q_3}\big)}\nn\\
&-(w-1)(q_3-2 i m) (2 m (w-1)+i q_3 (w+3))
e^{2 i \la \big(\f{\pi q_3}{|q_3| }+\tan^{-1}\f{2 |m|}{q_3}\big)}
\biggr)\ , \nn\\
G_2(k_s,q_3)=&0\ ,
\end{align}
where we have defined some parameters as given below for ease of presentation.
\begin{align}
g_1=& (w-1) (q_3+2 i m) e^{\f{2 i \pi  \la q_3}{|q_3| }}
(q_3 (w+3)-2 i m (w-1))\ ,\nn\\
&+ (w-1) (4 m^2 (w-1)-8 i m q_3+q_3^2 (w+3)) e^{4 i \la  \tan^{-1}\f{2|m|}{q_3}}\ ,\nn\\
& -2 (4 m^2 (w-1)^2+q_3^2 (w^2+2 w+5)) e^{i \la (\f{\pi
 q_3}{|q_3| }+2\tan ^{-1}\f{2|m|}{q_3})}\ ,\nn\\
b_1=&\ka q_3 ((w-1) (q_3+2 i m) e^{\f{i \pi  \la q_3}{|q_3|}}+(-q_3 (w+3)-2 i m (w-1)) e^{2 i
\la \tan ^{-1}\f{2|m|}{q_3}})\ , \nn\\
b_2=& e^{\f{i \pi  \la q_3}{|q_3|}} (q_3 (w+3)-2 i m (w-1))+i
(w-1) (2 m+i q_3) e^{2 i \la  \tan^{-1}\f{2 |m|}{q_3}}\ , \nn\\
\end{align}
 The solutions for C and D are
\begin{align}
C_1(p_s,k_s,q_3) =& \f{4 \pi (q_3+2 i m) (e^{2 i \la  \tan ^{-1}\f{2
|m|}{q_3}}-e^{2 i \la  \tan ^{-1}\f{2
\sqrt{k_s^2+m^2}}{q_3}}) e^{i \la  (\f{\pi 
q_3}{| q_3| }-2 \tan ^{-1}\f{2
\sqrt{m^2+p_s^2}}{q_3})}}{\ka  k_s^2 (i
(q_3+2 i m) e^{\f{i \pi  \la  q_3}{| q_3| }}+(2 m-i
q_3 \left(\f{w+3}{w-1}\right)) e^{2 i \la  \tan ^{-1}\f{2 |m|}{q_3}})}\ ,\nn\\
 C_2(p_s,k_s,q_3)=& \f{4 \pi  e^{2 i \la  (\tan ^{-1}\f{2
|m|}{q_3}-\tan^{-1}\f{2 \sqrt{m^2+p_s^2}}{q_3})}
((q_3+2 i m) e^{\f{i \pi  \la  q_3}{| q_3|}}-(q_3\left(\f{w+3}{w-1}\right)+2 i m) e^{2 i \la  \tan
^{-1}\f{2\sqrt{k_s^2+m^2}}{q_3}})}{\ka  (i (q_3+2 i m) e^{\f{i \pi\la  q_3}{| q_3| }}+(2 m-i q_3
\left(\f{w+3}{w-1}\right)) e^{2 i \la  \tan ^{-1}\f{2|m|}{q_3}})}\ , \nn
\end{align}
\begin{align}\label{solutionsmassiveintCD}
D_1(p_s,k_s,q_3)=&C_1(k_s,p_s,-q_3)\ , \nn\\
D_2(p_s,k_s,q_3)=&C_2(k_s,p_s,-q_3)\ . 
\end{align}
It is straightforward to show that the above solutions satisfy the various symmetry requirements
that follow from \eqref{map2}.

Although the solutions \eqref{solutionsmassiveintAB} and \eqref{solutionsmassiveintCD} are
quite complicated, a drastic simplification occurs at the $\mN=2$ point $w=1$
\begin{align}\label{solutionsN2}
A=&-\f{2 i \pi  e^{2 i \la  \big(\tan ^{-1}\f{2
\sqrt{k_s^2+m^2}}{q_3}-\tan^{-1}\f{2
\sqrt{m^2+p_s^2}}{q_3}\big)}}{\ka }\ , \nn \\
B=&\ 0 \ , \nn \\
C=&-\f{4 i \pi  e^{2 i \la \big(\tan ^{-1}\f{2
\sqrt{k_s^2+m^2}}{q_3}-\tan ^{-1}\f{2
\sqrt{m^2+p_s^2}}{q_3}\big)}}{\ka(k-p)_-}\ ,
\nn\\
D=&-\f{4 i \pi  e^{2 i \la  \big(\tan ^{-1}\f{2
\sqrt{k_s^2+m^2}}{q_3}-\tan ^{-1}\f{2
\sqrt{m^2+p_s^2}}{q_3}\big)}}{\ka(k-p)_-}\ .
\end{align}
It is satisfying that the complicated results of the general $\mN=1$ theory 
collapse to an extremely simple form at the $\mN=2$ point. 

\subsection{Onshell limit and the $S$ matrix}\label{onshell}

The explicit solution for the functions $A$, $B$, $C$ and $D$, presented in 
the previous subsection, completely determine $V$ in \eqref{eef}, and 
so the quadratic part of the exact (large $N$) IPI effective action. The
most general $ 2 \times 2$ $S$ matrix may now be obtained from \eqref{eef} as 
follows. We simply substitute the onshell expressions
\begin{align} \label{phmomos}
\Phi(p, \theta) = (2 \pi) \delta(p^2+m^2) 
\bigg[ & \theta(p^0) \bigg( a(\mathbf{p}) (1+m\te^2)+\te^\al
u_\al(\mathbf{p}) \al(\mathbf{p}) \bigg) \nn\\
&+ \theta(-p^0) \bigg(a^{c\dag}(-\mathbf{p}) (1+m\te^2)+\te^\al
v_\al(-\mathbf{p})\al^{c\dag}(-\mathbf{p})\bigg) \bigg]
\end{align}\
into \eqref{eef}
(here $a$ and $\alpha$ are the effectively free oscillators that create and 
destroy particles at very early or very late times; these oscillators obey the 
commutation relations \eqref{bcr}). Performing the integrals over 
$\theta^\alpha$ reduces  \eqref{eef} to a quartic form (let us call it $L$) 
in bosonic and fermionic oscillators. The $S$ matrix is obtained by sandwiching 
the resultant expression between the appropriate in and out states, and evaluating the 
resulting matrix elements using the commutation relations \eqref{bcr}. 

It may be verified that the quartic form in oscillators takes the form \footnote{The definition of
$A$ and $\tilde{A}$ reduces to the definition \eqref{casfield}  for $\phi=0$. While for $\ph=\pi$,
it reduces to \eqref{casfield} together with the identification $\te\rightarrow i \te$. With these
definitions $\tilde{A}=A^\dagger$ both at $\ph=0,\pi$.}
\begin{align} \label{qfo}
L= \sum_{\phi_i=0, \pi}\int\prod_{i=1}^4 & d\te_i \f{d^3 p_i}{((2\pi)^3)^4}  \de(p_i^2+m^2) S_M(p_1
,\ph_1, \theta_1, p_2,\ph_2, \theta_2, p_3,\ph_3, \theta_3, p_4,\ph_4, \theta_4)\nn\\
& \left(\delta_{\phi_i, 0} \theta(p^0_i) A(p_i,\phi_i, \theta_i) +
\delta_{\phi_i, \pi}\theta(-p^0_i)\tilde{A}(-p_i,\phi_i, \theta_i) \right) (2\pi)^3
\de^3(p_1+p_2+p_3+p_4)\nn\\
\text{where}\nn\\
& A(p_i,\phi_i, \theta_i)= a(\mbp_i)+\al(\mbp_i) e^{-\f{i\phi_i}{2}}\te_i\ ,\nn\\
& \tilde{A}(p_i,\phi_i,\theta_i)=a^\dagger(\mbp_i)+e^{-\f{i\phi_i}{2}}
\te_i\al^\dagger(\mbp_i)\ ,
\end{align}
where the one component fermionic variables $\theta_i$ are the fermionic 
variables that parameterize onshell superspace (see \S \ref{susyscat} ) and the master formula
is defined in \eqref{SUSYmaster}. Note that the phase variables 
$\phi_i$ are summed over two values $0$ and $\pi$; the symbol $\delta_{\phi, 0}$ is unity when
$\phi=0$ but zero when $\phi=\pi$ and $\delta_{\phi, \pi}$ has an analogous definition. 
 \eqref{qfo} compactly identifies the coefficient of every quartic form in oscillators. For
instance it asserts that the coefficient of $a_1 a_2 a_3^\dagger a_4^\dagger$ is the $S$ matrix for
scattering bosons with momentum $p_1, p_2$ to bosons with momenta $p_3, p_4$, while the the
coefficient of $\alpha_2 \alpha_4 a_1^\dagger a_3^\dagger $ is 
minus the $S$ matrix for scattering fermions with momentum $p_2, p_4$ to bosons with momentum $p_1,
p_3$, etc.

We can use the $\delta$ function in \eqref{qfo} to perform the integral over 
one of the four momenta; the integral over the remaining momenta may be 
recast as an integral over the momenta $p$ $k$ and $q$ employed in 
the previous section; specifically (see Fig
\ref{gaugeintSD} )
\beq
p_1=p+q \ , \ p_2= -k-q \ , \ p_3= -p \ , \ p_4=k\ .
\eeq
From the explicit results we get by substituting \eqref{phmomos} into 
\eqref{eef} we can read off all $S$ matrices at $q_\pm=0$. 

To start with, let us restrict our attention to the bosonic sector. From 
direct computation \footnote{Note that the onshell delta functions in the equations \eqref{qfob} 
and \eqref{qfof} ensure that
\beq\label{onshellcond}
p_3=k_3=-\f{q_3}{2}\ , \ p_s=k_s \ , \  k_s=\f{i}{2}\sqrt{q_3^2+4m^2} \ .
\eeq } we find that in this sector \eqref{qfo} reduces to 
\begin{align}\label{qfob}
 L_B=\sum_{\phi_i=0, \pi} \int & \f{d^3p}{(2\pi)^3}\f{dq_3}{(2\pi)}\f{d^3k}{(2\pi)^3}
\de((p+q)^2+m^2)\de((k+q)^2+m^2)\nn\\
&\de(p^2+m^2)\de(k^2+m^2) \TB (p, k, q_3)\nn\\
&\left(\delta_{\phi_i, 0}\theta(p^0) a(\mathbf{p+q}) +
\delta_{\phi_i, \pi}\theta(-p^0) a^\dagger(-\mbp-\mathbf{q}) \right)\nn\\
&\left(\delta_{\phi_i, 0}\theta(-k^0) a(\mathbf{-k-q})
+\delta_{\phi_i, \pi}\theta(k^0)a^\dagger(\mathbf{k+q}) \right)\nn\\
&\left(\delta_{\phi_i, 0}\theta(-p^0) a(-\mbp) +
\delta_{\phi_i, \pi}\theta(p^0)a^\dagger(\mbp) \right)\nn\\ 
&\left(\delta_{\phi_i, 0}\theta(k^0) a(\mathbf{k}) +
\delta_{\phi_i, \pi}\theta(-k^0)a^\dagger(-\mathbf{k}) \right)
\end{align}
while for the purely fermionic sector \eqref{qfo} reduces to
\begin{align}\label{qfof}
 L_F=\sum_{\phi_i=0, \pi} \int & \f{d^3p}{(2\pi)^3}\f{dq_3}{(2\pi)}\f{d^3k}{(2\pi)^3}
\de((p+q)^2+m^2)\de((k+q)^2+m^2)\nn\\
&\de(p^2+m^2)\de(k^2+m^2) \TF (p, k, q_3)\nn\\
&\left(\delta_{\phi_i, 0}\theta(p^0) \al(\mathbf{p+q}) +
\delta_{\phi_i, \pi}\theta(-p^0) \al^\dagger(-\mbp-\mathbf{q}) \right)\nn\\
&\left(\delta_{\phi_i, 0}\theta(-k^0) \al(\mathbf{-k-q})
+\delta_{\phi_i, \pi}\theta(k^0)\al^\dagger(\mathbf{k+q}) \right)\nn\\
&\left(\delta_{\phi_i, 0}\theta(-p^0) \al(-\mbp) +
\delta_{\phi_i, \pi}\theta(p^0)\al^\dagger(\mbp) \right)\nn\\ 
&\left(\delta_{\phi_i, 0}\theta(k^0) \al(\mathbf{k}) +
\delta_{\phi_i, \pi}\theta(-k^0)\al^\dagger(-\mathbf{k}) \right)
\end{align}
where \footnote{Our actual computations gave the functions $J_B$ and $J_F$ 
in the special case $q^\pm=0$. We obtained the answers reported in 
\eqref{TBcov} and \eqref{TFcov} by determining the unique covariant expression
that reduce to our answers for our special kinematics. While this procedure 
is completely correct (with standard conventions) for $J_B$, it is a bit 
inaccurate for $J_F$. The reason for this is that $J_F$ is Lorentz invariant
only upto a phase. As we have explained around \eqref{Fdefmas}, the 
phase of $J_F$ depends on the (arbitrary) phase of the $u$ and $v$ spinors 
of the particles in the scattering process. The accurate answer is obtained by covariantizing the
unambiguous $\Sf$ defined in \eqref{Fdef2}. $\SF$ is obtained by multiplying this result by the
quadrilinear term in spinor wavefunctions as defined in \eqref{Fnewdef}. This gives an explicit but
cumbersome expression for $\SF$, which agrees with the result presented above upto an overall 
convention dependent phase. This phase vanishes near identity scattering 
(where it could have interfered with identity), and we have dealt with 
this issue carefully in deriving the unitarity equation. In the 
equation above we have simply ignored the phase in order
to aid readability of formulas. } 
\begin{align}
  \TB=& \frac{4 i \pi}{\ka}\epsilon_{\mu\nu\rho}\f{q^\mu(p-k)^\nu(p+k)^\rho}{(p-k)^2}+
J_B(q,\la)\ , \label{TBcov}\\ 
  \TF=& \frac{4 i \pi}{\ka}\epsilon_{\mu\nu\rho}\f{q^\mu(p-k)^\nu(p+k)^\rho}{(p-k)^2}+
J_F(q,\la)\  , \label{TFcov}
\end{align}
where the $J$ functions\footnote{The $J$ functions are quite complicated and can be written in many
avatars. In this section we have written the most elegant form of the $J$ function, the other forms
are reported in Appendix \S\ref{japp}} are 
\begin{align}\label{Jfunm}
 J_B(q,\la)= &\f{4 \pi q}{\ka} \f{N_1 N_2+ M_1}{D_1 D_2}\ ,\nn\\
 J_F(q,\la)= &\f{4 \pi q}{\ka} \f{N_1 N_2+ M_2}{D_1 D_2}\ ,
\end{align}
where
\begin{align}\label{Jparm}
N_1= &\left(\left(\frac{2 |m|+i q}{2 |m|-i q}\right)^{-\lambda }  (w-1) (2 m+i q)+(w-1) (2 m-i
q)\right)\ ,\nn\\
N_2= &\left(\left(\frac{2 |m|+i q}{2 |m|-i q}\right)^{-\lambda }  (q (w+3)+2 i m (w-1))+(q (w+3)-2 i
m (w-1))\right)\ ,\nn\\
M_1=&-8 m q ((w+3) (w-1)-4 w) \left(\frac{2 |m|+i q}{2 |m|-i q}\right)^{-\lambda }\ ,\nn\\
M_2= & -8 m q (1 + w)^2 \left(\frac{2 |m|+i q}{2 |m|-i q}\right)^{-\lambda }\ ,\nn\\
D_1= &\left(i \left(\frac{2 |m|+i q}{2 |m|-i q}\right)^{-\lambda } (w-1) (2 m+i q)-2 i m (w-1)+q
(w+3)\right)\ , \nn\\
D_2= & \left( \left(\frac{2 |m|+i q}{2 |m|-i q}\right)^{-\lambda }  (-q (w+3)-2 i m (w-1))+(w-1)
(q+2 i m)\right)\ .
\end{align}

The equations \eqref{TBcov} and \eqref{TFcov} capture purely bosonic and 
purely fermionic $S$ matrices in all channels (particle-particle scattering 
in the symmetric and antisymmetric channels as well as particle-antiparticle
scattering in the adjoint channel) restricted to the kinematics $q_\pm=0$. 
Recall that supersymmetry (see \S \ref{susyscat}) determines all other scattering amplitudes 
in terms of the four boson and four fermion amplitudes, so the formulae 
 \eqref{TBcov} and \eqref{TFcov} are sufficient to determine all 
$2  \rightarrow 2 $ scattering processes restricted to our special kinematics. 
In other words $S_M$ in \eqref{qfo} is completely determined by 
\eqref{TBcov} and \eqref{TFcov} together with \eqref{SUSYmaster}.

\subsection{Duality of the $S$ matrix }\label{dsm}

Under the duality transformation (see \eqref{dtransf})
\beq
w'=\frac{3-w}{w+1},\lambda' = \lambda -\text{sgn}(\lambda ),m'= -m,\kappa'= -\kappa
\eeq
we have verified that 
\begin{align}\label{dualstat}
J_B(q,\ka',\la',w',m')= -J_F(q,\ka,\la,w,m)\ ,\nn\\
J_F(q,\ka',\la',w',m')= -J_B(q,\ka,\la,w,m)\ .
\end{align}
 provided \eqref{cond} is respected. In other words duality maps the purely bosonic and purely
fermionic $S$ matrices into one another. It follows that \eqref{TBcov} and \eqref{TFcov}
map to each other under duality upto a phase. As we have explained in subsection 
\S\ref{dsusy}, this result is sufficient to guarantee that the full $S$ matrix
(including, for instance, the $S$ matrix for Bose-Fermi scattering) is invariant 
under duality, once we interchange bosons with fermions. 

\subsection{\texorpdfstring{$S$}{S} matrices in various channels}\label{ons}

In this subsection we explicitly list the purely bosonic and purely 
fermionic $S$ matrices in every channel, as functions of the Mandelstam variables
of that channel. These results are, of course, easily extracted from 
\eqref{qfob} and \eqref{qfof}. There is a slight subtlety here; 
even though \eqref{TBcov} and \eqref{TFcov} are manifestly Lorentz 
invariant, it is not possible to write them entirely in terms of 
Mandelstam variables. \footnote{We define the Mandelstam variables as usual
\beq\label{mand}
s=-(p_1+p_2)^2\ , \ t= -(p_1-p_3)^2\ , \ u=-(p_1-p_4)^2\ .
\eeq} This is because (as was noted in \cite{Jain:2014nza}) 
$2+1$ dimensional kinematics allows for an additional 
$Z_2$ valued invariant (in addition to the Mandelstam variables) 
\beq\label{triad}
E(q,p-k,p+k)= \text{Sign}\left(\epsilon_{\mu\nu\rho}q^\mu(p-k)^\nu(p+k)^\rho\right)\ .
\eeq
\footnote{Note, in particular that the expression \eqref{triad} changes sign under the interchange of any two vectors.}
The sign of the first term in \eqref{TBcov} and \eqref{TFcov} 
is given by this new invariant as we will see in more detail below.

\subsubsection{U channel}
For particle-particle scattering
$$ P_i(p_1)+P_j(p_2)\rightarrow P_i(p_3)+P_j(p_4) $$
we have the direct scattering referred to as the U$_d$ (Symmetric) channel. \footnote{We adopt the
terminology of \cite{Jain:2014nza} in specifying scattering 
channels; we refer the reader to that paper for a more complete definition 
of the U$_d$, U$_e$, T, and S channels that we will repeatedly refer to 
below. } Our momenta assignments (see LHS of fig \ref{gaugeintSD}) are
\beq\label{udmom}
p_1=p+q\ , \  p_2=k \ , \ p_3=p \ , \ p_4=k+q \ .
\eeq
In terms of the Mandelstam variables 
\beq
s=-(p+q+k)^2 \ , \ t=-q^2 \ , \ u=-(p-k)^2\ ,
\eeq
the U$_d$ channel $T$ matrices for the boson-boson and fermion-fermion scattering are 
\begin{align}\label{Udsm}
 \TB^{U_d}&= E(q,p-k,p+k)\f{4\pi i}{k}\sqrt{\f{ts}{u}}+ J_B(\sqrt{-t},\la)\ ,\nn\\
 \TF^{U_d}&= E(q,p-k,p+k)\f{4\pi i}{k}\sqrt{\f{ts}{u}}+ J_F(\sqrt{-t},\la)\ .
\end{align}

For the exchange scattering, referred to as the U$_e$ (Antisymmetric) channel the momenta
assignments are (see LHS of fig \ref{gaugeintSD})
\beq\label{uemom}
p_1=k\ ,\  p_2=p+q \ , \ p_3=p \ , \  p_4=k+q \ .
\eeq
In terms of the Mandelstam variables 
\beq
s=-(p+q+k)^2 \ , \ t=-(p-k)^2 \ , \ u=-q^2\ ,
\eeq
the U$_e$ channel $T$ matrices for the boson-boson and fermion-fermion scattering are 
\begin{align}\label{Uesm}
 \TB^{U_e}&=E(q,p-k,p+k) \f{4\pi i}{k}\sqrt{\f{us}{t}}+ J_B(\sqrt{-u},\la)\ ,\nn\\
 \TF^{U_e}&=E(q,p-k,p+k) \f{4\pi i}{k}\sqrt{\f{us}{t}}+ J_F(\sqrt{-u},\la)\ .
\end{align}

\subsubsection{T channel}
For particle-antiparticle scattering $$ P_i(p_1)+A^j(p_2)\rightarrow P_i(p_3)+A^j(p_4) $$ $S$ matrix
in the adjoint channel is referred to as the T channel. The momentum assignments are (see LHS of fig
\ref{gaugeintSD})
\beq\label{tmom}
p_1=p+q \ , \ p_2=-k-q \ , \ p_3=p \ , \ p_4=-k \ .
\eeq
In terms of the Mandelstam variables 
\beq
s=-(p-k)^2 \ , \ t=-q^2 \ , \ u=-(p+q+k)^2\ ,
\eeq
the T channel $T$ matrices for the boson-boson and fermion-fermion scattering are 
\begin{align}\label{Tsm}
 \TB^T&= E(q,p-k,p+k)\f{4\pi i}{k}\sqrt{\f{tu}{s}}+ J_B(\sqrt{-t},\la)\ ,\nn\\
 \TF^T&= E(q,p-k,p+k)\f{4\pi i}{k}\sqrt{\f{tu}{s}}+ J_F(\sqrt{-t},\la)\ .
\end{align}
In particle-anti particle scattering there is also the singlet channel that we describe below.

\subsection{The Singlet (S) channel}\label{onss}
We now turn to the most interesting scattering process; the scattering of 
particles with antiparticles in the S (singlet) channel. In this channel 
the external lines on the LHS of Fig. \ref{gaugeintSD} are assigned 
positive energy (and so represent initial states) while those on the right
of the diagram are assigned negative energy (and so represent final states).
It follows that we must make the identifications
\beq\label{smom}
p_1=p+q\ ,\ p_2=-p \ , \ p_3= k+q \ , p_4=-k \ ,
\eeq
so that the Mandelstam variables for this scattering process are
\beq
s=-q^2 \ , \ t=-(p-k)^2 \ , \ u=-(p+k)^2\ .
\eeq
Note, in particular, that $s=-q^2$, and so is always negative when $q^\pm=0$. 
As we have been able to evaluate the offshell correlator $V$ (see 
\eqref{totform}) only for $q^\pm=0$, it follows that we cannot specialize 
our offshell computation to an onshell scattering process in the S channel 
in which $s \geq 4 m^2$. In other words we do not have a direct computation 
of S channel scattering in any frame. 

It is nonetheless tempting to simply assume that \eqref{TBcov} and 
\eqref{TFcov} continue to apply at every value of $q^\mu$ and not just 
when $q^\pm=0$; indeed this is what the usual assumptions of analyticity 
of $S$ matrices (and crossing symmetry in particular) would inevitably imply. 
Provisionally proceeding with this `naive' assumption, it follows upon 
performing the appropriate analytic continuation ($q^2 \rightarrow -s$ 
for positive $s$; see sec 4.4 of  \cite{Jain:2014nza}) that
\begin{align}\label{Ssmnaive}
 \TB^{S;\text{naive}}&= E(q,p-k,p+k)4\pi i\la\sqrt{\f{su}{t}}+ J_B(\sqrt{s},\la)\ ,\nn\\
 \TF^{S;\text{naive}}&= E(q,p-k,p+k)4\pi i\la\sqrt{\f{su}{t}}+ J_F(\sqrt{s},\la)\ ,
\end{align}
where
\begin{align}\label{SJfun}
 J_B(\sqrt{s},\la)= &-4 \pi i\la
\sqrt{s} \f{N_1 N_2+ M_1}{D_1 D_2}\ ,\nn\\
 J_F(\sqrt{s},\la)= & -4 \pi i\la
\sqrt{s} \f{N_1 N_2+ M_2}{D_1 D_2}\ ,
\end{align}
where
\begin{align}
N_1= &\left((w-1)(2 m+\sqrt{s})+(w-1)
(2 m-\sqrt{s})e^{i\pi\la}\left(\f{\sqrt{s}+2|m|}{\sqrt{s}-2|m|} \right)^\la \right)\ ,\nn\\
N_2= &\left((-i \sqrt{s} (w+3)+2 i m
(w-1))+(-i \sqrt{s} (w+3)-2 i m (w-1))e^{i\pi\la}\left(\f{\sqrt{s}+2|m|}{\sqrt{s}-2|m|}
\right)^\la\right)\ ,\nn\\
M_1=&8 m i \sqrt{s} ((w+3) (w-1)-4 w) e^{i\pi\la}\left(\f{\sqrt{s}+2|m|}{\sqrt{s}-2|m|} \right)^\la
\ ,\nn\\
M_2= & 8 m i \sqrt{s} (1 + w)^2 e^{i\pi\la}\left(\f{\sqrt{s}+2|m|}{\sqrt{s}-2|m|} \right)^\la\ ,
\nn\\
D_1= &\left(i (w-1) (2 m+\sqrt{s})-(2 i m
(w-1)+i\sqrt{s}(w+3))e^{i\pi\la}\left(\f{\sqrt{s}+2|m|}{\sqrt{s}-2|m|} \right)^\la \right)\ , \nn\\
D_2= & \left( (\sqrt{s} (w+3)-2 i m (w-1))+(w-1) (-i\sqrt{s}+2 i
m)e^{i\pi\la}\left(\f{\sqrt{s}+2|m|}{\sqrt{s}-2|m|} \right)^\la \right)\ .
\end{align}

Including the identity factors, the naive S channel $S$ matrix that follows from the usual rules
of crossing symmetry are
\begin{eqnarray}\label{swrong}
\SB^{S;\text{naive}}(\mbp_1,\mbp_2,\mbp_3,\mbp_4)&= I(\mbp_1,\mbp_2,\mbp_3,\mbp_4) + i
(2\pi)^3\de^3(p_1+p_2-p_3-p_4)\TB^{S;\text{naive}}(\mbp_1,\mbp_2,\mbp_3,\mbp_4)\ ,\nn\\
\SF^{S;\text{naive}}(\mbp_1,\mbp_2,\mbp_3,\mbp_4)&= I(\mbp_1,\mbp_2,\mbp_3,\mbp_4)+ i
(2\pi)^3\de^3(p_1+p_2-p_3-p_4)
\TF^{S;\text{naive}}(\mbp_1,\mbp_2,\mbp_3,\mbp_4)\ ,\nn\\
\end{eqnarray}
where the identity operator is defined in \eqref{identop}.

We pause here to note a subtlety. The quantity $\SF^{S;\text{naive}}$ quoted above 
equals the $S$ matrix in the S channel only upto phase. In order to obtain
the fully correct $S$ matrix we analytically continue the phase unambiguous
quantity $\Sf^{S;\text{naive}}$ \footnote{Indeed it does not make sense to 
analytically continue $\SF$ as the ambiguous phases of this quantity are 
not necessarily Lorentz invariant, and so are not functions only of the 
Mandelstam variables.}. The result of that continuation is given by 
\beq\label{ancon}
\Sf^{S;\text{naive}}= \frac{\SF^{S;\text{naive}}}{X(s)}
\eeq
where\footnote{The factor of $X(s)$ is the analytic continuation of (see \eqref{Fdef2})
\beq\label{Ad2}
\left(\bar{u}(p) u(p+q)\right) \left(\bar{v}(-k-q)v(-k)\right)= X(q) = - \f{q^2+4 m^2}{4 m^2}\ .
\eeq
The analytic continuation of the above formula is same as $- 4 Y(s)$ (see \eqref{Yy}.)
} 
\beq\label{AY}
X(s)=- \f{-s+4 m^2}{4 m^2}= -4Y(s)\ .
\eeq
The full four fermion amplitude in the S channel, including phase is then
given by 
$$A_F^{S;\text{naive}}= \Sf^{S;\text{naive}} X(p, k, q)$$
where\footnote{The spinor quadrilinear is as defined in \eqref{Fnewdef} with momentum assignments
corresponding to the S channel \eqref{smom}.}
\beq\label{Ad}
X(p,k,q)=\f{1}{4m^2} \left(u(p+q)u(-p)\right) \left(v(k+q)v(-k)\right) \ .
\eeq
It is not difficult to check that 
$$|X(p,k,q)|=X(s)\ .$$
It follows that the S channel 4 fermion amplitude agrees with $\SF$ upto 
a convention dependent phase. This phase factor may be shown to vanish 
near the identity momentum configuration ($p_1=p_3$, $p_2=p_4$) and so 
does not affect the interference with identity, and in general has no 
physical effect; it follows we would make no error if we simply regarded 
$\SF$ as the four fermion scattering amplitude. At any rate we have been 
careful to express the unitarity relation in terms of the phase unambiguous
quantity $\Sf$ given unambiguously by \eqref{Fdef2}.

The naive S channel $S$ matrix \eqref{swrong} is not duality \eqref{dtransf} 
invariant. In later section, we also show that it also does not obey the 
constraints of unitarity, leading to an apparent paradox. 

A very similar paradox was encountered in \cite{Jain:2014nza} where it was 
conjectured that the usual rules of crossing symmetry are modified in 
matter Chern-Simons theories. It was conjectured in  \cite{Jain:2014nza} that the correct
transformation rule under crossing symmetry for {\it any} matter Chern-Simons theory with
fundamental matter in the large $N$ limit is given by 
\begin{align}\label{scorrect}
\SB^S(\mbp_1,\mbp_2,\mbp_3,\mbp_4)&= I(\mbp_1,\mbp_2,\mbp_3,\mbp_4)+ i
(2\pi)^3\de^3(p_1+p_2-p_3-p_4)
\TB^S(\mbp_1,\mbp_2,\mbp_3,\mbp_4)\ ,\nn\\
\SF^S(\mbp_1,\mbp_2,\mbp_3,\mbp_4)&= I(\mbp_1,\mbp_2,\mbp_3,\mbp_4)+ i
(2\pi)^3\de^3(p_1+p_2-p_3-p_4)
\TF^S(\mbp_1,\mbp_2,\mbp_3,\mbp_4)\ ,
\end{align}
where
\begin{align}\label{Tcorrect}
\TB^S(\mbp_1,\mbp_2,\mbp_3,\mbp_4)= -i(\cos(\pi\la)-1) I(\mbp_1,\mbp_2,\mbp_3,\mbp_4) +
\f{\sin(\pi\la)}{\pi\la} \TB^{S;\text{naive}}(\mbp_1,\mbp_2,\mbp_3,\mbp_4)\ ,\nn\\
\TF^S(\mbp_1,\mbp_2,\mbp_3,\mbp_4)=-i(\cos(\pi\la)-1) I(\mbp_1,\mbp_2,\mbp_3,\mbp_4) +
\f{\sin(\pi\la)}{\pi\la} \TF^{S;\text{naive}}(\mbp_1,\mbp_2,\mbp_3,\mbp_4) \ ,
\end{align}
where \eqref{Ssmnaive} defines the $T$ matrices obtained from naive 
crossing rules. In the center of mass frame the conjectured $S$ matrix \eqref{scorrect} has the
form 
\begin{align}\label{Scorrectcm}
\SB^S(s,\te)&= 8 \pi \sqrt{s}\de(\te)+ i\TB^S(s,\te)\ ,\nn\\
\SF^S(s,\te)&= 8 \pi \sqrt{s}\de(\te)+ i\TF^S(s.\te)\ ,
\end{align}
where
\begin{align}\label{Tcorrectcm}
 \TB^S(s,\te) &=- 8 \pi i \sqrt{s}(\cos(\pi \la)-1) \de(\te) +
\f{\sin(\pi\la)}{\pi\la}\TB^{S;\text{naive}}(s,\te)\ ,\nn\\
 \TF^S(s,\te) &= - 8 \pi i \sqrt{s}(\cos(\pi \la)-1)
\de(\te)+\f{\sin(\pi\la)}{\pi\la}\TF^{S;\text{naive}}(s,\te)\ .
\end{align}
The naive analytically continued $T$ matrices are
\begin{align}\label{Tnaivecm}
 \TB^{S;\text{naive}}(s,\te) &=4 \pi i \la \sqrt{s} \cot(\te/2)+J_B(\sqrt{s},\la)\ ,\nn\\
 \TF^{S;\text{naive}}(s,\te) &=4 \pi i \la\sqrt{s} \cot(\te/2)+J_F(\sqrt{s},\la)\ ,
\end{align}
where the $J$ functions are as defined in \eqref{SJfun}. In other words the conjectured $S$ matrix
takes the following form
\begin{align}\label{Scorrectcmfin}
\SB^S(s,\te)&= 8 \pi \sqrt{s}\cos(\pi \la) \de(\te)+ i\f{\sin(\pi\la)}{\pi\la}\left(4 \pi i
\la\sqrt{s} \cot(\te/2)+J_B(\sqrt{s},\la)\right)\ ,\nn\\
\SF^S(s,\te)&= 8 \pi \sqrt{s}\cos(\pi \la) \de(\te)+ i\f{\sin(\pi\la)}{\pi\la}\left(4 \pi i
\la\sqrt{s} \cot(\te/2)+J_F(\sqrt{s},\la)\right)\ .
\end{align}

It was demonstrated in \cite{Jain:2014nza} that the conjecture \eqref{Tcorrect} yields an S channel
$S$ matrix that is both duality invariant and consistent with unitarity in the the systems under
study
in that paper. In this paper we will follow \cite{Jain:2014nza} to conjecture that
\eqref{Tcorrect} continues to define the correct S channel $S$ matrix for the theories under study.
In the next section we will demonstrate that \eqref{Tcorrect} obeys the nonlinear unitarity 
equations \eqref{unicondf1T} and \eqref{unicondf2T}. We regard this fact
as highly nontrivial evidence in support of the conjecture \eqref{Tcorrect}. 
As \eqref{Tcorrect} appears to work in at least two rather different classes of 
large N fundamental matter Chern-Simons theories (namely the purely bosonic 
and fermionic theories studied in  \cite{Jain:2014nza} and the supersymmetric
theories studied in this paper) it seems likely that \eqref{Tcorrect} applies
universally to all Chern-Simons fundamental matter theories, as suggested 
in \cite{Jain:2014nza}. 

\subsubsection{Straightforward non-relativistic limit}
The conjectured S channel $S$ matrix has a simple non-relativistic limit leading to the known
Aharonov-Bohm result (see section 2.6 of \cite{Jain:2014nza} for details). In this limit we take
(in the center of mass frame) $\sqrt{s}\to 2 m$ in the $T$ matrix \eqref{Tcorrectcm} with all
other parameters held fixed. In this limit we find
\begin{align}\label{TcorrectcmNR}
 \TB^S(s,\te) &=- 8 \pi i \sqrt{s}(\cos(\pi \la)-1) \de(\te) +
4 \sqrt{s}\sin(\pi\la)\left(i \cot(\te/2)-1\right)\ ,\nn\\
 \TF^S(s,\te) &= - 8 \pi i \sqrt{s}(\cos(\pi \la)-1)
\de(\te)+4 \sqrt{s}\sin(\pi\la)\left(i \cot(\te/2)+1\right)\ .
\end{align}
The non-relativistic limit also coincides with the $\mN=2$ limit of the $S$ matrix
\eqref{scorrect} as we show in the following subsection. In \S \ref{modscal} we describe a slightly
modified non-relativistic limit of the $S$ matrix.

\subsection{$S$ matrices in the \texorpdfstring{$\mN=2$}{N=2} theory}\label{onssN2}

As discussed in \S \ref{rnt} the $\mN=1$ theory \eqref{action} has an enhanced $\mN=2$
supersymmetric regime when the $\Phi^4$ coupling constant takes a special value $w=1$. We have
already seen that the momentum dependent functions in the offshell four point function simplify
dramatically \eqref{solutionsN2}, and so it is natural to expect that
the $S$ matrices at $w=1$ are much simpler than at generic $w$. This is
indeed the case as we now describe. 

By taking the limit $w \to 1$ in the $S$ matrix formulae presented in \eqref{TBcov} and
\eqref{TFcov}, we find that the four boson and four fermion
$\mN=2$ $S$ matrices take the very simple form \footnote{This is because the $J$ functions reported
in \eqref{TBcov} and \eqref{TFcov} have an extremely simple form at $w=1$ (see
\eqref{N2Jfunctions}).}
\begin{align}
 \TB^{\mN=2}=& \frac{4 i \pi}{\ka}\epsilon_{\mu\nu\rho}\f{q^\mu(p-k)^\nu(p+k)^\rho}{(p-k)^2}-
\f{8 \pi m}{\ka} \ , \label{TBcovN2}\\ 
  \TF^{\mN=2}=& \frac{4 i \pi}{\ka}\epsilon_{\mu\nu\rho}\f{q^\mu(p-k)^\nu(p+k)^\rho}{(p-k)^2}+
\f{8 \pi m}{\ka}\  . \label{TFcovN2}
\end{align}
The $S$ matrices above are simply those for tree level scattering. It follows 
that the tree level $S$ matrices in the three non-anyonic channels 
are not renormalized, at any order in the coupling constant, 
in the ${\cal N}=2$ theory.

There is an immediate (but rather trivial) check of this result. Recall that according to \S
\ref{manN2} the four boson and four fermion scattering amplitudes are not independent in the
$\mN=2$ theory; supersymmetry determines the former in terms of the latter. The precise relation is
derived in \ref{manN2} and is given by \eqref{f1F2relN2form} for particle-antiparticle scattering
and \eqref{f1F2relN2form1} for particle-particle scattering. It is easy to verify that
\eqref{TBcovN2} and \eqref{TFcovN2} trivially satisfy \eqref{f1F2relN2form} (or 
\eqref{f1F2relN2form1}) using \eqref{cdef},\eqref{cstdef} and appropriate momentum assignments for
the  channels of scattering discussed in section \S\ref{ons}. \footnote{As an example, in the T
channel (see \eqref{tmom}) we substitute the coefficients \eqref{cdef}, \eqref{cstdef}
into \eqref{f1F2relN2form} and evaluate it to get
\beq\label{f1F2relN2}
\SB=\SF \f{-2 m (k - p)_-+i q_3 (k+p)_-}{2 m
(k-p)_-+i q_3 (k+p)_-} \ .
\eeq
It is clear that the covariant form of the $S$ matrices given in \eqref{TBcovN2} and
\eqref{TFcovN2} trivially satisfy \eqref{f1F2relN2}. Similarly it can be easily checked that the
result \eqref{f1F2relN2} follows from \eqref{f1F2relN2form1} for particle-particle scattering.}

For completeness we now present explicit formulae for the $S$ matrices of the 
$\mN=2$ theory in the three non-anyonic channels.

For the U$_d$ channel
\begin{align}\label{UdsmN2}
 \TB^{U_d ; \mN=2}&= E(q,p-k,p+k)\f{4\pi i}{k}\sqrt{\f{ts}{u}}- \f{8 \pi m}{\ka}\ ,\nn\\
 \TF^{U_d ; \mN=2}&= E(q,p-k,p+k)\f{4\pi i}{k}\sqrt{\f{ts}{u}}+ \f{8 \pi m}{\ka}\ .
\end{align}
For the U$_e$ channel
\begin{align}\label{UesmN2}
 \TB^{U_e; \mN=2}&=E(q,p-k,p+k) \f{4\pi i}{k}\sqrt{\f{us}{t}}- \f{8 \pi m}{\ka}\ ,\nn\\
 \TF^{U_e; \mN=2}&=E(q,p-k,p+k) \f{4\pi i}{k}\sqrt{\f{us}{t}}+ \f{8 \pi m}{\ka}\ .
\end{align}
For the T channel
\begin{align}\label{TsmN2}
 \TB^{T; \mN=2}&= E(q,p-k,p+k)\f{4\pi i}{k}\sqrt{\f{tu}{s}}-\f{8 \pi m}{\ka}\ ,\nn\\
 \TF^{T; \mN=2}&= E(q,p-k,p+k)\f{4\pi i}{k}\sqrt{\f{tu}{s}}+ \f{8 \pi m}{\ka}\ .
\end{align}

Let us now turn to the singlet channel. As described in \S \ref{onss}, we cannot compute the S
channel $S$ matrix directly because of our choice of the kinematic regime $q_\pm=0$. The naive
analytic continuation of \eqref{TBcovN2} and \eqref{TFcovN2} to the S channel gives
\begin{align}\label{SsmnaiveN2}
 \TB^{S;\text{naive};\mN=2}&= E(q,p-k,p+k)4\pi i\la\sqrt{\f{su}{t}}-8 \pi m\la\ ,\nn\\
 \TF^{S;\text{naive};\mN=2}&= E(q,p-k,p+k)4\pi i\la\sqrt{\f{su}{t}}+8 \pi m\la\ .
\end{align}
Thus the naive S channel $S$ matrix for the $\mN=2$ theory is
\begin{align}\label{swrongN2}
\SB^{S;\text{naive};
\mN=2}(\mbp_1,\mbp_2,\mbp_3,\mbp_4)=& I(\mbp_1,\mbp_2,\mbp_3,\mbp_4) 
\nn\\&+ i (2\pi)^3\de^3(p_1+p_2-p_3-p_4)\TB^{S;\text{naive};\mN=2}(\mbp_1,\mbp_2,\mbp_3,\mbp_4)\
,\nn\\
\SF^{S;\text{naive}; \mN=2}(\mbp_1,\mbp_2,\mbp_3,\mbp_4) =& I(\mbp_1,\mbp_2,\mbp_3,\mbp_4)\nn\\
&+ i (2\pi)^3\de^3(p_1+p_2-p_3-p_4) \TF^{S;\text{naive};\mN=2}(\mbp_1,\mbp_2,\mbp_3,\mbp_4)\ .
\end{align}
As explained in the introduction \S\ref{intro}, this result is obviously non-unitary. 
Applying the modified crossing symmetry transformation rules \eqref{scorrect} we obtain our 
conjecture for the $\mN=2$ $S$ matrix in the singlet channel
\begin{align}\label{scorrectN2}
\SB^{S; \mN=2}(\mbp_1,\mbp_2,\mbp_3,\mbp_4)&= I(\mbp_1,\mbp_2,\mbp_3,\mbp_4)+ i
(2\pi)^3\de^3(p_1+p_2-p_3-p_4)
\TB^{S; \mN=2}(\mbp_1,\mbp_2,\mbp_3,\mbp_4)\ ,\nn\\
\SF^{S; \mN=2}(\mbp_1,\mbp_2,\mbp_3,\mbp_4)&= I(\mbp_1,\mbp_2,\mbp_3,\mbp_4)+ i
(2\pi)^3\de^3(p_1+p_2-p_3-p_4)
\TF^{S; \mN=2}(\mbp_1,\mbp_2,\mbp_3,\mbp_4)\ ,
\end{align}
where
\begin{align}\label{TcorrectN2}
\TB^{S; \mN=2}(\mbp_1,\mbp_2,\mbp_3,\mbp_4)= -i(\cos(\pi\la)-1) I(\mbp_1,\mbp_2,\mbp_3,\mbp_4) +
\f{\sin(\pi\la)}{\pi\la} \TB^{S;\text{naive}; \mN=2}(\mbp_1,\mbp_2,\mbp_3,\mbp_4)\ ,\nn\\
\TF^{S; \mN=2}(\mbp_1,\mbp_2,\mbp_3,\mbp_4)=-i(\cos(\pi\la)-1) I(\mbp_1,\mbp_2,\mbp_3,\mbp_4) +
\f{\sin(\pi\la)}{\pi\la} \TF^{S;\text{naive}; \mN=2}(\mbp_1,\mbp_2,\mbp_3,\mbp_4)\ .
\end{align}
In the center of mass frame the conjectured S channel $S$ matrix in the $\mN=2$ theory takes the
form
\begin{align}\label{ScorrectcmN2}
\SB^{S; \mN=2}(s,\te)&= 8 \pi \sqrt{s}\de(\te)+ i\TB^S(s,\te)\ ,\nn\\
\SF^{S; \mN=2}(s,\te)&= 8 \pi \sqrt{s}\de(\te)+ i\TF^S(s.\te)\ ,
\end{align}
where
\begin{align}\label{TcorrectcmN2}
 \TB^{S; \mN=2}(s,\te) &=- 8 \pi i \sqrt{s}(\cos(\pi \la)-1) \de(\te) +
\sin(\pi\la)(4 i \sqrt{s} \cot(\te/2)-8 m )\ ,\nn\\
 \TF^{S; \mN=2}(s,\te) &= - 8 \pi i \sqrt{s}(\cos(\pi \la)-1)
\de(\te)+\sin(\pi\la)(4 i \sqrt{s} \cot(\te/2)+8  m )\ .
\end{align}
Note that as $\sqrt{s}\to 2m$ \eqref{TcorrectcmN2} reproduces the straightforward non-relativistic
limit of the $\mN=1$ theory \eqref{TcorrectcmNR}.

In other words the conjectured S channel $S$ matrix for the $\mN=2$ theory takes the following form
in the center of mass frame
\begin{align}\label{ScorrectcmfinN2}
\SB^{S; \mN=2}(s,\te)&= 8 \pi \sqrt{s}\cos(\pi \la) \de(\te)+ i\sin(\pi\la)\left(4 i \sqrt{s}
\cot(\te/2)-8 m \right)\ ,\nn\\
\SF^{S; \mN=2}(s,\te)&= 8 \pi \sqrt{s}\cos(\pi \la) \de(\te)+ i\sin(\pi\la)\left(4 i \sqrt{s}
\cot(\te/2)+8 m \right)\ .
\end{align}
We explicitly show that the conjectured S channel $S$ matrix is unitary in the following
section.

\section{Unitarity}\label{unitarityx}

In this section, we first show that the $S$ matrices in the T and U channel 
obey the unitarity conditions \eqref{unicondf1T} and \eqref{unicondf2T} 
at leading order in the large $N$ limit. As the relevant unitarity equations 
are linear, the unitarity equation is a relatively weak consistency check of 
the $S$ matrices computed in this paper. 

We then proceed to demonstrate that the $S$ matrix \eqref{scorrect} also obeys
the constraints of unitarity. As the unitarity equation is nonlinear in the 
S channel, this constraint is highly nontrivial, we believe it provides an 
impressive consistency check of the conjecture \eqref{scorrect}.

\subsection{Unitarity in the T and U channel}

We begin by discussing the unitarity condition for the T (adjoint) 
and U (particle - particle) channels. Firstly we note that 
the $S$ matrices in these channels are $O(1/N)$. Therefore the 
LHS of \eqref{unicondf1T} and \eqref{unicondf2T} are $O(1/N^2)$. It follows 
that the unitarity equations \eqref{unicondf1T} and \eqref{unicondf2T} are obeyed at leading order
in the large $N$ limit provided
\begin{align}\label{unitcondTU}
 \TB(\mbp_1,\mbp_2,\mbp_3,\mbp_4)&=\TB^*(\mbp_3,\mbp_4,\mbp_1,\mbp_2)\ ,\nn\\
 \TF(\mbp_1,\mbp_2,\mbp_3,\mbp_4)&=\TF^*(\mbp_3,\mbp_4,\mbp_1,\mbp_2)\ .
\end{align}

The four boson and four fermion $S$ matrices in the T channel are given in terms of 
the universal functions in \eqref{TBcov} and \eqref{TFcov} after applying the 
momentum assignments \eqref{tmom}. It follows that \eqref{unitcondTU} holds in the
T channel provided 
\begin{align}\label{unitcondT}
 \TB^T(p+q,-k-q,p,-k)&=\TB^{T*}(p,-k,p+q,-k-q)\ ,\nn\\
 \TF^T(p+q,-k-q,p,-k)&=\TF^{T*}(p,-k,p+q,-k-q)\ .
\end{align}
This equation may be verified to be true (see below for some details).

Similarly the U$_d$ channel $S$ matrix is obtained via the momentum assignments \eqref{udmom}; 
It follows that \eqref{unitcondTU} is obeyed provided 
\begin{align}\label{unitcondUd}
 \TB^{U_d}(p+q,k,p,k+q)&=\TB^{U_d*}(p,k+q,p+q,k)\ ,\nn\\
 \TF^{U_d}(p+q,k,p,k+q)&=\TF^{U_d*}(p,k+q,p+q,k)\ ,
\end{align}
which can also be checked to be true. 

Finally in the U$_e$ channel it follows from the momentum assignments \eqref{uemom} that 
\eqref{unitcondTU} holds provided 
\begin{align}\label{unitcondUe}
 \TB^{U_e}(k,p+q,p,k+q)&=\TB^{U_e*}(p,k+q,k,p+q)\ ,\nn\\
 \TF^{U_e}(k,p+q,p,k+q)&=\TF^{U_e*}(p,k+q,k,p+q)\ ,
\end{align}
which we have also verified. 

The $T$ matrices for all the above channels of scattering are reported in 
\S\ref{ons}. Note that the starring of the $T$ matrices in \eqref{unitcondTU} also involves a
momentum exchange $p_1\Leftrightarrow p_3$ and $p_2\Leftrightarrow p_4$. It follows that under this
exchange $q \to-q$. \footnote{For instance in the T channel, we get the equations 
\begin{align}
p'+q'=p \ , \ p'=p+q \ , \ -k'-q'=-k\ , \ -k'=-k-q\ .
\end{align}
It follows that $q'=-q$.}

In verifying \eqref{unitcondT}, \eqref{unitcondUd} and \eqref{unitcondUe} we have used the fact
that the functions $J_B$ and $J_F$ are both invariant under the combined operation 
of complex conjugation accompanied by the flip $q\to-q$ (see \eqref{Jreal}). 
We also use the fact that in each case (T, U$_d$ and U$_e$)  the factor
$E(q,p-k,p+k)$ flips sign under the momentum exchange $p_1\Leftrightarrow p_3$ 
and $p_2\Leftrightarrow p_4$; the sign obtained from this process compensates 
the minus sign from complex conjugating the explicit factor of $i$. 
\footnote{The unitarity conditions in these channels are simply the statement
that the $S$ matrices are real. The reality of $S$ matrices is tightly connected
to the absence of two particle  branch cuts in the $S$ matrices in these 
channels at leading order in large $N$.}

\subsection{Unitarity in S channel}
The $S$ matrix in the S channel is of $O(1)$ and one has to use the full non-linear unitarity
conditions \eqref{umesheqB12} and \eqref{umesheqF12} . We reproduce them here for convenience. 
\begin{align}
\f{1}{8\pi\sqrt{s}}\int d\te\bigg(- Y(s) (\TB^S(s,\te)+4 Y(s) \Tf^S(s,\te))
(\TB^{S*}(s,-(\al-\te))+4 Y(s) \Tf^{S*}(s,-(\al-\te)))\nn\\
+ \TB^S(s,\te) \TB^{S*}(s,-(\al-\te)) \bigg)= i (\TB^{S*}(s,-\al)-\TB^S(s,\al))\ ,\label{umesheqB1}
\end{align}
\begin{align}
\f{1}{8\pi\sqrt{s}}\int d\te\bigg(Y(s) (\TB^S(s,\te)+4 Y(s) \Tf^S(s,\te))
(\TB^{S*}(s,-(\al-\te))+4 Y(s) \Tf^{S*}(s,-(\al-\te)))\nn\\ -16 Y(s)^2 \Tf^S(s,\te)
\Tf^{S*}(s,-(\al-\te))
\bigg)= i 4Y(s)(-\Tf^S(s,\al)+\Tf^{S*}(s,-\al))\label{umesheqF1}\ ,
\end{align}
where
\beq\label{Y}
Y(s)= \f{-s+4m^2}{16 m^2}
\eeq
is as defined in \eqref{Ydef}, and $\TB^S$ corresponds to the bosonic $T$ matrix while $\Tf^S$
corresponds to the phase unambiguous part of the fermionic $T$ matrix in the Singlet (S) channel
given in \eqref{Tcorrect} (also see \eqref{ancon}). In center of mass coordinates it takes the form
\beq
\Tf^S(s,\te)= -\f{\TF^S(s,\te)}{4 Y(s)}\ .
\eeq
Substituting the above into \eqref{umesheqB1} and \eqref{umesheqF1}, the conditions 
for unitarity may be rewritten as 
\begin{align}
\f{1}{8\pi\sqrt{s}}\int d\te\bigg(- Y(s) (\TB^S(s,\te)-\TF^S(s,\te))
(\TB^{S*}(s,-(\al-\te))-\TF^{S*}(s,-(\al-\te)))\nn\\
+ \TB^S(s,\te) \TB^{S*}(s,-(\al-\te)) \bigg)= i (\TB^{S*}(s,-\al)-\TB^S(s,\al))\ ,\label{umesheqB}
\end{align}
\begin{align}
\f{1}{8\pi\sqrt{s}}\int d\te\bigg(Y(s) (\TB^S(s,\te)-\TF^S(s,\te))
(\TB^{S*}(s,-(\al-\te))-\TF^{S*}(s,-(\al-\te)))\nn\\ -\TF^S(s,\te) \TF^{S*}(s,-(\al-\te))  \bigg)= i
(\TF^S(s,\al)-\TF^{S*}(s,-\al))\ .\label{umesheqF}
\end{align}
Let us pause to note that under duality $\TB\to \TF$  and vice versa; 
it follows then \eqref{umesheqB} and
\eqref{umesheqF} map to each other under duality. In other words 
the unitarity conditions are compatible with duality.

We will now verify that our S channel $S$ matrix is indeed compatible with 
unitarity. Let us recall that the angular  dependence of the $S$ matrix, in the 
center of mass frame is given by 
\begin{align}\label{SmatricesSchannel}
 \TB^S &= H_B T(\te)+W_B - i W_2 \de(\te)\ ,\nn\\
 \TF^S &= H_F T(\te)+W_F - i W_2 \de(\te)\ ,
\end{align}
 where 
$$T(\te)=i \cot(\te/2). $$ 
We will list the particular values of the coefficient functions $H_B(s)$ etc 
below; we will be able to proceed for a while leaving these functions 
unspecified. 

Substituting \eqref{SmatricesSchannel} in \eqref{umesheqB} and doing the
angle integrations\footnote{The angle integrations in \eqref{umesheqB} can be done by using the
formula \begin{equation}\label{cotcot}
 \int d\theta {\rm Pv} \cot \left(\frac{\theta}{2}\right)  {\rm Pv} \cot
\left(\frac{\alpha-\theta}{2}\right)= 2\pi  -4 \pi^2 \delta(\alpha),
\end{equation}
where ${\rm Pv}$ stands for principal value. See \eqref{Fo-cotcot} for a simple check of this
formula. } we find that \eqref{umesheqB} is obeyed if and only if
\begin{align}\label{unitarity1}
& H_B-H_B^*=\f{1}{8 \pi \sqrt{s}}(W_2 H_B^*- H_B W_2^*) \ ,\nn\\
& W_2+W_2^*= -\f{1}{8 \pi \sqrt{s}} (W_2 W_2^*+4 \pi^2 H_B H_B^*) \ ,\nn\\
&W_B-W_B^*= \f{1}{8 \pi \sqrt{s}} ( W_2 W_B^* -W_2^* W_B)-\f{i}{4\sqrt{s}} (H_B H_B^*-W_B W_B^*)-
 \f{iY}{4 \sqrt{s}} (W_B-W_F)(W_B^*-W_F^*)\ .
\end{align}
Similarly \eqref{umesheqF} is obeyed if and only if 
\begin{align}\label{unitarity2}
& H_F-H_F^*=\f{1}{8 \pi \sqrt{s}}(W_2 H_F^*- H_F W_2^*) \ ,\nn\\
& W_2+W_2^*= -\f{1}{8 \pi \sqrt{s}} (W_2 W_2^*+4 \pi^2 H_F H_F^*) \ ,\nn\\
&W_F-W_F^*= \f{1}{8 \pi \sqrt{s}} ( W_2 W_F^* -W_2^* W_F)-\f{i}{4\sqrt{s}} (H_F H_F^*-W_F W_F^*)-
 \f{iY}{4 \sqrt{s}} (W_B-W_F)(W_B^*-W_F^*)\ .
\end{align}
The first two equations of \eqref{unitarity1} and \eqref{unitarity2} are entirely identical to the
first two equations of equation 2.66 in \cite{Jain:2014nza} for the non-supersymmetric case. The
third equation has an additional contribution due to supersymmetry. Note that \eqref{unitarity1} and
\eqref{unitarity2} are compatible with duality under $H_B\to H_F$ and $W_B\to W_F$ and vice versa.

Let us now proceed to verify that the  equations \eqref{unitarity1} and \eqref{unitarity2} are
indeed obeyed; for this purpose we need to use the specific values of the coefficient functions in
\eqref{SmatricesSchannel}. These functions are easily read off from the formulae \eqref{Tcorrectcm}
(that we reproduce here for convenience)
\begin{align}\label{snew}
 \TB^S &=- 8 \pi i \sqrt{s}(\cos(\pi \la)-1) \de(\te) + \f{\sin(\pi\la)}{\pi\la}\biggl(4 \pi i
\la \sqrt{s} \cot(\te/2)+J_B(\sqrt{s},\la)
\biggr)\ ,\nn\\
 \TF^S &= - 8 \pi i \sqrt{s}(\cos(\pi \la)-1) \de(\te)+\f{\sin(\pi\la)}{\pi\la}\biggl(4 \pi i \la 
\sqrt{s} \cot(\te/2)+J_F(\sqrt{s},\la)\biggr)\ ,
\end{align}
from which we find 
\begin{align}\label{Ws}
W_B= J_B(\sqrt{s},\la) \f{\sin(\pi\la)}{\pi\la}\ ,\nn\\
W_F= J_F(\sqrt{s},\la) \f{\sin(\pi\la)}{\pi\la}\ ,
\end{align}
where the explicit form of the $J$ functions are given in \eqref{SJfun}. While we also identify
\beq\label{HBHFW2T}
H_B=H_F= 4 \sqrt{s} \sin(\pi \la), \ W_2= 8 \pi \sqrt{s}(\cos(\pi \la)-1), \ T(\te)=i \cot(\te/2)\ .
\eeq
Using the above relations it is very easy to see that the first two 
equations in each of \eqref{unitarity1} and \eqref{unitarity2} are satisfied. 
The first equation in each of \eqref{unitarity1} and \eqref{unitarity2} holds because $H_B$, $H_F$ 
and $W_2$ are all real. The second equation in each case boils down to a 
true trigonometric identity. 

The functions $W_B$ and $W_F$ occur only in the third equation in 
\eqref{unitarity1} and \eqref{unitarity2}. These equations assert 
two nonlinear identities relating the (rather complicated) $J_B$ and $J_F$ 
functions. We have verified by explicit computation that these identities  
are indeed obeyed. It follows that the conjectured $S$ matrix \eqref{scorrect} 
is indeed unitary. 

At the algebraic level, the satisfaction of the unitarity equation appears to 
be a minor miracle. A small mistake of any sort (a factor or two or an 
incorrect sign) causes this test to fail badly.  In particular, unitarity 
is a very sensitive test of the conjectured form \eqref{scorrect} of the $S$
matrix. Let us recall again that this conjecture was first made in 
\cite{Jain:2014nza}, where it was shown that it leads to a unitary 
$2 \rightarrow  2$ $S$ matrix. The supersymmetric $S$ 
matrices of this paper are more complicated than the $S$ matrices of the 
purely bosonic or purely fermionic theories of \cite{Jain:2014nza}. In 
particular the unitarity equation for four boson and four fermion $S$ matrices 
is different in this paper from the corresponding equations in 
\cite{Jain:2014nza} (the difference stems from the fact that two bosons can 
scatter not just to two bosons but also to two fermions, and this second 
process also contributes to the quadratic part of the unitarity equations). 
Nonetheless the prescription \eqref{scorrect} adopted from \cite{Jain:2014nza}
turns out to give results that obey the modified unitarity equation of this 
paper. In our opinion this constitutes a very nontrivial check of the 
crossing symmetry relation \eqref{scorrect} proposed in \cite{Jain:2014nza}.

The unitarity equation is satisfied for the arbitrary ${\cal N}=1$ susy theory, 
and so is, in particular obeyed for the ${\cal N}=2$ theory. Recall that 
the ${\cal N}=2$ theory has a particularly simple $S$ matrix \eqref{TcorrectcmN2}. In fact in the T
and U channels the ${\cal N}=2$ $S$ matrix is tree level exact at leading order 
in large $N$. According to the rules of naive crossing symmetry the 
S channel $S$ matrix would also have been tree level exact. This result is 
in obvious conflict with the unitarity equation: in the equation 
$-i(T-T^\dagger)= T T^\dagger$ the LHS vanishes at tree level while the RHS is 
obviously nonzero. The modified crossing symmetry rules \eqref{scorrect}
resolve this paradox in a very beautiful way. According to the rules 
\eqref{scorrect}, the $T$ matrix is not Hermitian even if $T^{naive}$ is; 
as the term in \eqref{scorrect} proportional to identity is imaginary. 
It follows from \eqref{scorrect} that both LHS and the RHS of the unitarity 
equation are nonzero; they are infact equal, as we now pause to explicitly 
demonstrate. In the ${\cal N}=2$ limit (see \eqref{TcorrectcmN2}) we have 
\begin{align}
& H_B=H_F= 4 \sqrt{s} \sin(\pi \la)\ ,\nn\\
&W_B=- 8m \sin(\pi \la)\ ,\nn\\
&W_F= 8m \sin(\pi \la)\ ,\nn\\
& W_2= 8 \pi \sqrt{s}(\cos(\pi \la)-1)\ .
\end{align}
The first equation in \eqref{unitarity1} is satisfied because everything is real. 
We have checked that the second equation is satisfied using a trigonometric 
identity. \footnote{This is the only equation in which the LHS and RHS are 
both nonzero. The LHS is the imaginary part of the coefficient of identity.}
The third equation works because we have
\beq
(H_B H_B^*-W_B W_B^*) = -16\sin^2(\pi \la) (-s +4 m^2)
\eeq
and 
\beq
Y (W_B-W_F)(W_B^*-W_F^*) = 16\sin^2(\pi \la) (-s +4m^2)
\eeq
the other terms don't matter because everything else is real. The same thing is true for
\eqref{unitarity2} since 
\beq
(H_F H_F^*-W_F W_F^*) = -16\sin^2(\pi \la) (-s +4 m^2)
\eeq
and thus the unitarity conditions are satisfied by the conjectured $S$ matrix \eqref{Tcorrect} in
the
$\mN=2$ theory as well.

\section{Pole structure of \texorpdfstring{$S$}{S} matrix in the S channel}

The S channel $S$ matrix studied in the last two sections turns out to have 
an interesting analytic structure. In this section we will demonstrate that 
the $S$ matrix has a pole whenever $w <-1$. As we demonstrate below the 
pole is at threshold at $w=-1$, migrates to lower masses as $w$ is further 
reduced until it actually occurs at zero mass at a critical value 
$w=w_c(\lambda) <-1$. As $w$ is further reduced, the squared mass of the pole 
increases again, until the pole mass returns to threshold at $w=-\infty$. 

In order to establish all these facts let us recall the structure of four 
boson and four fermion $S$ matrix in the S channel. The $S$ matrices take the 
form (see \eqref{SJfun})
\begin{equation} \label{ssd}
\TB^S= \frac{n_b}{d_1d_2}, ~~~~\TF^S=\frac{n_f}{d_1 d_2}\ ,
\end{equation}
where
\begin{align}\label{defdef}
 d_{1}=&-4 |m|^2 \left(\text{sgn}(\la) (w-1) \left(\left(\frac{1+y}{1-y}\right)^{\lambda
}-1\right)+y
\left(-w \left(\frac{1+y}{1-y}\right)^{\lambda }+w+\left(\frac{1+y}{1-y}\right)^{\lambda
}+3\right)\right)\ ,\nn\\
 d_{2}=&~\text{sgn}(\la) (w-1) \left(\left(\frac{1+y}{1-y}\right)^{\lambda }-1\right)+y \left(w
\left(\left(\frac{1+y}{1-y}\right)^{\lambda }-1\right)+3 \left(\frac{1+y}{1-y}\right)^{\lambda
}+1\right)\ ,
\end{align}
\begin{align}
n_b= & -32 |m|^3 y \sin (\pi  \lambda ) \biggl(8~\text{sgn}(\la) (w+1) y
\left(\frac{1+y}{1-y}\right)^{\lambda }\nn\\&+(w-1) (\text{sgn}(\la)-y)
\left(\frac{1+y}{1-y}\right)^{2 \lambda} (\text{sgn}(\la) (w-1)+(w+3) y)\nn\\&-(w-1)
(\text{sgn}(\la)+y)
(\text{sgn}(\la) (w-1)-(w+3) y)\biggr)\ ,\nn\\
n_f= & 32 |m|^3 y \sin (\pi  \lambda ) \biggl(8~\text{sgn}(\la) (w+1) y
\left(\frac{1+y}{1-y}\right)^{\lambda }\nn\\
&-(w-1) (\text{sgn}(\la)-y)\left(\frac{1+y}{1-y}\right)^{2 \lambda} (\text{sgn}(\la) (w-1)+(w+3)
y)\nn\\&+(w-1) (\text{sgn}(\la)+y) (\text{sgn}(\la) (w-1)-(w+3) y)\biggr)\ ,
\end{align}
where $y=\sqrt{s}/2|m|$.
Through this discussion we assume that $\lambda m>0$ (recall this condition 
was needed for duality invariance).

The denominators $d_1$, $d_2$ and the numerators are all polynomials of 
$y$ and the quantity
$$X= \left( \frac{1+y}{1-y} \right)^\lambda\ .$$
Most of the interesting scaling behaviors we will encounter below 
are a consequence of the dependence of all quantities on $X$. 
Note that $d_1$ and $d_2$ are linear functions of $X$ while $n_b$ and 
$n_f$ are quadratic functions of $X$. It is consequently possible to 
recast $n_b$ and $n_f$ in the form 
\begin{equation} \begin{split} \label{numform}
n_b&= a_b d_1 d_2 + b_b d_1 + c_b d_2 \ ,\nn \\ 
n_f&= a_f d_1 d_2 + b_f d_1 + c_f d_2\ .\\
\end{split}
\end{equation}
Here $a_b$, $b_b$, $c_b$, $a_f$, $b_f$ and $c_f$ are polynomials of $y$ (but 
are independent of $X$) and are given by 
\begin{align} \label{abcval}
a_b & = y\ ,\nn\\
b_b & = (w-1)(\text{sgn}(\la)+y)^2\ ,\nn\\
c_b & = -4 |m|^2 (\text{sgn}(\la)-y) (\text{sgn}(\la) (w-1)-(w+3) y)\ ,\nn\\
a_f & = y \ ,\nn\\
b_f & = -(w-1) \left(1-y^2\right)\ ,\nn\\
c_f & = 4 |m|^2 (\text{sgn}(\la)+y) (\text{sgn}(\la) (w-1)-(w+3) y)\ .
\end{align}

In order to study the poles of the $S$ matrix we need to investigate 
the zeroes of the functions $d_1$ and $d_2$. Let us first consider the 
case $\lambda >0$. In this case it turns out that $d_1(y)$ has a zero 
for $w \in (-\infty, w_c]$, while $d_2(y)$ has a zero in the range 
$w \in [w_c, -1]$
where 
\begin{equation}\label{defwc}
w_c(\lambda)= 1 - \frac{2 }{|\lambda|}\ .
\end{equation}
At $w= - \infty$ the zero of $d_1$ occurs at $y=1$. 
As $w$ is increased the $y$ value of the zero decreases, until it reaches 
$y=0$ at $w=w_c$. At larger values of $w$, $d_1$ no longer has a zero. 
However $d_2(y)$ develops a zero. The zero of $d_2(y)$ starts out at $y=0$ 
when $w=w_c$, and then increases, reaching $y=1$ at $w=-1$. At larger 
values of $w$ neither $d_1$ nor $d_2$ have a zero.

When $\lambda<0$ we have an identical situation except that the roles 
of $d_1$ and $d_2$ are reversed. $d_2(y)$ has a zero 
for $w \in (-\infty, w_c]$, while $d_1(y)$ has a zero in the range 
$w \in [w_c, -1]$. At $w= - \infty$ the zero of $d_2$ occurs at $y=1$. 
As $w$ is increased the $y$ value of the zero decreases, until it reaches 
$y=0$ at $w=w_c$. At larger values of $w$, $d_2$ no longer has a zero. 
However $d_1(y)$ develops a zero. The zero of $d_1(y)$ starts out at $y=0$ 
when $w=w_c$, and then increases, reaching $y=1$ at $w=-1$. At larger 
values of $w$ neither $d_1$ nor $d_2$ have a zero.

In summary our $S$ matrix has a pole for $w \in (-\infty, -1]$. The pole 
lies at threshold at the end points of this range, and becomes massless 
at $w=w_c$. There are clearly three special values of $w$ in this range: 
$w=-1$, $w=w_c$ and $w= -\infty$. In the rest of this section we examine 
the neighborhood of three special points in turn. 

\subsection{Behavior near $w=-1 -\de w$}

In this subsection we study the pole in the neighborhood of $w=-1$. 
When $w\rightarrow -1-\delta w$ with $0<\de w<<1$, we also expand $y\rightarrow 1-\delta
y$ (where $0< \de y << 1$) and find that
\begin{align}\label{behm11}
 d_1\sim&4 |m|^2\biggl((\text{sgn}(\la)-1)\left(\de w - 2\left(\frac{2}{\de y}\right)^{\lambda
}\right) +2 (\text{sgn}(\la)+1)\biggr)\ ,\nn\\
d_2\sim & (\text{sgn}(\lambda )+1)\left(2 -\left(\frac{2}{\de y}\right)^{\lambda }
\de w \right)- 2 \left(\f{2}{\de y}\right)^\la
(\text{sgn}(\la)-1)\ ,\nn\\
a_b \sim &  1-\de y\ ,\nn\\
b_b \sim &  -(2+\de w)(\text{sgn}(\la)+1-\de y)^2\ ,\nn\\
c_b \sim & 4 |m|^2 (\text{sgn}(\la)-1+\de y) (\text{sgn}(\la) (2+\de w)+(2-\de w) (1-\de
y))\ ,\nn\\
a_f \sim &  1-\de y \ ,\nn\\
b_f \sim &  2\de y(2+\de w) (2-\de y)\ ,\nn\\
c_f \sim &  -4 |m|^2 (\text{sgn}(\la)+1-\de y) (\text{sgn}(\la) (\de w+2)+(2-\de w) (1-\de y))\ .
\end{align}
Let us first consider the case $\lambda >0$. In this case $d_1$ equals 
$16m^2$ at leading order and so does not have a zero for $\delta w$ and 
$\delta y$ small. On the other hand 
$$d_2 \propto  \left(2 -\left(\frac{2}{\de y}\right)^{\lambda }
\de w \right)$$
and so vanishes when 
\beq \label{scwm1}
\frac{\de w}{2}= \left(\frac{\de y}{2}\right)^{|\lambda|}, ~~~
\frac{\de y}{2}=\left(\frac{\de w}{2}\right)^{\f{1}{|\lambda|}}\ .
\eeq

When $\lambda<0$, $d_2$ is a monotonic function that never vanishes. However 
$d_1$ vanishes provided the condition \eqref{scwm1} is met. It follows that
the $S$ matrix has a pole when \eqref{scwm1} is satisfied for both signs of 
$\lambda$.

The pole in the $S$ matrix occurs due to the vanishing of the denominator 
$d_1d_2$. As this denominator is the same for both the boson boson 
$ \rightarrow$ boson boson and the fermion fermion $ \rightarrow$ 
fermion fermion $S$ matrices, both these scattering processes have
a pole at the value of $y$ listed in \eqref{scwm1}. The residue of this 
pole is, however, significantly different in the four boson and four 
fermion scattering processes. Let us first consider the four boson scattering 
term. The residue of the pole is determined by $b_b$ evaluated at  
\eqref{scwm1} (in the case $\lambda>0$) and $c_b$ evaluated at 
the same pole (in the case $\lambda <0$). In either case we find the structure
of the pole for four boson scattering to be
\begin{equation}\label{polebm1}
\TB\sim \f{\left(\f{\de y}{2}\right)^{|\la|}}{\de w- 2\left(\f{\de y}{2}\right)^{|\la|}}\ .
\end{equation} 
In a similar manner the residue of the pole for four fermion scattering is determined by $b_f$
evaluated at  
\eqref{scwm1} (in the case $\lambda>0$) and $c_f$ evaluated at 
the same pole (in the case $\lambda <0$). In either case we find that 
\begin{equation}\label{polefm1}
\TF\sim \f{\left(\f{\de y}{2}\right)^{1+|\la|}}{\de w- 2\left(\f{\de y}{2}\right)^{|\la|}}\ .
\end{equation} 
Notice that while the residue of the pole for four fermion scattering 
is suppressed compared to the residue of the same pole for 
four boson scattering by a factor of $(\delta w)^\frac{1}{|\lambda|}$.

\subsection{Pole near $y=0$}
There exists a critical value, $w=w_c(\lambda)$, at which both $d_1$ and $d_2$ have zeroes at
$y=0$. In order to locate $w_c$ we expand $d_1$ and $d_2$ about
$y=0$. To linear order we find 
\begin{align}
 d_1=d_2\sim & y (\la  \text{sgn}(\la) (w-1)+2)\ .
\end{align}
Clearly $d_1$ and $d_2$ have a common zero at $y=0$ provided 
\beq\label{wcr}
w=w_c(\lambda)= 1-\frac{2}{|\la|}\ .
\eeq

In order to study this pole in the neighborhood of $w=w_c$ we set 
$w=w_c+\de w$ (with $|\de w|<1$) near $y=\de y$ (with $0<\de y<<1$); expanding
in $\delta w$ and $\delta y$ we find 
\begin{align}\label{beh0}
d_1\sim & \f{8|m|^2 \de y \left(\de w \la +2 \de y
(1-|\la|)\right)}{\text{sgn}(\la)}\ ,\nn\\
d_2\sim & \f{\de y\left(\de w \la -2 \de y (1-|\la|)\right)}{\text{sgn}(\la)}\ ,\nn\\
n_b \sim &- \f{512 |m|^3 \sin(\pi\la)\de y^2 (-1+|\lambda| )}{\la}\ ,\nn\\
n_f \sim & \f{512 |m|^3 \sin(\pi\la)\de y^2 (-1+|\lambda| )}{\la}\ .
\end{align}

The product $d_1d_2$ vanishes when 
\beq\label{scbzer}
\de y = \f{|\la \de w|}{2(1-|\la|)}, ~~~ {\rm i.e.}~~~ \de y^2= \frac{\lambda^2 \delta w ^2}
{4 (1-|\lambda|)^2}\ .
\eeq
\footnote{ $d_1 d_2$ also vanishes quadratically at $\delta y=0$. 
Note however that both $n_b$ and $n_f$ are proportional to $\delta y^2$. 
Consequently the factors of $\delta y^2$ cancel between the numerator and 
denominator.} The residue of the pole at $y=0$ for any sign of $\la$ is given by substituting
\eqref{scbzer} into the functions $n_b$ and $n_f$ in \eqref{beh0}.
We find the pole structure of the bosonic $S$ matrix near $y=0$ to be
\begin{align}\label{polewcb}
 \TB\sim-\f{64 |m| \sin(\pi\la)(-1+|\lambda| )}{|\la|\left(\de w^2 \la^2 -4 \de y^2
(1-|\la|)^2\right)}\ .
\end{align}
In a similar manner we find the pole structure of the fermion $S$ matrix near $y=0$ to be
\begin{align}\label{polewcf}
\TF\sim\f{64 |m| \sin(\pi\la)(-1+|\lambda| )}{|\la|\left(\de w^2 \la^2 -4 \de y^2
(1-|\la|)^2\right)} \ .
\end{align}

\subsection{Behavior at $w$ $\rightarrow$ $-\infty$ }

We now turn to the analysis of the pole structure at $w\to -\infty$. This is easily achieved by
setting $w=-\frac{1}{\de w}$ with $0< \de w<<1$ and $y\rightarrow 1-\delta y$ with
$0<\delta y<<1$ . The various functions \eqref{defdef} in the $S$ matrix \eqref{ssd} have the
behavior
\begin{align}\label{behinf}
 d_1\sim & \f{4|m|^2}{\de w}\left(
(\de y+\text{sgn}(\la)-1)\left(1-\left(\f{2}{\de y}\right)^{\la}\right)+(\text{sgn}(\la)+3)
\de w\right)\ ,\nn\\
d_2 \sim & \f{1}{\de w}\left((-\de y+\text{sgn}(\la)+1)-\left(\frac{2}{\de y}\right)^{\la }
((\text{sgn}(\la)-3) \de w +\text{sgn}(\la)+1)\right)\ ,\nn\\
a_b \sim &  1-\de y\ ,\nn\\
b_b \sim &  -\f{1}{\de w}(\text{sgn}(\la)+1-\de y)^2\ ,\nn\\
c_b \sim & -4 |m|^2 (\text{sgn}(\la)-1+\de y) (-\text{sgn}(\la) (1+\f{1}{\de w})-(3-\f{1}{\de w})
(1-\de y))\ ,\nn\\
a_f \sim &  1-\de y\ , \nn\\
b_f \sim &  \f{\de y}{\de w} (2-\de y)\ ,\nn\\
c_f \sim & 4 |m|^2 (\text{sgn}(\la)+1-\de y) (-\text{sgn}(\la) (1+\f{1}{\de w})-(3-\f{1}{\de w})
(1-\de y))\ .
\end{align}
Let us first consider the case $\lambda >0$. In this case $d_2$ is a monotonic function that never
vanishes and so does not have a zero for $\delta w$ and 
$\delta y$ small. On the other hand 
$$d_1 \propto  \left(\de w -\f{1}{2}\left(\frac{\de y}{2}\right)^{1-|\lambda| }\right)$$
and so vanishes when 
\beq\label{scwinf}
\de w= \f{1}{2} \left(\f{\de y}{2}\right)^{1-|\la|}, ~~~ \de y= \left(\f{4 \de
w}{2^{|\la|}}\right)^{\f{1}{1-|\la|}}\ .
\eeq
When $\lambda<0$, $d_1$ is a constant $-8 m^2$. However 
$d_2$ vanishes provided the condition \eqref{scwinf} is met. It follows that
the $S$ matrix has a pole when \eqref{scwinf} is satisfied for both signs of 
$\lambda$.

The pole in the $S$ matrix occurs due to the vanishing of the denominator 
$d_1d_2$. As this denominator is the same for both the boson boson 
$ \rightarrow$ boson boson and the fermion fermion $ \rightarrow$ 
fermion fermion $S$ matrices, both these scattering processes have
a pole at the value of $y$ listed in \eqref{scwinf}. The residue of this 
pole is different in the four boson and four fermion scattering processes as before. Let us first
consider the four boson scattering term. The residue of the pole is determined by $c_b$ evaluated at
\eqref{scwinf} (in the case $\lambda>0$) and $b_b$ evaluated at 
the same pole (in the case $\lambda <0$). In either case we find the structure
of the pole for four boson scattering to be
\begin{equation}\label{polebinf}
\TB\sim \f{\left(\f{\de y}{2}\right)^{2-|\la|}}{\de w-\f{1}{2}\left(\f{\de y}{2}\right)^{1-|\la|}}\
.
\end{equation} 
In a similar manner the residue of the pole for four fermion scattering is determined by $c_f$
evaluated at \eqref{scwm1} (in the case $\lambda>0$) and $b_f$ evaluated at 
the same pole (in the case $\lambda <0$). In either case we find that 
\begin{equation}\label{polefinf}
\TF\sim \f{\left(\f{\de y}{2}\right)^{1-|\la|}}{\de w-\f{1}{2}\left(\f{\de y}{2}\right)^{1-|\la|}}\
.
\end{equation} 
Notice that the residue of the pole for four boson scattering is supressed by 
a factor of $(\delta w)^\frac{1}{1-|\lambda|}$ compared to the residue for 
four fermion scattering.

 \subsection{Duality invariance}\label{dualpol}
It is most interesting to note that the statements and results obtained in the above
sections (\eqref{scwm1}, \eqref{wcr} and \eqref{scwinf}) are all duality invariant.
This is most transparent from the observation that under the duality transformation
\eqref{dtransf}\footnote{Under duality transformation $d_1$ and $d_2$ transform into one another
upto an overall non-zero factor. This overall factor is cancelled by an identical contribution from
the duality transform of the numerator.}
\begin{align}
 d_1\leftrightarrow d_1\ ,\nn\\
 d_2\leftrightarrow d_2\ .
\end{align} 
Hence the zeroes of $d_1$ and $d_2$ (\eqref{scwm1} and \eqref{scwinf}) should map to
themselves, and $w_c$ \eqref{wcr} should be duality invariant. Also recollect that under duality
the bosonic and fermionic $S$ matrices map to one another. Thus it is natural to expect that the
pole in the bosonic $S$ matrix at $w=-1$ \eqref{scwm1} should map to the pole of the fermionic $S$
matrix at $w=-\infty$ \eqref{scwinf} and vice versa. Since both the bosonic and fermionic $S$
matrices have a pole at $w=w_c$ \eqref{wcr} at $y=0$, this pole should be self dual.

Upon using \eqref{dtransf} on \eqref{wcr} it is straightforward to see that it is duality invariant.
The slightly non-trivial part is the mapping of the two scaling regimes \eqref{scwm1} and
\eqref{scwinf}. It is straightforward to obtain the identification from $w=-\infty$ to $w=-1$
from \eqref{dtransf}
\beq
-\f{1}{\de w_\infty}= \f{3- (-1-\de w_{-1})}{1+(-1-\de w_{-1})}\sim -\f{4}{\de w_{-1}}
\eeq
Using the above result in \eqref{scwinf} and applying \eqref{dtransf} for $\la$ it is easy to check
that \eqref{scwm1} follows (and vice versa).

\subsection{Scaling limit of the \texorpdfstring{$S$}{S} matrix} \label{modscal}
In this subsection we discuss a particularly interesting near-threshold limit of the S-matrix. It
was shown in \cite{Dandekar:2014era} that in this limit the $S$ matrices for the boson-boson and
fermion-fermion reduce to the ones that are obtained by solving the Schrodinger equation with
Amelino-Camelia-Bak boundary conditions \cite{AmelinoCamelia:1994we,Kim:1996rz}. In this subsection
we illustrate that the analysis of \cite{Dandekar:2014era} applies for our results as well. We
consider the near threshold region
\begin{equation}
 y=1+ \frac{k^2}{2m^2}
\end{equation} 
with $k<<1$ and
\begin{equation}
 w=-1-\delta w
\end{equation} 
where $0<\delta w<<1$. In the limit
\begin{equation}
 k\rightarrow 0,~~\delta w \rightarrow 0 ,~~,\frac{k^2}{4m^2}\left(\frac{\delta
w}{2}\right)^{-\frac{1}{|\lambda|}}=\rm{fixed}
\end{equation}
the $J$ function in the bosonic $S$ matrix (\eqref{ssd}) reduces to \footnote{Here we work in
the regime $\sqrt{s}>2m$ i.e $y>1$ and hence the appearance of the factors of $e^{i\pi \la}$.}
\begin{equation}\label{anred}
 J_B=8|m \sin(\pi\lambda)|
\frac{1+e^{i\pi|\lambda|}\frac{A_{R}}{k^{2|\lambda|}}}{1-e^{i\pi|\lambda|}\frac{A_{R}}{k^{
2|\lambda|}}}.
\end{equation}
where
\beq\label{anr}
A_{R}=\frac{4^{|\lambda|}}{2}|m|^{2|\lambda|}\delta w\ .
\eeq
Comparing our Lagrangian \eqref{compact} with that of eq 1.1 of \cite{Dandekar:2014era} we make the
parameter identifications
$$\delta w=\frac{\delta b_4}{8|m|\pi\lambda}\ .$$
Substituting $\de w$ in \eqref{anr} we see that \eqref{anred} matches exactly with eq 1.12 of
\cite{Dandekar:2014era}. 

\subsection{Effective theory near $w=w_c$?}

As we have explained above, our theory develops a massless bound state 
at $w=w_c$; the mass of this bound state scales like $w - w_c$ in units of 
the mass of the scattering particles. \footnote{We expect all of these results 
to continue to hold at finite $N$ at least when $N$ is large; in the rest 
of the discussion we assume that $N$ is finite, and so the interactions 
between two bound state particles is not parametrically suppressed.}
When $w-w_c \ll 1$ there is a separation of scales between the new bound 
state and all other excitations in our theory. In this regime the effective 
dynamics of the nearly massless particles should be governed by an autonomous
quantum field theory that makes no reference to UV degrees of freedom. 
It seems likely that the superfield that creates the bound states is 
a real ${\cal N}=1$ superfield. The fixed point that governs the dynamics of 
this field presumably has a single relevant deformation; as it was possible 
to approach this theory with a single fine tuning (setting $w=w_c$).
 These considerations suggest that the dynamics of the light bound state 
is governed by an ${\cal N}=1$ Wilson-Fisher theory built out of a single 
real superfield. If this suggestion is correct it would imply that the long distance dynamics of
the light bound states is independent of $\la$. Given that the bound states are gauge neutral this
possibility does not seem absurd to us. It would be interesting to study this suggestion in future
work.

\section{Discussion}

In this paper we have presented computations and conjectures for the all 
orders $S$ matrix in the most general renormalizable ${\cal N}=1$ Chern-Simons matter theory with a
single fundamental matter multiplet. Our 
results are consistent with unitarity if we assume that the
usual results of crossing symmetry are modified in precisely the manner
proposed in \cite{Jain:2014nza}, whereas the usual crossing symmetry rules are inconsistent with
unitarity. We view this fact as a nontrivial 
consistency check of the crossing symmetry rules proposed in   
\cite{Jain:2014nza}.

The `particle - antiparticle' $S$ matrix in the singlet channel conjectured in 
this paper has an interesting analytic structure. In a certain range of 
superpotential parameters the $S$ matrix has a bound state pole; a one 
parameter tuning of superpotential parameters can be used to set the 
pole mass to zero. We find the existence of a massless bound state in a 
theory whose elementary excitations are all massive fascinating. 
It would be interesting to further investigate the low energy dynamics of these massless bound
states. It would also be interesting to investigate if these bound states are `visible' in the 
explicit results for the partition functions of Chern-Simons matter theories. 

As we have explained in the previous section, our singlet sector 
particle - antiparticle $S$ matrix has a simple non-relativistic limit. 
It would be useful to reproduce this scattering amplitude from the solution 
of a manifestly supersymmetric Schrodinger equation. 

The results of this paper suggest many natural extensions and questions. 
First it would be useful to generalize the computations of this paper to the  mass deformed 
${\cal N}=3$ and especially to the mass deformed ${\cal N}=6$ susy 
gauge theories (the later is necessarily a $U(N) \times U(M)$ theory; the 
methods of this paper are likely to be useful in the limit $N \to \infty$ 
with $M$ held fixed). This generalization should allow us to make contact 
with earlier studies of scattering in ABJ theory \cite{Agarwal:2008pu,Bargheer:2012cp,Bianchi:2011fc,Chen:2011vv,Bianchi:2011dg,Bianchi:2014iia,
Bianchi:2012cq} that were performed arbitrary values of $M$ and $N$ but perturbatively (to given
loop order) in $\lambda$. 

At the ${\cal N}=2$ point the $S$ matrices presented in this paper 
are tree level exact in the three non anyonic channels, and depend on 
$\lambda$ in a very simple way in the singlet channel. It is possible that 
this very simple result can be deduced in a more structural manner using only general principles and
${\cal N}=2$ supersymmetry. It would be interesting if this were the case. 

As an intermediate step in the computation of the $S$ matrix we evaluated 
the off shell four point function of four superfields. This four 
point correlator was rather complicated in the general ${\cal N}=1$ theory, 
but extremely simple at the ${\cal N}=2$ point. The four point correlator 
(or sum of ladder diagrams) is a useful intermediate piece in the evaluation
of two, three and four point functions of gauge invariant operators
\cite{Aharony:2012nh,GurAri:2012is,Frishman:2014cma,Bedhotiya:2015uga}. 
The simplicity of the ${\cal N}=2$ results suggest that it would be rather 
easy to explicitly evaluate such correlators, at least in special kinematic 
limits. Such computations could be used as independent checks of duality
as well as well as inputs into ${\cal N}=2$ generalizations of the 
Maldacena-Zhiboedov solutions of Chern-Simons fundamental matter theories
\cite{Maldacena:2011jn,Maldacena:2012sf}. 

All of the computations in this paper have been performed under the 
assumption $\lambda m \geq 0$. Atleast naively all of the checks of duality 
(including earlier checks involving the partion function) fail when 
$\lambda m <0$. It would be interesting to understand why this is the case. 
It is possible that our theory undergoes a phase transition as $\lambda m$
changes sign (see \cite{Aharony:2012ns,Jain:2013gza} for related discussion). It would be
interesting to understand this better.  

We believe that the results of this paper put the crossing symmetry 
relations conjectured in \cite{Jain:2014nza} on a firm footing. It would be
interesting to find a rigorous proof of these crossing relations, and even 
more interesting to hit upon a plausible generalization of these 
relations to finite $N$ and $k$. From a traditional 
perturbative point of view the modified crossing symmetry rules 
are presumably related to infrared divergences. It thus seems possible that 
one route to a proof and generalization of these relations lies in a 
detailed study of the infrared divergences of the relevant Feynman graphs.
We hope to return to several of these questions in the future.

\acknowledgments
We would like to thank S. Ananth, Y. Dandekar, T. Hartman, Y. T Huang, R. Loganayagam, 
M. Mandlik, D. Son and S. Wadia for useful discussions. We would also 
like to thank O Aharony and S. Prakash for helpful comments on the manuscript. 
K.I. and V.U. would like to thank Warren Siegel for helpful correspondence.
K.I. and S.Mazumdar would like to acknowledge IOPB and IIAS - Hebrew 
University for hospitality while this work was in
progress. K.I. would also like to thank ICTP, Trieste for 
hospitality while this work was in progress. S.M. would like to acknowledge 
the hospitality of HRI Allahabad, IISER Kolkata, the University of
Swansea, and the University of Cambridge while this work was in progress.  
The work of S.J. was supported in part by the grant DOE 0000216873. The work 
of K.I, S.M., S. Mazumdar and V.U. was supported in part by a joint UGC-ISF grant 
F. No. 6-20/2014(IC).The work of S.Y. was supported by the Israeli Science Foundation under
grant 352/13 and 504/13. We would all like to acknowledge our debt to the people of
India for their generous and steady support for research in basic sciences.

\appendix
\section{Notations and conventions}\label{appendix}
\subsection{Gamma matrices}\label{gamma}
In this section, we list the various notations and conventions used in this paper. We follow those
of \cite{Gates:1983nr}. We list them here for convenience. 

The metric signature is $\et_{\mu\nu}=\{-,+,+\}$. In three dimensions the Lorentz group is
$SL(2,\mathbb{R})$ and it acts on two component real spinors $\psi^\al$, where $\al$ are the spinor
indices. A vector is represented by either a real and symmetric spinor $V_{\al\be}$ or a symmetric
traceless spinor $V_\al^\be$, where $V_{\al\be}=V_\mu\ga^\mu_{\al\be}$. We will choose our gamma
matrices in the real and symmetric form
\cite{Dumitrescu:2011iu}
\beq\label{realsymm}
 \ga^\mu_{\al\be}=\{\mathbb{I},\si^3,\si^1\} \ .
\eeq
The charge conjugation matrix $C_{\al\be}$ is used to raise and lower the spinor indices
\beq\label{raiselower}
C_{\al\be}=-C_{\be\al}=\begin{bmatrix}
              0 & -i \\
	      i & 0
             \end{bmatrix}=-C^{\al\be} \ .
\eeq
In the above, note that $C_{\be\al}= C^T$ and $C^{\al\be}=(C^T)^{-1}$. It follows that
\beq
C_{\al\ga}C^{\ga\be}=-\de_\al^{\ \be} \ ,
\eeq
where $\de_{\al}^{ \ \be}$ is the usual identity matrix. The spinor indices are raised and lowered
using the NW-SE convention
\beq
\psi^\al=C^{\al\be}\psi_\be \ , \psi_\al=\psi^\be C_{\be\al} \ .
\eeq
We also use the notation $\psi^2=\f{1}{2}\psi^\al\psi_\al=i\psi^+\ps^-$. Note that $\psi^2$ is
Hermitian. Since $\ps^\al$ is real, it is clear that $\ps_\al$ is imaginary since the charge
conjugate matrix is imaginary. 

The Clifford algebra is defined using the matrices $(\ga^\mu)_\al^{\ \be}$ and these can be obtained
by raising the indices using $C^{\al\be}$ as illustrated above
\beq
(\ga^\mu)_\al^{ \ \be}=\{\si^2,-i\si^1,i\si^3\} \ .
\eeq
Note that these matrices are purely imaginary. Choosing the $\ga^\mu_{\al\be}$ as real and
symmetric always yields this and vice versa. Our $\mu=0,1,3$, since at some point we will do an
euclidean rotation from the $\mu=0$ direction to $\mu=2$. It is clear that $(\ga^0)^2=1,
(\ga^1)^2=-1, (\ga^3)^2=-1$, therefore with our metric conventions the Clifford algebra is satisfied
by
\beq\label{clifford}
(\ga^\mu)_\al^{ \ \ta}(\ga^{\nu})_\ta^{\ \be}+(\ga^\nu)_\al^{ \ \ta}(\ga^{\mu})_\ta^{\
\be}=-2\et^{\mu\nu}\de_\al^{\ \be} \ .
\eeq
Another very useful relation is 
\beq
[\ga^\mu,\ga^\nu]=-2i \ep^{\mu\nu\rh}\ga_\rh \ , \ep^{013}=-1 \ .
\eeq
For completion we also note that
\beq
(\ga^\mu)^{\al\be}=\{\mathbb{I},\si^3,\si^1\} \ .
\eeq
As a consequence of the Clifford algebra \eqref{clifford}, we get a minus sign in the trace
\beq
k_{\al}^{\ \be} k_\be^{ \ \al}=-2 k^2 \ .
\eeq
The Euclidean counterpart of \eqref{clifford} is obtained by the standard Euclidean rotation
$\ga^0\rightarrow i \ga^2$
\beq
(\ga^\mu)_\al^{ \ \be}=\{i \si^2,-i\si^1,i\si^3\} \ ,\ \mu=2,1,3,
\eeq
and they satisfy the Euclidean Clifford algebra
\beq\label{cliffordeuclidean}
(\ga^\mu)_\al^{ \ \ta}(\ga^{\nu})_\ta^{\ \be}+(\ga^\nu)_\al^{ \ \ta}(\ga^{\mu})_\ta^{\
\be}=-2\de^{\mu\nu}\de_\al^{\ \be} \ .
\eeq
where $\de_{\mu\nu}=(+,+,+)$.
\subsection{Superspace}\label{superspace}
The two component Grassmann parameters $\te$ that appear in various places in superspace have
the properties
\begin{align}\label{grassmann}
 & \int d\te=0 \ , \int d\te \te=1 \ ,\int d^2\te \te^2=-1 \ , \int d^2\te \te^\al
\te^\be=C^{\al\be} \ ,\nn \\ 
& \f{\p\te^\al}{\p\te^\be}=\de_\be^{\ \al} \ ,  C^{\al\be}\f{\p}{\p\te^\be}\f{\p}{\p\te^\al}
\te^2=-2 \ ,\te^\al \te^\be=-C^{\al\be}\te^2 \ , \te_\al \te_\be=-C_{\al\be}\te^2 \ .
\end{align}
The definition of the delta function in superspace follows from the relation 
\beq
\int d^2\te \te^2 =-1 \implies \de^2(\te)=-\te^2 \ .
\eeq
Formally we write
\beq\label{gdelta}
\de^2(\te_1-\te_2)=-(\te_1-\te_2)^2= -(\te_1^2+\te^2_2-\te_1\te_2).
\eeq
The superspace derivatives are defined as
\begin{align}
 D_\al=\f{\p}{\p\te^\al}+ i \te^\be \p_{\al\be} \ , D^\al=C^{\al\be}D_\be \ .
\end{align}
We will mostly use the momentum space version of the above in which we replace
$i\p_{\al\be}\rightarrow k_{\al\be}$
\begin{align}\label{superderivativemom}
 D_\al=\f{\p}{\p\te^\al}+ \te^\be k_{\al\be} \ .
\end{align}
Note that the choice of the real and symmetric basis in \ref{realsymm} makes the momentum operator
Hermitian. The superspace derivatives satisfy the algebra
\beq\label{Dalgmom}
\{D_\al,D_{\be}\}=2 k_{\al\be} \ .
\eeq
The tracelessness of $(\ga^\mu)_\al^{ \ \be}$ implies that 
\beq
\{D^\al,D_\al\}=0 \ .
\eeq
Care has to be taken when integrating by parts with superderivatives due to their anticommuting
nature. From the expression for $D_\al$ we can construct
\beq\label{Dsq}
D^2=\f{1}{2}D^\al D_\al=\f{1}{2}\bigg(C^{\be\al}\f{\p}{\p\te^\al}\f{\p}{\p\te^\be}+2\te^\al
k_\al^{ \ \be}\f{\p}{\p\te^\be}+2\te^2 k^2\bigg) \ .
\eeq
From the above it is easy to verify 
\begin{align}\label{Dsqsq}
&(D^2)^2=-k^2 ,\nn\\ 
&D^2 D_\al=-D_\al D^2=k_{\al\be}D^\be\nn\\
&D^\al D_\be D_\al=0 \ .
\end{align}
using the properties given in \eqref{grassmann}. Yet another extremely useful relation is the
action of the superderivative square \eqref{Dsq} on the delta function \eqref{gdelta}
\beq\label{expformula}
D^2_{\te_1,k}\de^2(\te_1-\te_2)=1-\te_1^\al \te_2^\be k_{\al\be} -\te_1^2\te_2^2k^2=
\exp(-\te_1^\al \te_2^\be k_{\al\be}) \ .
\eeq
We will often suppress the spinor indices in the exponential with the understanding that the spinor
indices are contracted as indicated above. Some useful formulae are
\begin{align}\label{usefulformulae}
& \de^2(\te_1-\te_2)\de^2(\te_2-\te_1)=0 \ ,\nn\\
& \de^2(\te_1-\te_2)D^\al_{\te_2,k}\de^2(\te_2-\te_1)=0\ ,\nn\\
& \de^2(\te_1-\te_2)D^2_{\te_2,k}\de^2(\te_2-\te_1)=\de^2(\te_1-\te_2) \ ,
\end{align}
and the transfer rule
\beq\label{transferrule}
D_\al^{\te_1,p}\de^2(\te_1-\te_2)=-D_\al^{\te_2,-p}\de^2(\te_2-\te_1)\ .
\eeq

The supersymmetry generators
\beq\label{susygen}
Q_\al^{\te,k}=i\bigg(\f{\p}{\p\te^\al}-\te^\be k_{\al\be}\bigg) \  ,
\eeq
satisfy the anticommutation relations
\begin{align}
& \{Q_\al,Q_\be\}=2 k_{\al\be} \ , \nn\\
& \{Q_\al,D_\be\}=0 \ .
\end{align}
It is also clear that the transfer rule \eqref{transferrule} is the statement that the delta
function of $\te$ is a supersymmetric invariant. 

\subsection{Superfields}\label{superfieldexp}
The scalar superfield $\Ph(x,\te)$ contains a complex scalar $\ph$, a complex fermion $\ps^\al$,
and a complex auxiliary field $F$. The vector superfield $\Ga^\al(x,\te)$ consists of the gauge
field $V_{\al\be}$, the gaugino $\la^\al$, an auxiliary scalar $B$ and an auxiliary fermion
$\ch^\al$. The following superfield expansions are used repeatedly in several
places. We list them here for easy reference. 
\begin{align}
&\Ph = \ph+\te\ps-\te^2 F \ , \nn \\
&\bar{\Ph} =\bar{\ph}+\te\bar{\ps}-\te^2 \bar{F} \ ,\nn\\
&\bar{\Ph}\Ph
= \bar{\ph}\ph+\te^\al(\bar{\ph}\ps_\al+\bar{\ps}_\al\ph)-\te^2(\bar{F}\ph+\bar{\ph}F+\bar{\ps}\ps)\
, \nn\\
&D_\al\Ph =\ps_\al-\te_\al F+ i\te^2\p_\al^{\ \be}\ps_\be+i\te^\be\p_{\al\be}\ph \ , \nn\\
&D^\al\bar{\Ph}D_\al\Ph\big|_{\te^2} =\te^2(2\bar{F}F+2i\bar{\ps}^\al\p_\al^{ \ \be}\ps_\be-2
\p\bar{\ph}\p\ph)\ ,\nn\\
& D^2_{q,\te}(\bar{\Ph}\Ph)=(\bar{\ph}F+\bar{F}\ph+\bar{\ps}\ps)+\te^\al q_\al^{ \
\be}(\bar{\ph}\ps+\bar{\ps}\ph)+\te^2 q^2 (\bar{\ph}\ph)^2 \ , \nn\\
&\Ga^\al=\ch^\al-\te^\al B+i \te^\be A_\be^{ \ \al}-\te^2(2\la^\al-i\p^{\al\be}\ch_\be) \ .
\end{align}

\section{A check on the constraints of supersymmetry on \texorpdfstring{$S$}{S} matrices}
\label{manN1ex}

In \S\ref{susyscat} we demonstrated that the manifestly supersymmetric scattering of any $\mN=1$
theory in three dimensions is described by two independent functions. In this section, we directly
verify this result in theories whose 
offshell effective action takes the form \eqref{eef} with the function 
$V$ that takes the particular supersymmetric form \eqref{totform} 
(and so is determined by four unspecified functions $A$, $B$ $C$ and $D$). 

We wish to use \eqref{eef} to study scattering. In order to do this we 
evaluate \eqref{eef} with the fields $\Phi$ and ${\bar \Phi}$ in that action 
chosen to be the most general linearized onshell solutions to the 
equations of motion. In this appendix we focus on a particular scattering 
process - scattering in the adjoint channel. At leading order in the large 
$N$ limit we can focus on this channel by choosing the solution for 
$\Phi_m$ and ${\bar \Phi}^m$ in \eqref{eef} to be positive energy solutions 
(representing initial states), while ${\bar \Phi}^m$ and $\Phi_n$ are expanded
in negative energy solutions (representing final states). The negative 
and positive energy solutions are both allowed to be an arbitrary linear
combination of bosonic and fermionic solutions. Plugging these solutions into 
\eqref{eef} yields a functional of the coefficients of the bosonic and fermionic
solutions in the four superfields in \eqref{eef}. The coefficients of 
various terms in this functional are simply the $S$ matrices. For instance 
the coefficient of the term proportional to the product of four bosonic modes is the four boson
scattering amplitude, etc. 

Let us schematically represent the scattering process we study by 
\begin{eqnarray}\label{tso}
 \left(
\begin{array}{c}
 \Phi (\te_1,p_1) \\
 \bar{\Ph}(\te_2,p_2) \\
\end{array}
\right)\to \left(
\begin{array}{c}
 \bar{\Ph}(\te_3,p_3) \\
 \Phi (\te_4,p_4) \\
\end{array}
\right)\nn
\end{eqnarray}
where the LHS represents the in-states and the RHS represents the out-states. 
The momentum assignments in \eqref{eef} are
\beq
p_1=p+q \ , \ p_2=-k-q \ , \ p_3=p \ , \ p_4=-k \ .
\eeq

In component form \eqref{tso} encodes the following $S$ matrices
 \begin{eqnarray}
 \SB:\left(
\begin{array}{c}
 \phi (p_1) \\
 \bar{\ph}(p_2) \\
\end{array}
\right)\to \left(
\begin{array}{c}
 \bar{\ph}(p_3) \\
 \phi (p_4) \\
\end{array}
\right) \ , \ 
 \SF:\left(
\begin{array}{c}
 \psi (p_1) \\
 \bar{\ps}(p_2) \\
\end{array}
\right)\to \left(
\begin{array}{c}
 \bar{\ps}(p_3) \\
 \psi (p_4) \\
\end{array}
\right)\nn
\end{eqnarray}

\begin{eqnarray}
H_1:\left(\begin{array}{c}
 \phi(p_1) \\
 \bar{\phi}(p_2) \\
\end{array}\right)
\rightarrow 
\left( \begin{array}{c}
 \bar{\psi}(p_3) \\
 \psi(p_4) \\
\end{array}\right)\ , \ 
H_2: \left(
\begin{array}{c}
 \ps (p_1) \\
 \bar{\ps}(p_2) \\
\end{array}
\right)\to \left(
\begin{array}{c}
 \bar{\ph}(p_3) \\
 \ph (p_4) \\
\end{array}
\right)\nn
\end{eqnarray}
\begin{eqnarray}
H_3: \left(
\begin{array}{c}
 \phi (p_1) \\
 \bar{\ps}(p_2) \\
\end{array}
\right)\to \left(
\begin{array}{c}
 \bar{\ph}(p_3) \\
 \psi (p_4) \\
\end{array}
\right)\ , \ 
H_4: \left(
\begin{array}{c}
 \psi (p_1) \\
 \bar{\ph}(p_2) \\
\end{array}
\right)\to \left(
\begin{array}{c}
 \bar{\ps}(p_3) \\
 \phi (p_4) \\
\end{array}
\right)\nn
\end{eqnarray}
\begin{eqnarray}\label{tsall}
H_5: \left(
\begin{array}{c}
 \phi (p_1) \\
 \bar{\ps}(p_2) \\
\end{array}
\right)\to \left(
\begin{array}{c}
 \bar{\ps}(p_3) \\
 \phi (p_4) \\
\end{array}
\right)\ , \ 
H_6:\left(
\begin{array}{c}
 \psi (p_1) \\
 \bar{\ph}(p_2) \\
\end{array}
\right)\to \left(
\begin{array}{c}
 \bar{\ph}(p_3) \\
 \psi (p_4) \\
\end{array}
\right)
\end{eqnarray}
These $S$ matrix elements are all obtained by the process spelt out above
in terms of the four unknown functions $A, B, C, D$ (which we will take 
to be arbitrary and unrelated).  The functions A,B,C,D are to be evaluated at
the onshell conditions that follow from taking the momenta onshell, but that 
will play no role in what follows. 

It is not difficult to demonstrate that the 
boson-boson $\rightarrow$ boson boson and the 
fermion-fermion $\rightarrow$ fermion fermion
 $S$ matrices are given in terms of the functions $A$, $B$, $C$ and $D$ by 
\footnote{For the T channel we have used
\begin{align}\label{fermionsolns}
& v_\al(-k)=
\begin{pmatrix}
-\sqrt{k_+}\\ \f{(q_3-2 i m)}{2 \sqrt{k_+}} 
\end{pmatrix}
\ , \ \bar{v}^\al(-k-q)=
\begin{pmatrix} 
 -\f{2m+i q_3}{2\sqrt{k_+}} & i \sqrt{k_+}
\end{pmatrix}\nn\\
& u_\al(p+q)=
\begin{pmatrix}
 -i\sqrt{p_+} \\ \f{2m-iq_3}{2\sqrt{p_+}}
\end{pmatrix}
\ , \ 
\bar{u}^\al(p)=
\begin{pmatrix}
-\f{(2 i m+ q_3)}{2\sqrt{p_+}} & -\sqrt{p_+}
\end{pmatrix}
\end{align}}
\begin{align}
\SB &=(-4 i A m+ 4 B m^2-B q_3^2-q_3 (C k_-+ D p_-)) \ , \nn\\
\SF &=(B C^{\beta \alpha} C^{\delta \gamma}-i C\  C^{\beta \alpha }C^{+\gamma }
C^{+\delta }+ i D C^{\delta \gamma } C^{+\alpha } C^{+\beta })\bar{u}_{\al}(p_3)u_{\beta
}(p_1)v_\ga(p_2)\bar{v}_\de(p_4)\nn\\
&=-B (4 m^2+q_3^2)+ C k_- (2 i m-q_3)-D p_-(q_3+2 i m)\ .
\end{align}

The $S$ matrices for the remaining processes in \eqref{tsall} are 
also easily obtained: we find 
\beq\label{sbsf}
H_i= a_i \SB+ b_i \SF\ 
\eeq
where the coefficients are given by
\begin{align}
&a_1=\frac{\left(4 m^2+q_3^2\right) (q_3 (p-k)_-+2 i m
(k+p)_-)}{32 m k_- p_-\sqrt{k_+ p_+}}
\ &, \ &b_1=\frac{\left(4 m^2+q_3^2\right) (q_3 (k-p)_-+2 i m
(k+p)_-)}{32 m k_- p_-\sqrt{k_+ p_+}}\nn\\
&a_2=\frac{\left(4 m^2+q_3^2\right) (q_3 (p-k)_-+2 i m
(k+p)_-)}{32 m k_- p_-\sqrt{k_+ p_+}}
\ &, \ &b_2=\frac{\left(4 m^2+q_3^2\right) (q_3 (k-p)_-+2 i m
(k+p)_-)}{32 m k_- p_-\sqrt{k_+ p_+}}\nn\\
&a_3=-\f{2m+i q_3}{4m}\ &, \ &b_3=\f{2m+i q_3}{4m}\nn\\
&a_4=\f{2m-i q_3}{4m}\ &, \ &b_4=-\f{2m-i q_3}{4m}\nn\\
&a_5=\frac{\left(4 m^2+q_3^2\right) (q_3 (k+p)_--2 i m
(k-p)_-)}{32 m k_- p_-\sqrt{k_+ p_+}}
\ &, \ &b_5=-\frac{i \left(4 m^2+q_3^2\right) (2 m (k-p)_--i q_3
(k+p)_-)}{32 m k_- p_-\sqrt{k_+ p_+}}\nn\\
&a_6=\frac{i \left(4 m^2+q_3^2\right) (2 m (k-p)_-+i q_3
(k+p)_-)}{32 m k_- p_-\sqrt{k_+ p_+}}
\ &, \ &b_6=\frac{\left(4 m^2+q_3^2\right) (q_3 (k+p)_-+2 i m
(k-p)_-)}{32 m k_- p_-\sqrt{k_+ p_+}}
\end{align}
The above set of coefficients match with the coefficients directly evaluated 
from \eqref{cdef} and
\eqref{cstdef}. This is a consistency check of the results of \S\ref{susyscat}.

For the $\mN=2$ theory the $S$ matrix \eqref{SUSYSmatrix} should also obey an additional constraint
(see \S\ref{manN2}) that relates $\SB$ and $\SF$ through \eqref{f1F2relN2form}. For the T channel
this relation was evaluated in \eqref{f1F2relN2}, substituting this in \eqref{sbsf} it is easy to
verify that the $\te_2 \te_3$ and $\te_1\te_4$ terms in \eqref{SUSYSmatrix}
\begin{eqnarray}
H_5: \left(
\begin{array}{c}
 \phi (p_1) \\
 \bar{\ps}(p_2) \\
\end{array}
\right)\to \left(
\begin{array}{c}
 \bar{\ps}(p_3) \\
 \phi (p_4) \\
\end{array}
\right)\ , \ 
H_6:\left(
\begin{array}{c}
 \psi (p_1) \\
 \bar{\ph}(p_2) \\
\end{array}
\right)\to \left(
\begin{array}{c}
 \bar{\ph}(p_3) \\
 \psi (p_4) \\
\end{array}
\right)
\end{eqnarray}
vanish for the $\mN=2$ theory. This is consistent with the fact that the corresponding terms in the
tree level component Lagrangian \eqref{compact} vanish at the $\mN=2$ point $w=1$. 

\section{Manifest ${\cal N}=2$ supersymmetry invariance}\label{manN2}

In this appendix we discuss the general constraints on the $S$ matrix obtained 
by imposing ${\cal N}=2$ supersymmetry. In subsection \ref{susyscat} we have 
already solved the constraints coming from ${\cal N}=1$ supersymmetry. 
As an ${\cal N}=2$ theory is in particular also ${\cal N}=1$ supersymmetric, 
the results of this appendix will be a specialization of those of 
subsection \ref{susyscat}.

In the case of ${\cal N}=2$, we have to recall the notion of chirality. 
A `chiral' (antichiral) ${\cal N}=2$ superfield $\Phi$ is defined as
\begin{equation}\label{chidef}
  {\bar D}_\alpha\Phi = 0, ~~~~ D_\alpha{\bar\Phi} = 0.
\end{equation}
We define the following:
\begin{align}\label{opdefs}
 \theta_\alpha  = \frac{1}{\sqrt{2}}(\theta^{(1)}_\alpha - i \theta^{(2)}_\alpha), ~~~~~
{\bar\theta}_\alpha  = \frac{1}{\sqrt{2}}(\theta^{(1)}_\alpha + i \theta^{(2)}_\alpha).
\end{align}
Where the superscripts $(1)$ and $(2)$ indicate the two (real) copies of the ${\cal N}=1$
superspace. With these definitions, we can define the supercharges
\begin{align}\label{qdefs}
 Q_\alpha  &= \frac{1}{\sqrt{2}}(Q^{(1)}_\alpha + i Q^{(2)}_\alpha) =
i\left(\frac{\partial}{\partial\theta^\alpha} - i{\bar\theta}^\beta\partial_{\beta\alpha}\right),\\
{\bar Q}_\alpha &= \frac{1}{\sqrt{2}}(Q^{(1)}_\alpha - i Q^{(2)}_\alpha) =
i\left(\frac{\partial}{\partial{\bar\theta}^\alpha} - i\theta^\beta\partial_{\beta\alpha}\right).
\end{align}
Likewise, we can define the supercovariant derivative operators
\begin{align}\label{ddefs}
 D_\alpha  &= \frac{1}{\sqrt{2}}(D^{(1)}_\alpha + i D^{(2)}_\alpha) =
\left(\frac{\partial}{\partial\theta^\alpha} + i{\bar\theta}^\beta\partial_{\beta\alpha}\right),\\
{\bar D}_\alpha &= \frac{1}{\sqrt{2}}(D^{(1)}_\alpha - i D^{(2)}_\alpha) =
\left(\frac{\partial}{\partial{\bar\theta}^\alpha} + i\theta^\beta\partial_{\beta\alpha}\right).
\end{align}
The solutions to the constraints \eqref{chidef} for (off-shell) chiral and anti-chiral fields are
\begin{align}\label{chisol}
 \Phi = \phi + \sqrt{2}\theta\psi - \theta^2F +
i\theta{\bar\theta}\partial\phi-i\sqrt{2}\theta^2({\bar\theta}\slashed{\partial}\psi) +
\theta^2{\bar\theta}^2\partial^2\phi, \\
 {\bar\Phi} = {\bar\phi} + \sqrt{2}{\bar\theta}{\bar\psi} - {\bar\theta}^2{\bar F} -
i\theta{\bar\theta}\partial{\bar\phi}-i\sqrt{2}{\bar\theta}^2(\theta\slashed{\partial}{\bar\psi}) +
\theta^2{\bar\theta}^2\partial^2{\bar\phi}.
\end{align}
Here $\theta{\bar\theta}\partial\phi = \theta^\alpha{\bar\theta}^\beta\partial_{\alpha\beta}$ and
${\bar\theta}\slashed{\partial}\psi = {\bar\theta}^\alpha\partial_\alpha^{~\beta}\psi_\beta$ and so
on. 

In the context of the current paper the chiral matter superfield transforms 
in the fundamental representation of the gauge group while the antichiral matter superfield 
transforms in the antifundamental representation of the gauge group. It follows that it is 
impossible to add a gauge invariant quadratic superpotential to our action 
(recall that an ${\cal N}=2$ superpotential can only depend on chiral 
multiplets) in order to endow our fields with mass. However it is possible 
to make the matter fields massive while preserving ${\cal N}=2$ supersymmetry; 
the fields can be made massive using a $D$ term. 

As our theory has no superpotential, it follows that $F={\bar F} = 0$ 
on shell. We are interested in the action of supersymmetry on the 
on-shell component fields $\phi$
(${\bar\phi}$) which are defined as
\begin{align}\label{onshdefphi}
 \phi(x) = \int \frac{d^2p}{(2\pi)^2\sqrt{2p^0}}\left[a({\mathbf p})e^{i p\cdot x} +
a^{c\dagger}({\mathbf p})e^{-i p\cdot x}\right], \\ {\bar\phi}(x) = \int
\frac{d^2p}{(2\pi)^2\sqrt{2p^0}}\left[a^c({\mathbf p})e^{i p\cdot x} + a^\dagger({\mathbf p})e^{-i
p\cdot x}\right].
\end{align}
Likewise, for $\psi$ ($\psi^\dagger$) we have
\begin{align}\label{onshdefpsi}
 \psi(x) = \int \frac{d^2p}{(2\pi)^2\sqrt{2p^0}}\left[u_\alpha({\mathbf p}) \alpha({\mathbf p})e^{i
p\cdot x} + v_\alpha({\mathbf p})\alpha^{c\dagger}({\mathbf p})e^{-i p\cdot x}\right], \\
\psi^\dagger(x) = \int \frac{d^2p}{(2\pi)^2\sqrt{2p^0}}\left[u_\alpha({\mathbf p})\alpha^c({\mathbf
p})e^{i p\cdot x} + v_\alpha({\mathbf p})\alpha^\dagger({\mathbf p})e^{-i p\cdot x}\right].
\end{align}
In order to obtain this action we used the transformation properties listed in
equations F.16-F.20 of \cite{Jain:2012qi} and then specialized to the onshell limit. \footnote{Note
that the action of $Q_\alpha$ on the chiral field $\Phi$ is 
different from the action on the anti-chiral field ${\bar\Phi}$. Similar 
remarks apply for ${\bar Q}_\alpha$.} The results may be summarized as follows. 
 As before, we define the (super) creation and annihilation operators
\begin{align}\label{Adefs}
 A({\mathbf p}) &= a({\mathbf p}) + \alpha({\mathbf p})\theta, ~~~~~~~~ A^c({\mathbf p}) =
a^c({\mathbf p}) + \alpha^c({\mathbf p})\theta, \\
 A^{\dagger}({\mathbf p}) &= a^\dagger({\mathbf p}) + \theta\alpha^\dagger({\mathbf p}), ~~~~~
A^{c\dagger}({\mathbf p}) = a^{c\dagger}({\mathbf p}) + \theta\alpha^{c\dagger}({\mathbf p}).
\end{align}
The action of $Q_\alpha$ (and ${\bar Q}_\alpha$) on $A$ and $A^\dagger$ is
\begin{align}\label{qact1}
  [Q_\alpha, A({\mathbf p})] &= -i\sqrt{2} u_\alpha({\mathbf p})
\frac{\overrightarrow{\partial}}{\partial\theta}, ~~~~ [{\bar Q}_\alpha, A({\mathbf p})] =
i\sqrt{2}u^*_\alpha({\mathbf p})\theta, \nn\\
  [Q_\alpha, A^\dagger({\mathbf p})] &= i\sqrt{2}v^*_\alpha({\mathbf p})\theta, ~~~~ [{\bar
Q}_\alpha, A^\dagger({\mathbf p})] = i\sqrt{2}v_\alpha({\mathbf
p})\frac{\overrightarrow{\partial}}{\partial\theta}.
\end{align}
Similarly, the action of $Q_\alpha$ (and ${\bar Q}_\alpha$) on $A^c$ and $A^{c\dagger}$ is
\begin{align}\label{qact2}
  [Q_\alpha, A^c({\mathbf p})] &= i\sqrt{2} u^*_\alpha({\mathbf p})\theta, ~~~~ [{\bar Q}_\alpha,
A^c({\mathbf p})] = -i\sqrt{2}u_\alpha({\mathbf p})\frac{\overrightarrow{\partial}}{\partial\theta},
\nn\\
  [Q_\alpha, A^{c\dagger}({\mathbf p})] &= i\sqrt{2}v_\alpha({\mathbf
p})\frac{\overrightarrow{\partial}}{\partial\theta}, ~~~~ [{\bar Q}_\alpha, A^{c\dagger}({\mathbf
p})] = i\sqrt{2}v^*_\alpha({\mathbf p})\theta.
\end{align}
It is clear from \eqref{qact1} that $(Q_\alpha+{\bar Q}_\alpha)/\sqrt{2}$ produces the action of the
first supercharge $Q^{(1)}_\alpha$, which we have seen earlier. That this action produces the
correct differential operator given earlier is obvious as well. Therefore, in order to obtain the
second supercharge $Q^{(2)}_\alpha$, we simply operate with the other linear combination
$(Q_\alpha-{\bar Q}_\alpha)/i\sqrt{2}$.

Note that for the ${\cal N}=1$ case, it doesn't matter if we used $A^\dagger$ or
$A^{c\dagger}$ for the initial states ($A$ or $A^c$ for the final states), as is clear from
\eqref{qact2}. This agrees with the fact that the linear combination 
$(Q_\alpha+{\bar Q}_\alpha)/\sqrt{2}$ produces the same equation on all 
$S$ matrix elements. 
However other linear combinations of the two ${\cal N}=2$ supersymmetries 
act differently on $A$ and $A^c$, and so the constraints of ${\cal N}=2$ 
supersymmetry are different depending on which scattering processes we 
consider. 

\subsection{Particle - antiparticle scattering}

Let us first study the invariance of the
following $S$ matrix element
\begin{equation}\label{elem}
S({\mathbf p}_1, \theta_1, {\mathbf p}_2, \theta_2, {\mathbf p}_3, \theta_3, 
{\mathbf p}_4, \theta_4)= 
\langle 0 | A_4( {\mathbf p}_4, \theta_4) A^c_3( {\mathbf p}_3, \theta_3)
A_2^\dagger({\mathbf p}_2, \theta_2) A_1^{c\dagger}({\mathbf p}_1, \theta_1) 
| 0 \rangle .
\end{equation}
In the context of our paper, this is the $S$ matrix for particle - antiparticle 
scattering. 
The full ${\cal N}=2$ invariance of the $S$ matrix is expressed as
\begin{align}\label{n2inv}
 \left(\sum_{i=1}^4 Q^i_\alpha({\mathbf p}_i,\theta_i)\right)S({\mathbf p}_i,\theta_i) = 0,
\text{~and~}  \left(\sum_{i=1}^4 {\bar Q}^i_\alpha({\mathbf p}_i,\theta_i)\right)S({\mathbf
p}_i,\theta_i) = 0.
\end{align}
The above conditions \eqref{n2inv} produce the following constraints for the $S$ matrix element
\eqref{elem}
\begin{align}\label{n2inv1}
 \left(\sum_{i=1}^4 Q^i_\alpha({\mathbf p}_i,\theta_i)\right)&S({\mathbf p}_i,\theta_i) = 0
\Rightarrow\nn \\&\left(i v_\alpha({\mathbf
p}_1)\frac{\overrightarrow{\partial}}{\partial\theta_1}+i v^*({\mathbf p}_2)\theta_2 + i
u^*_\alpha({\mathbf p}_3)\theta_3- i u_\alpha({\mathbf
p}_4)\frac{\overrightarrow{\partial}}{\partial\theta_4}\right)S({\mathbf p}_i,\theta_i) = 0,\nn\\
 \left(\sum_{i=1}^4 {\bar Q}^i_\alpha({\mathbf p}_i,\theta_i)\right)&S({\mathbf p}_i,\theta_i) = 0
\Rightarrow\nn \\&\left(i v^*_\alpha({\mathbf p}_1)\theta_1 + i v_\alpha({\mathbf
p}_2)\frac{\overrightarrow{\partial}}{\partial\theta_2}-i u_\alpha({\mathbf
p}_3)\frac{\overrightarrow{\partial}}{\partial\theta_3}+i u^*_\alpha({\mathbf
p}_4)\theta_4\right)S({\mathbf p}_i,\theta_i) = 0.
\end{align}
We check in what follows that the combination
\begin{equation}\label{oldn1eq}
  \left(\frac{1}{\sqrt{2}}\sum_{i=1}^4 Q^i_\alpha({\mathbf p}_i,\theta_i) + {\bar
Q}^i_\alpha({\mathbf p}_i,\theta_i)\right)S({\mathbf p}_i,\theta_i) = 0
\end{equation}
produces the same equation (and therefore solution) of ${\cal N}=1$ which we have already found. We
easily find that this gives
\begin{align}\label{firstn1eqn}
 \left(i v_\alpha({\mathbf p}_1)\frac{\overrightarrow{\partial}}{\partial\theta_1} \right.&\left.+ i
v_\alpha({\mathbf p}_2)\frac{\overrightarrow{\partial}}{\partial\theta_2} - i u_\alpha({\mathbf
p}_3)\frac{\overrightarrow{\partial}}{\partial\theta_3} - i u_\alpha({\mathbf
p}_4)\frac{\overrightarrow{\partial}}{\partial\theta_4}\right.\nn\\&\bigg.+ i v^*_\alpha({\mathbf
p}_1)\theta_1 + i v^*_\alpha({\mathbf p}_2)\theta_2 + i u^*_\alpha({\mathbf p}_3)\theta_3 + i
u^*_\alpha({\mathbf p}_4)\theta_4\bigg)S({\mathbf p}_i,\theta_i) = 0. 
 \end{align}
Now, we turn to the other linear combination, which is
\begin{equation}\label{newn1eq}
  \left(\frac{1}{i\sqrt{2}}\sum_{i=1}^4 Q^i_\alpha({\mathbf p}_i,\theta_i) - {\bar
Q}^i_\alpha({\mathbf p}_i,\theta_i)\right)S({\mathbf p}_i,\theta_i) = 0.
\end{equation}
This readily gives the differential equation
\begin{align}\label{secn1eqn}
 \left(i v_\alpha({\mathbf p}_1)\frac{\overrightarrow{\partial}}{\partial\theta_1} \right.&\left.- i
v_\alpha({\mathbf p}_2)\frac{\overrightarrow{\partial}}{\partial\theta_2} + i u_\alpha({\mathbf
p}_3)\frac{\overrightarrow{\partial}}{\partial\theta_3} - i u_\alpha({\mathbf
p}_4)\frac{\overrightarrow{\partial}}{\partial\theta_4}\right.\nn\\&\bigg.- i v^*_\alpha({\mathbf
p}_1)\theta_1 + i v^*_\alpha({\mathbf p}_2)\theta_2 + i u^*_\alpha({\mathbf p}_3)\theta_3 - i
u^*_\alpha({\mathbf p}_4)\theta_4\bigg)S({\mathbf p}_i,\theta_i) = 0. 
\end{align}
The equation \eqref{firstn1eqn} is the same as it was for the $\mN=1$ theory, whereas the second
equation \eqref{secn1eqn} must be obeyed by the same $S$ matrix in the $\mN=2$ point. Thus
\eqref{secn1eqn1} is an additional constraint obeyed by the $\mN=2$ $S$ matrix \eqref{SUSYSmatrix}.
It
follows that \eqref{secn1eqn} gives a relation between $\SB$ and $\SF$ 
\begin{align}\label{f1F2relN2form}
\SB\left(C_{12} v_\al(\mbp_1)-C_{23} u_\al(\mbp_3)+C_{24}
u_\al(\mbp_4)+v^*_\al(\mbp_2)\right)
=\SF(C^*_{13}u_\al(\mbp_4)+C^*_{14}u_\al(\mbp_3)+C_{34}^*v_\al(\mbp_1))
\end{align}
Thus, the $\mN=2$ $S$ matrix for particle-antiparticle scattering consists of only one independent
function, with the other related by \eqref{f1F2relN2form}.

\subsection{Particle - particle scattering}
Now, consider the other $S$ matrix element (which was considered in the previous ${\cal N}=1$
computation)
\begin{equation}\label{elem1}
S({\mathbf p}_1, \theta_1, {\mathbf p}_2, \theta_2, {\mathbf p}_3, \theta_3, 
{\mathbf p}_4, \theta_4)= 
\langle 0 | A_4( {\mathbf p}_4, \theta_4) A_3( {\mathbf p}_3, \theta_3)
A_2^\dagger({\mathbf p}_2, \theta_2) A_1^{\dagger}({\mathbf p}_1, \theta_1) 
| 0 \rangle .
\end{equation}
The conditions \eqref{n2inv} produce the following for the $S$ matrix element \eqref{elem1}
\begin{align}\label{n2inv2}
 \left(\sum_{i=1}^4 Q^i_\alpha({\mathbf p}_i,\theta_i)\right)&S({\mathbf p}_i,\theta_i) = 0
\Rightarrow\nn \\&\left(i v^*_\alpha({\mathbf p}_1)\theta_1+i v^*({\mathbf p}_2)\theta_2 - i
u_\alpha({\mathbf p}_3)\frac{\overrightarrow{\partial}}{\partial\theta_3}- i u_\alpha({\mathbf
p}_4)\frac{\overrightarrow{\partial}}{\partial\theta_4}\right)S({\mathbf p}_i,\theta_i) = 0,\nn\\
 \left(\sum_{i=1}^4 {\bar Q}^i_\alpha({\mathbf p}_i,\theta_i)\right)&S({\mathbf p}_i,\theta_i) = 0
\Rightarrow\nn \\&\left(i v_\alpha({\mathbf p}_1)\frac{\overrightarrow{\partial}}{\partial\theta_1}
+ i v_\alpha({\mathbf p}_2)\frac{\overrightarrow{\partial}}{\partial\theta_2}+i u^*_\alpha({\mathbf
p}_3)\theta_3+i u^*_\alpha({\mathbf p}_4)\theta_4\right)S({\mathbf p}_i,\theta_i) = 0.
\end{align}
For the combination \eqref{oldn1eq} we get
\begin{align}\label{firstn1eqn1}
 \left(i v_\alpha({\mathbf p}_1)\frac{\overrightarrow{\partial}}{\partial\theta_1} \right.&\left.+ i
v_\alpha({\mathbf p}_2)\frac{\overrightarrow{\partial}}{\partial\theta_2} - i u_\alpha({\mathbf
p}_3)\frac{\overrightarrow{\partial}}{\partial\theta_3} - i u_\alpha({\mathbf
p}_4)\frac{\overrightarrow{\partial}}{\partial\theta_4}\right.\nn\\&\bigg.+ i v^*_\alpha({\mathbf
p}_1)\theta_1 + i v^*_\alpha({\mathbf p}_2)\theta_2 + i u^*_\alpha({\mathbf p}_3)\theta_3 + i
u^*_\alpha({\mathbf p}_4)\theta_4\bigg)S({\mathbf p}_i,\theta_i) = 0,
 \end{align}
and for the combination \eqref{newn1eq} we have
\begin{align}\label{secn1eqn1}
 \left(-i v_\alpha({\mathbf p}_1)\frac{\overrightarrow{\partial}}{\partial\theta_1} \right.&\left.-
i v_\alpha({\mathbf p}_2)\frac{\overrightarrow{\partial}}{\partial\theta_2} - i u_\alpha({\mathbf
p}_3)\frac{\overrightarrow{\partial}}{\partial\theta_3} - i u_\alpha({\mathbf
p}_4)\frac{\overrightarrow{\partial}}{\partial\theta_4}\right.\nn\\&\bigg.+ i v^*_\alpha({\mathbf
p}_1)\theta_1 + i v^*_\alpha({\mathbf p}_2)\theta_2 - i u^*_\alpha({\mathbf p}_3)\theta_3 - i
u^*_\alpha({\mathbf p}_4)\theta_4\bigg)S({\mathbf p}_i,\theta_i) = 0. 
\end{align}

Similar to the particle-anti particle case discussed in the previous section. The equation
\eqref{firstn1eqn1} is the same as it was for the $\mN=1$ theory, whereas the second
equation \eqref{secn1eqn1} must be obeyed by the same $S$ matrix in the $\mN=2$ point. It
follows that \eqref{secn1eqn1} gives a relation between $\SB$ and $\SF$ 
\begin{align}\label{f1F2relN2form1}
\SB\left(C_{13} u_\al(\mbp_3)+C_{14} u_\al(\mbp_4)+C_{12}
v_\al(\mbp_2)+v^*_\al(\mbp_1)\right)
=\SF(C^*_{24}u_\al(\mbp_3)-C^*_{23}u_\al(\mbp_4)+C_{34}^*v_\al(\mbp_2))
\end{align}
The $\mN=2$ $S$ matrix for particle-particle scattering consists of only one independent
function, with the other related by \eqref{f1F2relN2form1}.

Thus in the $\mN=2$ theory the $S$ matrix is only made of one independent function. Note that the
results of this section are true for \emph{any} three dimensional $\mN=2$ theory.
It simply follows from the supersymmetric ward identity \eqref{n2inv} and is independent
of the details of the theory.  

\section{Identities for \texorpdfstring{$S$}{S} matrices in onshell superspace}\label{sp}

In this subsection we demonstrate that the product of two supersymmetric 
$S$ matrices is supersymmetric. In other words we demonstrate that 
\begin{align}\label{susyinvdef}
 \left(\sum_{i=1}^4 Q_\alpha^i({\mathbf p}_i,\theta_i)\right)S_1\star S_2 = 0.
\end{align}
provided $S_1$ and $S_2$ independently obey the same equation.

This can be analyzed as follows. We have the invariance (differential) equation for $S_1$ and $S_2$
\begin{multline}\label{susyinvS}
\left(\overrightarrow{Q}_{{\tilde v}({\mathbf p}_1)} + \overrightarrow{Q}_{{\tilde v}({\mathbf
p}_2)}+\overrightarrow{Q}_{u({\mathbf p}_3)} + \overrightarrow{Q}_{u({\mathbf
p}_4)}\right)S_i({\mathbf p}_1, \theta_1, {\mathbf p}_2, \theta_2, {\mathbf p}_3, \theta_3, 
{\mathbf p}_4, \theta_4) = 0\\
\text{with~} p_1 + p_2 = p_3 + p_4.
\end{multline}
where the left-acting supercharges $\overrightarrow{Q}_{{\tilde v}({\mathbf p})}$ are defined as
\beq\label{tQdef}
 \overrightarrow{Q}_{{\tilde v}({\mathbf p})} = i\left(v_\alpha({\mathbf
p})\overrightarrow{\f{\partial}{\partial\theta}} + v_\alpha^*({\mathbf p})\theta\right)
\eeq
in contrast to \eqref{Qactdef}, because we're acting from the left. It may be easily checked that
this indeed produces the correct action of $Q$ on $A^\dagger$. The reader is reminded that the
(left- acting) supercharges $\overrightarrow{Q}_{u({\mathbf p})}$ are defined as
\beq\label{tQdef1}
 \overrightarrow{Q}_{u({\mathbf p})} = i\left(-u_\alpha({\mathbf
p})\overrightarrow{\f{\partial}{\partial\theta}} + u_\alpha^*({\mathbf p})\theta\right).
\eeq
Note that
\begin{align}\label{star}
 (\overrightarrow{Q}_{{\tilde v}({\mathbf p})})^* =
 \overrightarrow{Q}_{u({\mathbf p})}\  ,\nn\\(\overrightarrow{Q}_{u({\mathbf p})})^* =
\overrightarrow{Q}_{{\tilde v}({\mathbf p})}\ .
\end{align}
We have used the fact that while complex conjugating, the grassmannian derivatives acting from the
left act from the right (and vice-versa) and to bring any such right acting derivative to the left
involves introducing an extra minus sign.
Armed with the definitions above, we can rewrite \eqref{susyinvS} as (all differential operators
henceforth, unless noted otherwise, are taken to act from the left)
\begin{multline}\label{susyinvS1}
\left(Q^*_{u({\mathbf p}_1)} + Q^*_{u({\mathbf p}_2)}+Q_{u({\mathbf p}_3)} + Q_{u({\mathbf
p}_4)}\right)S_i({\mathbf p}_1, \theta_1, {\mathbf p}_2, \theta_2, {\mathbf p}_3, \theta_3, 
{\mathbf p}_4, \theta_4) = 0\ .\\
\end{multline}
The next step is to observe that
\begin{multline}\label{susyinvid}
\left(Q^*_{u({\mathbf p}_1)} + Q^*_{u({\mathbf p}_2)}+Q_{u({\mathbf p}_3)} + Q_{u({\mathbf
p}_4)}\right)\exp(\theta_1\theta_3 + \theta_2\theta_4)2p_3^0(2\pi)^2\delta^{(2)}({\mathbf
p}_1-{\mathbf p}_3)\\2p_4^0(2\pi)^2\delta^{(2)}({\mathbf p}_2-{\mathbf p}_4) = 0
\end{multline}
after we set ${\mathbf p}_1 = {\mathbf p}_3$ and ${\mathbf p}_2 = {\mathbf p}_4$. We now act on
\eqref{sost} with
\begin{multline}\label{susyinvuni}
 \left(Q^*_{u({\mathbf p}_1)} + Q^*_{u({\mathbf p}_2)}+Q_{u({\mathbf p}_3)} + Q_{u({\mathbf
p}_4)}\right)\int d\Gamma\bigg[S_1({\mathbf p}_1, \theta_1, {\mathbf p}_2, \theta_2, {\mathbf k}_3,
\phi_1, 
{\mathbf k}_4, \phi_2)\\\exp(\phi_1\phi_3+\phi_2\phi_4)2k_1^0(2\pi)^2\delta^{(2)}({\mathbf
k}_3-{\mathbf k}_1)2k_2^0(2\pi)^2\delta^{(2)}({\mathbf k}_4-{\mathbf k}_2)\\S_2({\mathbf k}_1,
\phi_3, {\mathbf k}_2, \phi_4, {\mathbf p}_3, \theta_3, 
{\mathbf p}_4, \theta_4)\bigg]\ .
\end{multline}
Proceeding with \eqref{susyinvuni}, one finds
\begin{multline}\label{susyinvuni1}
-\int d\Gamma \bigg[\left(Q_{u({\mathbf k}_3)} + Q_{u({\mathbf k}_4)}\right)S_1({\mathbf p}_1,
\theta_1, {\mathbf p}_2, \theta_2, {\mathbf k}_3, \phi_1, 
{\mathbf k}_4,
\phi_2)\exp(\phi_1\phi_3+\phi_2\phi_4)\bigg.\\\left.2k_1^0(2\pi)^2\delta^{(2)}({\mathbf
k}_3-{\mathbf k}_1)2k_2^0(2\pi)^2\delta^{(2)}({\mathbf k}_4-{\mathbf k}_2)S_2({\mathbf k}_1,
\phi_3, {\mathbf k}_2, \phi_4, {\mathbf p}_3, \theta_3, 
{\mathbf p}_4, \theta_4)\right.\\+\left.S_1({\mathbf p}_1, \theta_1, {\mathbf p}_2, \theta_2,
{\mathbf k}_3, \phi_1, 
{\mathbf k}_4, \phi_2)\exp(\phi_1\phi_3+\phi_2\phi_4)2k_1^0(2\pi)^2\delta^{(2)}({\mathbf
k}_3-{\mathbf k}_1)\right.\\
\bigg.2k_2^0(2\pi)^2\delta^{(2)}({\mathbf k}_4-{\mathbf k}_2)\left(Q^*_{u({\mathbf k}_1)} +
Q^*_{u({\mathbf k}_2)}\right)S_2({\mathbf k}_1,
\phi_3, {\mathbf k}_2, \phi_4, {\mathbf p}_3, \theta_3, 
{\mathbf p}_4, \theta_4)\bigg].
\end{multline}
We next integrate by parts keeping in mind that only the derivative parts of the $Q$ change sign (as
a consequence of the integration by parts). This gives
\begin{multline}\label{susyinvuni2}
\int d\Gamma \bigg[S_1({\mathbf p}_1, \theta_1, {\mathbf p}_2, \theta_2, {\mathbf k}_3, \phi_1, 
{\mathbf k}_4, \phi_2)\bigg.\\\left.\left({\tilde Q}_{u({\mathbf k}_3)} + {\tilde Q}_{u({\mathbf
k}_4)}+{\tilde Q}^*_{u({\mathbf k}_1)} + {\tilde Q}^*_{u({\mathbf
k}_2)}\right)\exp(\phi_1\phi_3+\phi_2\phi_4)2k_1^0(2\pi)^2\delta^{(2)}({\mathbf
k}_3-{\mathbf k}_1)\right.\\\bigg.2k_2^0(2\pi)^2\delta^{(2)}({\mathbf k}_4-{\mathbf
k}_2)S_2({\mathbf k}_1,
\phi_3, {\mathbf k}_2, \phi_4, {\mathbf p}_3, \theta_3, 
{\mathbf p}_4, \theta_4)\bigg].
\end{multline}
Here, by ${\tilde Q}_{u(p)}$ and ${\tilde Q}^*_{u(p)}$ we mean
\begin{align}\label{newQdef}
 {\tilde Q}_{u(p)} = i\left(u_\alpha({\mathbf p})\overrightarrow{\f{\partial}{\partial\theta}} +
u_\alpha^*({\mathbf p})\theta\right)\ , \\
 {\tilde Q}^*_{u(p)} = i\left(u^*_\alpha({\mathbf p})\overrightarrow{\f{\partial}{\partial\theta}} -
u_\alpha({\mathbf p})\theta\right)\ .
\end{align}
It can be easily checked (just like \eqref{susyinvid}) that (on setting ${\mathbf k}_3 = {\mathbf
k}_1$ and ${\mathbf k}_4 = {\mathbf k}_2$)
\begin{multline}\label{susyinvid2}
 \left({\tilde Q}_{u({\mathbf k}_3)} + {\tilde Q}_{u({\mathbf k}_4)}+{\tilde Q}^*_{u({\mathbf k}_1)}
+ {\tilde Q}^*_{u({\mathbf
k}_2)}\right)\exp(\phi_1\phi_3+\phi_2\phi_4)2k_1^0(2\pi)^2\delta^{(2)}({\mathbf
k}_3-{\mathbf k}_1)\\2k_2^0(2\pi)^2\delta^{(2)}({\mathbf k}_4-{\mathbf k}_2)=0,
\end{multline}
completing the proof.

\section{Details of the unitarity equation}\label{unitdet}
In this section, we simplify the unitarity equations \eqref{unicond1} and \eqref{unicond2}. We
define
$$Z({\mathbf p}_i) = \f{1}{4m^2}v^*({\mathbf
p}_1)v^*({\mathbf p}_2)~ v({\mathbf p}_3)v({\mathbf p}_4)$$
and rewrite \eqref{unicond1} and \eqref{unicond2} as
\begin{IEEEeqnarray}{l}\label{newunicond1}
\int d\Gamma^\prime\left[\SB({\mathbf p}_1,{\mathbf p}_2,{\mathbf k}_3,{\mathbf k}_4)\SB^*({\mathbf
p}_3,{\mathbf p}_4,{\mathbf k}_3,{\mathbf k}_4)\right.\nn\\\left.-Y({\mathbf p}_3,{\mathbf
p}_4)\left(\SB({\mathbf p}_1, {\mathbf p}_2, {\mathbf k}_3,
{\mathbf k}_4)\SB^*({\mathbf p}_3, {\mathbf p}_4, {\mathbf k}_3,
{\mathbf k}_4)\right.\right.\nn\\\left.\left.+4Y({\mathbf p}_3,{\mathbf p}_4)\left(\SB({\mathbf
p}_1, {\mathbf p}_2, {\mathbf k}_3,
{\mathbf k}_4)\SF^*({\mathbf p}_3, {\mathbf p}_4, {\mathbf k}_3,
{\mathbf k}_4)+\SF({\mathbf p}_1, {\mathbf p}_2, {\mathbf k}_3,
{\mathbf k}_4)\SB^*({\mathbf p}_3, {\mathbf p}_4, {\mathbf k}_3,
{\mathbf k}_4)\right)\right.\right.\nn\\\left.\left.+16Y^2({\mathbf p}_3,{\mathbf p}_4) \SF({\mathbf
p}_1,{\mathbf p}_2,{\mathbf k}_3,{\mathbf k}_4)\SF^*({\mathbf p}_3,{\mathbf p}_4,{\mathbf
k}_3,{\mathbf k}_4)\right)\right] = 2p_3^0(2\pi)^2\delta^{(2)}({\mathbf p}_1-{\mathbf
p}_3)2p_4^0(2\pi)^2\delta^{(2)}({\mathbf p}_2-{\mathbf p}_4)\nn\\
\end{IEEEeqnarray}
and 
\begin{IEEEeqnarray}{l}\label{newunicond2}
Z({\mathbf p}_i)\int d\Gamma^\prime\left[-4Y({\mathbf p}_3,{\mathbf p}_4)\SF({\mathbf p}_1,{\mathbf
p}_2,{\mathbf k}_3,{\mathbf k}_4)\SF^*({\mathbf p}_3,{\mathbf p}_4,{\mathbf k}_3,{\mathbf
k}_4)\right.\nn\\\left.+\left(4Y^2({\mathbf p}_3,{\mathbf p}_4)\SF({\mathbf p}_1, {\mathbf p}_2,
{\mathbf k}_3,
{\mathbf k}_4)\SF^*({\mathbf p}_3, {\mathbf p}_4, {\mathbf k}_3,
{\mathbf k}_4)\right.\right.\nn\\\left.\left.+Y({\mathbf p}_3,{\mathbf p}_4)\left(\SB({\mathbf p}_1,
{\mathbf p}_2, {\mathbf k}_3,
{\mathbf k}_4)\SF^*({\mathbf p}_3, {\mathbf p}_4, {\mathbf k}_3,
{\mathbf k}_4)+\SF({\mathbf p}_1, {\mathbf p}_2, {\mathbf k}_3,
{\mathbf k}_4)\SB^*({\mathbf p}_3, {\mathbf p}_4, {\mathbf k}_3,
{\mathbf k}_4)\right)\right.\right.\nn\\\left.\left.+\f{1}{4}\SB({\mathbf p}_1,{\mathbf
p}_2,{\mathbf k}_3,{\mathbf k}_4)\SB^*({\mathbf p}_3,{\mathbf p}_4,{\mathbf k}_3,{\mathbf
k}_4)\right)\right] = -2p_3^0(2\pi)^2\delta^{(2)}({\mathbf p}_1-{\mathbf
p}_3)2p_4^0(2\pi)^2\delta^{(2)}({\mathbf p}_2-{\mathbf p}_4).\nn\\
\end{IEEEeqnarray}
Since the factor $Z({\mathbf p}_i)$ depends only on the external momenta ${\mathbf p}_i$, we may
evaluate it on the delta functions and this simply yields $Z({\mathbf p}_i) = 4Y({\mathbf
p}_3,{\mathbf p}_4)$. We finally arrive at
\begin{IEEEeqnarray}{l}\label{newunicond3}
\int d\Gamma^\prime\bigg[\SB({\mathbf p}_1,{\mathbf p}_2,{\mathbf k}_3,{\mathbf k}_4)\SB^*({\mathbf
p}_3,{\mathbf p}_4,{\mathbf k}_3,{\mathbf k}_4)\bigg.\nn\\\left.-Y({\mathbf p}_3,{\mathbf
p}_4)\bigg(\SB({\mathbf p}_1, {\mathbf p}_2, {\mathbf k}_3,
{\mathbf k}_4)\SB^*({\mathbf p}_3, {\mathbf p}_4, {\mathbf k}_3,
{\mathbf k}_4)\bigg.\right.\nn\\\left.\left.+4Y({\mathbf p}_3,{\mathbf p}_4)\big(\SB({\mathbf p}_1,
{\mathbf p}_2, {\mathbf k}_3,
{\mathbf k}_4)\SF^*({\mathbf p}_3, {\mathbf p}_4, {\mathbf k}_3,
{\mathbf k}_4)+\SF({\mathbf p}_1, {\mathbf p}_2, {\mathbf k}_3,
{\mathbf k}_4)\SB^*({\mathbf p}_3, {\mathbf p}_4, {\mathbf k}_3,
{\mathbf k}_4)\big)\right.\right.\nn\\\bigg.\bigg.+16Y^2({\mathbf p}_3,{\mathbf p}_4) \SF({\mathbf
p}_1,{\mathbf p}_2,{\mathbf k}_3,{\mathbf k}_4)\SF^*({\mathbf p}_3,{\mathbf p}_4,{\mathbf
k}_3,{\mathbf k}_4)\bigg)\bigg] = 2p_3^0(2\pi)^2\delta^{(2)}({\mathbf p}_1-{\mathbf
p}_3)2p_4^0(2\pi)^2\delta^{(2)}({\mathbf p}_2-{\mathbf p}_4)\nn\\
\end{IEEEeqnarray}
and
\begin{IEEEeqnarray}{l}\label{newunicond4}
\int d\Gamma^\prime\bigg[-16Y^2({\mathbf p}_3,{\mathbf p}_4)\SF({\mathbf p}_1,{\mathbf p}_2,{\mathbf
k}_3,{\mathbf k}_4)\SF^*({\mathbf p}_3,{\mathbf p}_4,{\mathbf k}_3,{\mathbf
k}_4)\bigg.\nn\\\left.+Y({\mathbf p}_3,{\mathbf p}_4)\bigg(\SB({\mathbf p}_1, {\mathbf p}_2,
{\mathbf k}_3,
{\mathbf k}_4)\SB^*({\mathbf p}_3, {\mathbf p}_4, {\mathbf k}_3,
{\mathbf k}_4)\bigg.\right.\nn\\\left.\left.+4Y({\mathbf p}_3,{\mathbf p}_4)\big(\SB({\mathbf p}_1,
{\mathbf p}_2, {\mathbf k}_3,
{\mathbf k}_4)\SF^*({\mathbf p}_3, {\mathbf p}_4, {\mathbf k}_3,
{\mathbf k}_4)+\SF({\mathbf p}_1, {\mathbf p}_2, {\mathbf k}_3,
{\mathbf k}_4)\SB^*({\mathbf p}_3, {\mathbf p}_4, {\mathbf k}_3,
{\mathbf k}_4)\big)\right.\right.\nn\\\bigg.\bigg.+16Y^2({\mathbf p}_3,{\mathbf p}_4) \SF({\mathbf
p}_1,{\mathbf p}_2,{\mathbf k}_3,{\mathbf k}_4)\SF^*({\mathbf p}_3,{\mathbf p}_4,{\mathbf
k}_3,{\mathbf k}_4)\bigg)\bigg] \nn= -2p_3^0(2\pi)^2\delta^{(2)}({\mathbf p}_1-{\mathbf
p}_3)2p_4^0(2\pi)^2\delta^{(2)}({\mathbf p}_2-{\mathbf p}_4).\nn\\
\end{IEEEeqnarray}
The above equations can be more compactly written as \eqref{newunicond5} and
\eqref{newunicond6} respectively (since $p_3\cdot p_4 = p_1\cdot p_2$).

\section{Going to supersymmetric Light cone gauge}\label{susylcgauge}

In this brief appendix we will demonstrate that (upto the usual problem with 
zero modes) it is always possible to find a super gauge transformation 
that takes us to the supersymmetric lightcone gauge $ \Gamma_- =0$ 

Let us start with a gauge configuration that obeys our gauge condition 
$\Gamma_-=0$. Starting with this gauge configuration, we will now demonstrate 
that we can perform a gauge transformation that will take $\Gamma_-$ to 
any desired value, say ${\tilde \Gamma}_-$.

Performing the gauge transformation \eqref{gaugetransformations} we find 
that the new value of $\Gamma_-$ is simply $D_- K$. Let 
\beq
K=M+\te\ze-\te^2 P ,
\eeq
where $M, \ze^\al, P$ are gauge parameters. 
It follows that  
\begin{equation}
D_- K= \ze_- - \te_- (\p_{-+}M +P)+\te_+ \p_{--}M-i\te_+\te_-(\p_{-+}\ze_--\p_{--}\ze_+) \ 
\end{equation}
Now let us suppose that 
\begin{equation}
 - {\tilde \Ga}_-=\chi_- - \te_- (B+A_{+-})+\te_+A_{--}+i\te_+\te_-(2\la_-+\p_{--}\ch_+-\p_{-+}\chi_-) \nn
\end{equation}
We need to find $K$ so that 
$$D_-K= {\tilde \Gamma}_-$$
Equating coefficients on the two sides of this equation we find 
\begin{align}\label{gaugechoice}
 \ch_-+\ze_- & =0 \ ,\nn \\
B+ A_{+-}+P+\p_{-+}M &=0 \ , \nn\\
A_{--}+\p_{--}M &=0 \ ,\nn\\
2 \la_-+\p_{--}(\ch_+ +\ze_+) -\p_{-+}(\ch_-+\ze_-)& =0 \ ,
\end{align}
which are then solved to get,
\begin{align}
 \ze_- & =-\ch_- \ , \nn\\ 
 \ze_+ & =- 2\p_{--}^{-1} \la_- - \ch_+ \ , \nn\\
 M & =-\p_{--}^{-1}A_{--} \ , \nn\\
P & = -B-A_{+-}+\p_{-+}(\p_{--}^{-1}A_{--}) \ .
\end{align}
Substituting the above expressions in the expansion for K, we can write
\beq
K=-\p_{--}^{-1} A_{--}-i \te_- (2 \p_{--}^{-1}\la_-+ \ch_+)+i\te_+ \ch_- +
i\te_+\te_-(\p_{-+}\p_{--}^{-1}A_{--}-B-A_{+-}) \ .
\eeq
It can be checked that the form of $K$ obtained above follows from
\beq
K= i\p_{--}^{-1}D_{-}\Ga_- \ ,
\eeq
which is a supersymmetric version of the gauge transformation used to 
generate an arbitrary $A_-$ starting from usual lightcone gauge.

\section{Details of the self energy computation}\label{selfieApp}
In this subsection, we will demonstrate that the self energy $\Si(p,\te_1,\te_2)$ is a constant
independent of the momenta $p$. As discussed in \S \ref{selfie2} $\Si(p,\te_1,\te_2)$ obeys the
integral equation
\begin{align}\label{inte1}
\Sigma(p, \theta_1, \theta_2) =&  \  2\pi\la w \int \f{d^3r}{(2\pi)^3}
\de^2(\te_1-\te_2)P(r, \theta_1,\theta_2)\nn\\
&- 2\pi\la \int \f{d^3
r}{(2\pi)^3}D^{\te_2,-p}_-D^{\te_1,p}_-\left(\f{\de^2(\te_1-\te_2)}{(p-r)_{--}}P(r,
\theta_1,\theta_2)\right)\nn\\
&+2\pi\la
\int\f{d^3r}{(2\pi)^3}\f{\de^2(\te_1-\te_2)}{(p-r)_{--}}D_-^{\te_1,r}D_-^{\te_2,-r}P(r,\te_1,\te_2,
)\ .
\end{align}
We will now simplify the second and third terms in \eqref{inte1}. In \S \ref{selfie2} we already
observed that the general form of the propagator is of the form given by \eqref{pform}. Using the
formulae \eqref{expformula} and \eqref{usefulformulae} we can write \eqref{pform} as
\beq\label{pform2}
P(p,\te_1,\te_2)=\left( C_1(p) D^2_{\te_1,p} +C_2(p)\right)\de^2(\te_1-\te_2)
\eeq

In the second term of \eqref{inte1} we have to evaluate 
\beq\label{inte12ndterm}
C_1(p)D_-^{\te_2,-p}D_-^{\te_1,p} \left(\de^2(\te_1-\te_2)D^2_{\te_1,p}\de^2(\te_1-\te_2)\right)\ ,
\eeq
since the product of $\de^2(\te_1-\te_2)$ vanishes. We further use the formulae
\eqref{usefulformulae} and then the transfer rule \eqref{transferrule} to get
\begin{align}\label{inte12ndans}
-C_1(p)D_-^{\te_2,-p}D_-^{\te_2,-p}\de^2(\te_1-\te_2)&=p_{--} C_1(p)\de^2(\te_1-\te_2) \nn\\
&= p_{--}\de^2(\te_1-\te_2)P(r,\te_1,\te_2) \ ,
\end{align}
where we have used the algebra \eqref{Dalgmom} in the first line and \eqref{usefulformulae} in the
second.

Let us now proceed to simplify the third term in \eqref{inte1}. We need to evaluate
\begin{align}\label{inte13rdterm}
\de^2(\te_1-\te_2) D_-^{\te_1,r}D_-^{\te_2,-r}
\left(C_1(p)D^2_{\te_1,r}\de^2(\te_1-\te_2)+C_2(p)\de^2(\te_1-\te_2)\right)\nn\\
=C_1(p)\de^2(\te_1-\te_2) D_-^{\te_1,r}D_-^{\te_2,-r} D^2_{\te_1,r}\de^2(\te_1-\te_2)\ ,
\end{align}
where we have used the transfer rule \eqref{transferrule} and the fact that the product of
$\de^2(\te_1 -\te_2)$ vanishes. We further simplify
\begin{align}\label{inte13rdans}
 C_1(p)\de^2(\te_1-\te_2) D_-^{\te_1,r}D_-^{\te_2,-r} D^2_{\te_1,r}\de^2(\te_1-\te_2)&=
-C_1(p)\de^2(\te_1-\te_2)r_-^\be D_-^{\te_1,r} D_\be^{\te_2,-r}\de^2(\te_1-\te_2)\nn\\
&= C_1(p)\de^2(\te_1-\te_2)r_-^+D_-^{\te_1,r} D_+^{\te_1,r}\de^2(\te_1-\te_2)\nn\\
&=C_1(p)\de^2(\te_1-\te_2)(-ir_-^+) D^2_{\te_1,r}\de^2(\te_1-\te_2)\nn\\
&= r_{--} \de^2(\te_1-\te_2) P(r,\te_1,\te_2)\ ,
\end{align}
where in the first line we have used \eqref{Dsqsq}, in the second line 
the expression is nonzero for $\be=-$ and we have used the transfer rule \eqref{transferrule},
while the third line follows from the identity $-i D^2=D_-D_+$ and the last line follows from the
arguments used before.

Thus, using the results \eqref{inte13rdans} and \eqref{inte12ndans} in \eqref{inte1} we get the
final form as given in \eqref{intea}
\begin{align}\label{intea1}
\Sigma(p, \theta_1, \theta_2) =&  \  2\pi\la w \int \f{d^3r}{(2\pi)^3}
\de^2(\te_1-\te_2)P(r, \theta_1,\theta_2)\nn\\
&- 2\pi\la \int \f{d^3
r}{(2\pi)^3}\f{p_{--}}{(p-r)_{--}}\de^2(\te_1-\te_2)P(r,\theta_1,\theta_2)\nn\\
&+2\pi\la
\int\f{d^3r}{(2\pi)^3}\f{r_{--}}{(p-r)_{--}}\de^2(\te_1-\te_2)P(r,\te_1,\te_2)\ .
\end{align}
From the above it is clear that the momentum dependence cancels between the second and third terms
and we get
\begin{align}
 \Sigma(p, \theta_1, \theta_2) =& \ 2\pi\la
(w-1)\int\f{d^3r}{(2\pi)^3}\de^2(\te_1-\te_2)P(r,\te_1,\te_2)\ .
\end{align}

\section{Details relating to the evaluation of the offshell 
four point function}

\subsection{Supersymmetry constraints on the offshell four point function} 
\label{susy4pt}
\begin{figure}[h]
 \begin{center}
\includegraphics{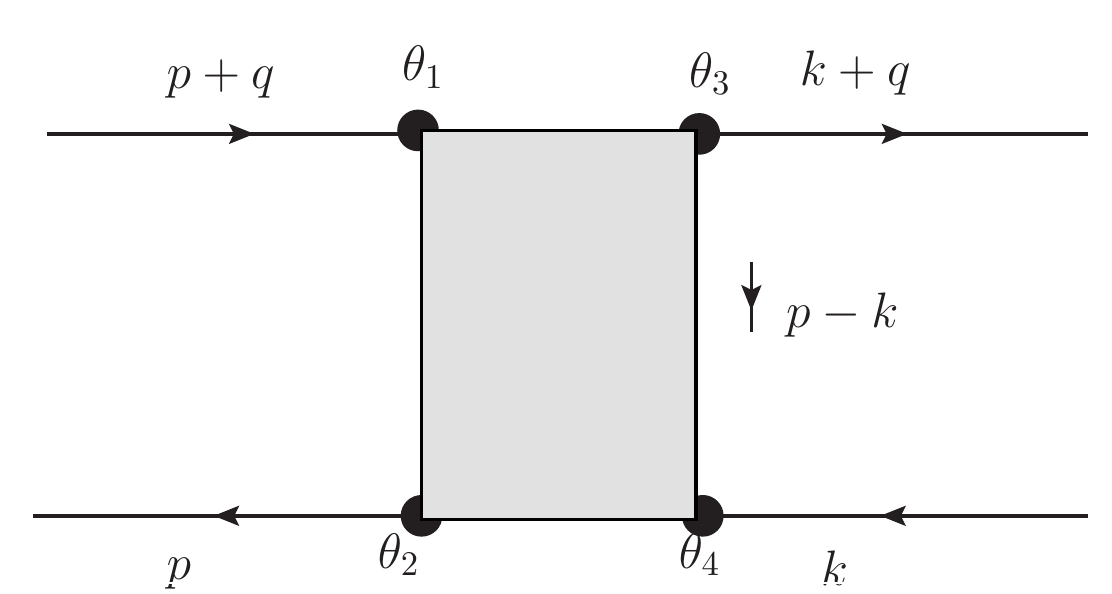}
  \caption{\label{4ptfungen} Four point function in superspace}
 \end{center}
\end{figure}

In this section we will constrain the most general form of the four point function using
supersymmetry (see fig \ref{4ptfungen}).  
Supersymmetric invariance of the four point function in superspace \eqref{4ptamp}
implies that
\beq\label{4ptsusy11}
(Q_{\te_1,p+q}+Q_{\te_2,-p}+Q_{\te_3,-k-q}+Q_{\te_4,k})V(\te_1,\te_2,\te_3,\te_4,p,q,k)=0 \ .
\eeq
This can be simplified using \eqref{susygen} and written as
\beq\label{4ptsusy22}
\sum_{i=1}^{4}
\bigg(\f{\p}{\p\te_i^\al}-p_{\al\be}(\te_1-\te_2)^\be-q_{\al\be}(\te_1-\te_3)^\be-k_{\al\be}
(\te_4-\te_3)^\be\bigg)V(\te_1,\te_2,\te_3,p,q,k)=0 \ .
\eeq
We can make the following variable changes to simplify the equation (we suppress spinor indices for
simplicity in notation)
\begin{align}
& X=\sum_{i=1}^4 \te_i \ ,\nn\\
& X_{12}=\te_1-\te_2 \ , \nn \\
& X_{13}=\te_1-\te_3 \ , \nn \\
& X_{43}=\te_4-\te_3 \ .
\end{align}
The inverse coordinates are
\begin{align}
& \te_1= \f{1}{4} (X+X_{12}+2X_{13}-X_{43}) \ , \nn\\
& \te_2= \f{1}{4} (X-3X_{12}+2X_{13}-X_{43}) \ , \nn\\
& \te_3= \f{1}{4} (X+X_{12}-2X_{13}-X_{43}) \ , \nn\\
& \te_4= \f{1}{4} (X+X_{12}-2X_{13}+3X_{43}) \ .
\end{align}
In terms of the new coordinates, the derivatives are then expressed as
\begin{align}
& \f{\p}{\p\te_1}=\f{\p}{\p X}+\f{\p}{\p X_{12}}+\f{\p}{\p X_{13}}\ , \nn\\
& \f{\p}{\p\te_2}=\f{\p}{\p X}-\f{\p}{\p X_{12}} , \nn\\
& \f{\p}{\p\te_3}=\f{\p}{\p X}-\f{\p}{\p X_{13}}-\f{\p}{\p X_{43}}\ , \nn\\
& \sum_{i=1}^4\f{\p}{\p\te_i}= 4\f{\p}{\p X} \ .
\end{align}
Using the above, one can rewrite \eqref{4ptsusy22} as
\beq
(4 \f{\p}{\p X}-p.X_{12}-q.X_{13}-k.X_{43})V(X,X_{12},X_{13},X_{43},p,q,k)=0 ,
\eeq
where $p.X_{12}=p_{\al\be}X_{12}^\be$. The above equation can be thought of as a differential
equation in the variables $X_{ij}$ and is solved by
\begin{align}\label{4ptsusyinvariantgen1}
V(\te_1,\te_2,\te_3,\te_4,p,q,k)=\exp\bigg(\f{1}{4}X.(p.X_{12}+q.X_{13}+k.X_{43})\bigg)F(X_{12},X_{
13},X_{43},p,q,k) \ .
\end{align}
This is the most general form of a four point function in superspace that is invariant under
supersymmetry.

\subsection{Explicitly evaluating $V_0$}\label{4pttree}
\begin{figure}[h]
\begin{center}
\includegraphics[width=16cm]{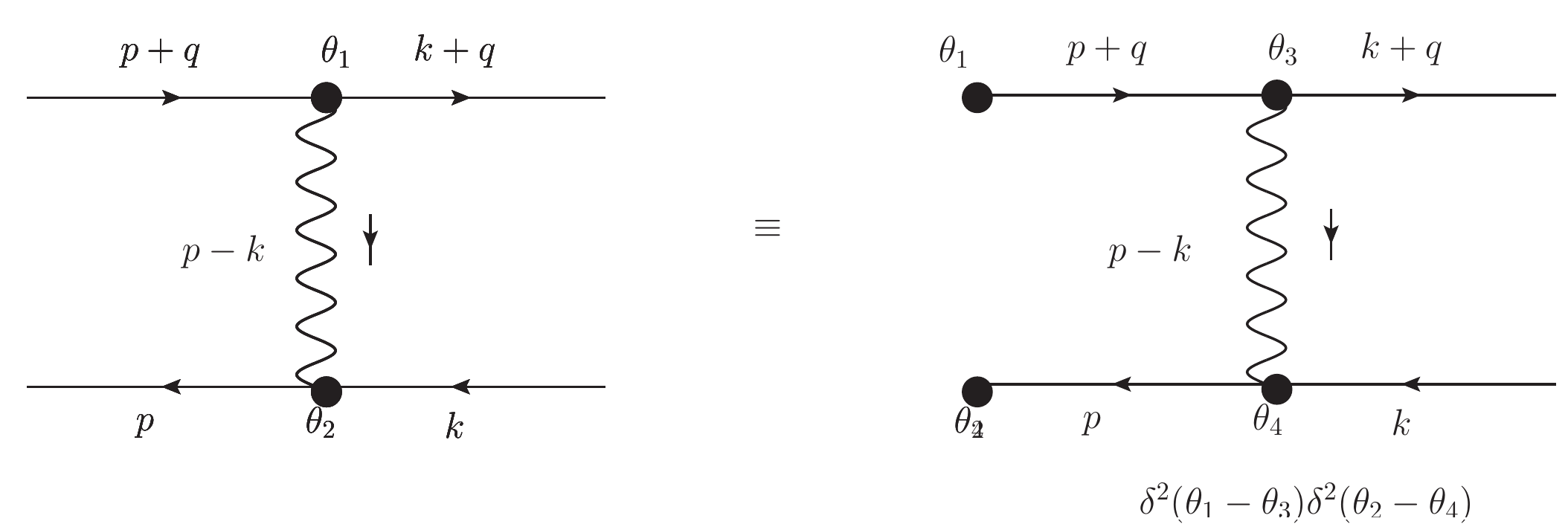}
\caption{\label{gaugetree}Four point function for gauge interaction: Tree diagram}
\end{center}
\end{figure}

In this subsection, we will compute the tree level diagram for the four point function due to the
gauge superfield interaction. (see fig \ref{gaugetree}). In fig \ref{gaugetree} the two diagrams are
equivalent ways to represent the same process. 

\begin{align}\label{V0gauge}
V_0(\te_1,\te_2,\te_3,\te_4,p,q,k)^{gauge}=&\f{-2\pi}{\ka(p-k)_{--}} (D_-^{\te_2,p}-D_-^{\te_4,k}
)(D_-^{\te_1,p+q}-D_-^{\te_3,-(k+q)})(\de^2_{13}\de^2_{24}\de^2_{12})\ ,
\end{align}
where $\de_{ij}^2=\de^2(\te_i-\te_j)$. \footnote{Note that each vertex factor in Fig \ref{gaugetree}
has a factor of D, resulting in two powers of D in \eqref{V0gauge}.}

It can be explicitly checked that (see \eqref{Xdef} for definition of $X_{ij}$)
\begin{align}
 (D_-^{\te_2,p}-
D_-^{\te_4,k})(D_-^{\te_1,p+q}-D_-^{\te_3,-(k+q)})(\de^2_{13}\de^2_{24}\de^2_{12})=&\exp\bigg(\f{1}{
4 }X.(p.X_{12}+q.X_{13}+k.X_{43})\bigg)\nn\\ &F_{tree}(X_{12},X_{13},X_{43}) \ ,
\end{align}
where
\beq\label{Ftree}
F_{tree}=2i X_{12}^+X_{13}^+X_{43}^+ (X_{12}^-+X_{34}^-) \ .
\eeq
Thus the final result for the tree level diagram is given by
\begin{align}\label{4pttreefinal}
V_0(\te_1,\te_2,\te_3,\te_4,p,q,k)^{gauge}=&-\f{4\pi i}{\ka(p-k)_{--}}\exp\big(\f{1}{
4 }X_{1234}.(p.X_{12}+q.X_{13}+k.X_{43})\big)\nn\\
&\  X_{12}^+X_{13}^+X_{43}^+ (X_{12}^-+X_{34}^-) \ .
\end{align}
It is clear that the shift invariant function \eqref{Ftree} has the general structure of
\eqref{totform}, with the appropriate identification
\beq
A(p,q,k)=-\f{4\pi i}{\ka} \f{1}{(p-k)_{--}}\ , \ B(p,q,k)=-\f{4\pi i}{\ka} \f{1}{(p-k)_{--}}
\eeq
Note that the  Fig \ref{gaugetree} has the $\mathbb{Z}_2$ symmetry \eqref{map}. It is
straightforward to check that \eqref{4pttreefinal} is invariant under \eqref{map}. 

\subsection{Closure of the ansatz \eqref{totform}}\label{associativity}

In this section, we establish the consistency of the ansatz 
\eqref{totform} as a solution of the integral equation 
\eqref{int4pt}. Consistency is established by plugging the ansatz 
\eqref{totform} into the RHS of this integral equation, and verifying
that the resultant $\theta$ structure is once again of the form given 
in \eqref{totform}. In other words we will show that the dependence of  
\begin{align}
\int \f{d^3r}{(2\pi)^3} d^2\te_a d^2\te_b
d^2\te_A d^2\te_B \bigg(& NV_0(\te_1,\te_2,\te_a,\te_b,p,q,r)P(r+q,\te_a,\te_A) \nn\\
& P(r,\te_B,\te_b)V(\te_A,\te_B,\te_3,\te_4,r,q,k)\bigg)
\end{align}
on $\theta_1$, $\theta_2$, $\theta_3$ and $\theta_4$ is given by the 
form \eqref{totform}  with appropriately identified functions 
$A$, $B$, $C$ $D$. 

The algebraic closure described above actually follows from a more general 
closure property that we now explain. Note that the tree level four point 
function $V_0$ \eqref{V0} is itself of the form \eqref{totform}. The more 
general closure property (which we will explain below) is that the expression  
\begin{align}\label{mult}
V_{12} =V_1 \star V_2 \equiv \int \f{d^3r}{(2\pi)^3} d^2\te_a d^2\te_b
d^2\te_A d^2\te_B \bigg(& V_1(\te_1,\te_2,\te_a,\te_b,p,q,r)P(r+q,\te_a,\te_A) \nn\\
& P(r,\te_B,\te_b)V_2(\te_A,\te_B,\te_3,\te_4,r,q,k)\bigg) 
\end{align}
takes the form \eqref{totform} whenever $V_1$ and $V_2$ are both also of 
the form \eqref{totform}. In other words \eqref{mult} defines a 
closed multiplication rule on expressions of the form \eqref{totform}.

The explicit verification of the closure described the last paragraph
follows from straightforward
algebra. Let \footnote{We have used the notations $X_{12ab}=\te_1+\te_2+\te_a+\te_b$
and $X_{AB34}=\te_A+\te_B+\te_3+\te_4$.}
\begin{align}
V_1(\te_1,\te_2,\te_a,\te_b,p,q,r)=\exp\big(\f{1}{4}X_{12ab}.(p.X_{12}+q.X_{1a}+r.X_{ba}
)\big)F_1(X_{12},X_{1a},X_{ba},p,q,r)
\end{align}
where 
\begin{align}
F_1(X_{12},X_{1a},X_{ba},p,q,r)=X_{AB}^+ X_{43}^+ \bigg(& A_1(p,r,q) X_{12}^- X_{ba}^-X_{1a}^+
X_{1a}^-+B_1(p,r,q) X_{12}^- X_{ba}^- \nn\\
& + C_1(p,r,q)  X_{12}^- X_{1a}^+ + D_1(p,r,q) X_{1a}^+ X_{ba}^-\bigg) \ .
\end{align}
and
\begin{align}
V_2(\te_A,\te_B,\te_3,\te_4,r,q,k)=\exp\big(\f{1}{4}X_{AB34}.(r.X_{AB}+q.X_{A3}+k.X_{43})\big)F_2(X_
{AB},X_{A3},X_{43},r,q,k)\ ,
\end{align}
where
\begin{align}
F_2(X_{AB},X_{A3},X_{43},r,q,k)=X_{AB}^+ X_{43}^+ \bigg(& A_2(r,k,q) X_{AB}^- X_{43}^-X_{A3}^+
X_{A3}^-+B_2(r,k,q) X_{AB}^- X_{43}^- \nn\\
& + C_2(r,k,q)  X_{AB}^- X_{A3}^+ + D_2(r,k,q) X_{A3}^+ X_{43}^-\bigg) \ .
\end{align}
Evaluating the integrals over $\te_a, \te_b, \te_A, \te_B$, we find that $V_{12}$ in \eqref{mult} 
is of the form \eqref{totform} with
\begin{align}\label{convolutionrules}
A_{12}=-\f{1}{4}q_{3}\int d^3\mathcal{R} \biggl(&
(C_{1}C_{2}k_{-}-D_{1}D_{2}p_{-}+2B_{2}C_{1}q_{3}-2B_{1}D_{2}q_{3 })r_-
\nn\\&+2A_{2}(D_{1}p_{-}+2B_{1}q_{3}+2C_{1}r_{-})+2A_{1}(C_{2}k_{-}+2B_{2}q_{3}+
2D_{2}r_{-})\biggr),\nn
\end{align}
\begin{align}
B_{12}=-\f{1}{4}\int
d^3\mathcal{R}
\biggl(&(2A_{2}-C_{2}k_{-})(2A_{1}+D_{1}p_{-})+4B_{1}B_{2}q_{3}^2+3C_{1}D_{2}r_{-}^{2}\nn\\
&+(2A_{2}C_{1}-2A_{1}D_{2}-C_{1}C_{2}k_{-}-D_{1}D_{2}p_{-}+4B_{2}C_{1}q_{3}+4B_{1}D_{2}q_{3})r_
{-}\biggr) ,\nn
\end{align}
\begin{align}
C_{12}=&-\f{1}{2}\int d^3\mathcal{R} C_{2}q_{3}(2A_{1}+D_{1}p_{-}+2B_{1}q_{3}+3C_{1}r_{-})\ , \nn\\
D_{12}=&-\f{1}{2}\int d^3\mathcal{R}D_{1}q_{3}(-2A_{2}+C_{2}k_{-}+2B_{2}q_{3}+3D_{2}r_{-}) \ .
\end{align}
where
$$d^3\mathcal{R}=\f{d^3r}{(2\pi)^3}\f{1}{(r^2+m^2)((r+q)^2+m^2)}$$

It follows from \eqref{mult} that 
\begin{equation}\label{assocp}
 (V_1 \star V_2) \star V_3=V_1 \star( V_2 \star V_3)
\end{equation}
as both expressions in \eqref{assocp} are given by the same integral 
(the expressions differ only in the order in which the $\theta$ and 
internal momentum integrals are performed). In other words the product 
defined above is associative. We have directly checked that the 
explicit multiplication formula \eqref{convolutionrules} defines an 
associative product rule.

\subsection{Consistency check of the integral equation}\label{consistencycheck}
In this section, we demonstrate that the
integral equations \eqref{integraleqmassiveA}-\eqref{integraleqmassiveD} are consistent with the
$\mathbb{Z}_2$ symmetry \eqref{map}. First we note that the $\mathbb{Z}_2$ invariance \eqref{map} of
\eqref{totform} imposes the following conditions on the unknown functions of momenta
\begin{align}\label{constraints}
& A(p,k,q)=A(k,p,-q) \ , B(p,k,q)=B(k,p,-q) \ ,\nn\\
& C(p,k,q)=-D(k,p,-q) \ , D(p,k,q)=-C(k,p,-q) \ .
\end{align}
These conditions can be written in the form of a matrix given by
\beq\label{map3}
E(p,k,q)=T E(k,p,-q)\ ,
\eeq
where 
\begin{align}
T=\left(\begin{array}{cccc} 
1&0&0&0\\
0&1&0&0\\
0&0&0&-1\\
0&0&-1&0\end{array}\right)\ , E(p,k,q)= \left(\begin{array}{c} A \\ B \\ C \\ D \end{array}\right)
\end{align}

The integral equations \eqref{integraleqmassiveA}-\eqref{integraleqmassiveD} can be written in
differential form by taking derivatives of $p_+$ and using the formulae in Appendix
\S \ref{rintegrals} \footnote{Taking derivatives with respect to $p_+$ eliminates the $r_\pm$
integrals because of the delta functions. The remaining $r_3$ integrals can be easily performed
(see Appendix \S \ref{rintegrals}).}
\begin{align}\label{ppeq1}
 \p_{p_+}E(p,k,q)= S(p,k,q) +H(p,k_-,q) E(p,k,q) \ 
\end{align}
where $S(p,k,q)$ is a source term. The equation for $k_+$ can be obtained from the above equation as
follows
\begin{align}\label{kpeq1}
 \p_{k_+}E(p,k,q) &= T\p_{k_+} E(k,p,-q)\ ,\nn \\
&= T S(k,p,-q)+ T H(k,p_-,-q) E(k,p,-q)\ ,\nn\\
&= T S(k,p,-q)+ T H(k,p_-,-q)T E(p,k,q)\ ,
\end{align}
where we have used \eqref{map3}. Applying $k_+$, $p_+$ derivative on \eqref{ppeq1} and \eqref{kpeq1}
respectively and taking the difference we get
\begin{align}
 \p_{k_+}S(p,k,q)+ H(p,k_-,q)\bigg(T S(k,p,-q)+ T H(k,p_-,-q)T E(p,k,q)\bigg)\nn\\
=T\p_{p_+}S(k,p,-q)+TH(k,p_-,-q)T\bigg(S(p,k,q)+H(p,k_-,-q)E(p,k,q)\bigg)\ .
\end{align}
Comparing coefficients of $E(p,k,q)$ in the above equation we get the condition
\beq\label{consistencycondition1}
[H(p,k_-,q),T H(k,p_-,-q)T]=0 \ .
\eeq
For the integral equations \eqref{integraleqmassiveA}-\eqref{integraleqmassiveD}, the $H(p,k_-,q)$
are given by

\beq\label{Hpq1}
H(p,k_-,q_3)=\f{1}{a(p_s,q_3)}\left(
\begin{array}{cccc}
  (6 q_3-4 i m)p_- & 2 q_3 (2 i m+q_3) p_- & (2 im+q_3)k_- p_-  & -(2 i m+q_3)p_-^2  \\
 4 p_- & 4 q_3 p_-  & -2 k_- p_- & 2 p_-^2 \\
 0 & 0 & 8q_3 p_-  & 0 \\
 8 i m-4 q_3 & 4 q_3 (q_3-2 i m) & 2 (q_3-2 i m)k_-  & (4 i m+6 q_3) p_- \\
\end{array}
\right)
\eeq
where
\beq
a(p_s,q_3)=\f{\sqrt{m^2+p_s^2} \left(4 m^2+q_3^2+4 p_s^2\right)}{2\pi}\ .
\eeq
The matrix $T H(k,p_,-q_3)T$ is
\beq\label{THTpq1}
T H(k,p_,-q_3)T=\f{1}{a(k_s,q_3)}\left(
\begin{array}{cccc}
  -(4 i m+6 q_3)k_- & 2 q_3 (q_3-2 i m)k_-  & -
(q_3-2 i m)k_-^2 & (q_3-2 i m)k_- p_-  \\
 4 k_- & -4 q_3 k_-  & -2 k_-^2 & 2 k_- p_- \\
 -8 i m-4 q_3 & 4 (-2 i m-q_3) q_3 & (4 i m-6 q_3)k_-  & -(4 i m+2 q_3) p_- \\
 0 & 0 & 0 & -8q_3  k_-  \\
\end{array}
\right)\ ,
\eeq
It is straightforward to check that \eqref{Hpq1} and \eqref{THTpq1} commute. Thus the system of
differential equations \eqref{ppeq1} obey the integrability conditions
\eqref{consistencycondition1}. Thus the differential equations \eqref{ppeq1} will have
solutions that respect the $\mathbb{Z}_2$ symmetry. 

 \subsection{Useful formulae}\label{rintegrals}
The Euclidean measure for the basic integrals are
\begin{align}
 \int  \f{(d^3 r)_E}{(2\pi)^3} =\f{1}{(2\pi)^3}\int r_s  dr_ s dr_3
d\te \ ,
\end{align}
where $r_s^2=r_+r_-=r_1^2+r_2^2$ and $r^2=r_s^2+r_3^2$. Here the integration limits are $-\infty
\leq r_3 \leq \infty$, $0\leq r_s \leq \infty$.
Most often we encounter integrals of the type,
\beq
H(q)=\int \f{d^3 r}{(2\pi)^3} \f{1}{(r^2+m^2)((r+q)^2+m^2)}= \f{1}{4\pi |q_3|}\tan^{-1}
\left(\left|\f{q_3}{2 m}\right|\right)
\eeq
where we have set $q_\pm=0$. Another frequently appearing integral is 
\beq
\int \f{d^3 r}{(2\pi)^3} \f{1}{r^2+m^2}= -\f{|m|}{4\pi}
\eeq
where we have regulated the divergence using dimensional regularization.

In the integral equations \eqref{integraleqmassiveA}-\eqref{integraleqmassiveD}, there are no
explicit functions of $r_3$ appearing in the integral equations and the $r_3$ integral can be
exactly done
\beq \label{intr3}
\int_{-\infty}^\infty  \f{dr_3}{(r_s^2+r_3^2+m^2)(r_s^2 + (r_3 + q_3)^2 + m^2)}=
\f{2\pi}{\sqrt{r_s^2+m^2}(4 m^2 + q_3^2 + 4 r_s^2)} \ .
\eeq
The results for the angle integrals are
\begin{align}\label{intangle}
 & \int_0^{2\pi} \f{d\te}{(r-p)_- \ (k-r)_- }=\f{2\pi}{(k-p)_-}
\bigg(\f{k_+}{k_s^2}\te[k_s-r_s]-\f{p_+}{p_s^2}\te[p_s-r_s] \bigg) \ ,\nn\\
 & \int_0^{2\pi} \f{d\te \ r_-}{(r-p)_- \ (k-r)_- }=\f{2\pi}{(k-p)_-}
\bigg(\te[k_s-r_s]-\te[p_s-r_s] \bigg)\ ,\nn\\
 & \int_0^{2\pi} \f{d\te \ r_-^2}{(r-p)_- \ (k-r)_- }=-\f{2\pi}{(k-p)_-}
\bigg(k_-(1-\te[k_s-r_s])-p_-(1-\te[p_s-r_s]) \bigg) \ .
\end{align}
while the $r_s$ integrals are done with the limits from 0 to $\infty$. We will also make use of the
formula
\beq \label{deroz}
\p_{\bar{z}}\bigg(\f{1}{z}\bigg)= 2\pi \de^2(z,\bar{z})
\eeq
to derive the differential form of the integral equations.

For doing the angle integrations in \eqref{umesheqB} we used the formula \eqref{cotcot}
\begin{equation}
 \int d\theta {\rm Pv} \cot \left(\frac{\theta}{2}\right)  {\rm Pv} \cot
\left(\frac{\alpha-\theta}{2}\right)= 2\pi  -4 \pi^2 \delta(\alpha),
\end{equation}
where ${\rm Pv}$ stands for principal value. This formula is easily
verified by calculating the Fourier coefficients as follows
\begin{equation}
\begin{split}
\int \frac{d \alpha}{2\pi} e^{-i \alpha }
 \int d\theta {\rm Pv} \cot \left(\frac{\theta}{2}\right)  
{\rm Pv} \cot \left(\frac{\alpha-\theta}{2}\right)
=&\oint \frac{d \omega}{2\pi \omega} \omega^{-m}
 \oint \frac{dz}{z} {\rm Pv} 
\left(\frac{z+1}{z-1}\right)
{\rm Pv} \left(\frac{z+\omega}{\omega-z}\right)
\\
=& \begin{cases}
 -i\oint dz \,{\rm Pv} \left(\frac{z+1}{z-1}\right)
z^{-m-1}
= -2\pi
& (m >0)
\\
 0 &   (m = 0)
\\
 i\oint dz \,{\rm Pv} \left(\frac{z+1}{z-1}\right)
z^{-m-1}
= -2\pi
& (m < 0)
   \end{cases}
\end{split}
\label{Fo-cotcot}
\end{equation}
where $z = e^{i \theta}$ and $\omega = e^{i \alpha}$.
By comparing 
\eqref{Fo-cotcot}
with Fourier coefficients of delta function,
\begin{equation}
\delta(\alpha)  = \frac{1}{2\pi}\sum_{m=-\infty}^{\infty} 
e^{i m \alpha},
\label{eq:delta-coef}
\end{equation}
we can immediately check \eqref{cotcot}. 

\section{Properties of the \texorpdfstring{$J$}{J} functions}\label{japp}

The $J$ functions are given by 
\begin{align}\label{Jfuns}
J_B(q_3,\la)&=\f{4 \pi q_3}{\ka}\f{n_1+n_2+n_3}{d_1+d_2+d_3}\ ,\nn\\
J_F(q_3,\la)&=\f{4 \pi q_3}{\ka}\f{-n_1+n_2+n_3}{d_1+d_2+d_3}\ ,
\end{align}
where the parameters are
\begin{align}\label{smatrixparameters}
n_1= & 16 m q_3 (w+1) e^{i \la\left(2 \tan ^{-1}\f{2|m|}{q_3}+\pi 
\text{sgn}(q_3)\right)}\ ,\nn\\
n_2= & (w-1) (q_3+2 i m) (2 m (w-1)+i q_3 (w+3)) \left(-e^{2 i \pi  \la 
\text{sgn}(q_3)}\right)\ , \nn\\ 
n_3= &(w-1) (2 m+i q_3) (q_3 (w+3)+2 i m (w-1)) e^{4 i \la  \tan
^{-1}\f{2|m|}{q_3}})\ ,\nn\\
 d_1=& (w-1) \left(4 m^2 (w-1)-8 i m q_3+q_3^2 (w+3)\right) e^{4 i \lambda  \tan
^{-1}\frac{2 |m|}{q_3}}\ ,\nn\\
d_2=&(w-1) \left(4 m^2 (w-1)+8 i m q_3+q_3^2 (w+3)\right) e^{2 i \pi  \lambda 
\text{sgn}(q_3)}\ , \nn\\
d_3=&-2 \left(4 m^2 (w-1)^2+q_3^2 (w (w+2)+5)\right) e^{i \lambda  \left(2 \tan
^{-1}\frac{2| m|}{q_3}+\pi  \text{sgn}(q_3)\right)}\ .
\end{align}
Both the $J$ functions \eqref{Jfuns} are even functions of $q_3$
\beq\label{Jeven}
J_B(q_3,\la)=J_B(-q_3,\la)\ , \ J_F(q_3,\la)=J_F(-q_3,\la)\ .
\eeq

Therefore in \eqref{Jfuns} we can replace $q_3$ with $|q_3|$ and rewrite them as
\begin{align}\label{finalJfunction}
 &J_B(|q_3|,\la)=\f{4 \pi
|q_3|}{\ka}\f{(\tilde{n}_1+\tilde{n}_2+\tilde{n}_3)}{(\tilde{d}_1+\tilde{d}_2+\tilde{d}_3)}\ ,\nn\\
 &J_F(|q_3|,\la)=\f{4 \pi
|q_3|}{\ka}\f{(-\tilde{n}_1+\tilde{n}_2+\tilde{n}_3)}{(\tilde{d}_1+\tilde{d}_2+\tilde{d}_3)}\ ,
\end{align}
where
\begin{align}\label{smatrixparameters2}
\tilde{n}_1= & 16 m |q_3| (w+1) e^{i \la\left(2 \tan ^{-1}\f{2|m|}{|q_3|}+\pi 
\right)}\ ,\nn\\
\tilde{n}_2= & (w-1) (|q_3|+2 i m) (2 m (w-1)+i |q_3| (w+3)) \left(-e^{2 i \pi  \la 
}\right)\ , \nn\\ 
\tilde{n}_3= &(w-1) (2 m+i |q_3|) (|q_3| (w+3)+2 i m (w-1)) e^{4 i \la  \tan
^{-1}\f{2|m|}{|q_3|}})\ ,\nn\\
 \tilde{d}_1=& (w-1) \left(4 m^2 (w-1)-8 i m |q_3|+|q_3|^2 (w+3)\right) e^{4 i \lambda  \tan
^{-1}\frac{2 |m|}{|q_3|}}\ ,\nn\\
\tilde{d}_2=&(w-1) \left(4 m^2 (w-1)+8 i m |q_3|+|q_3|^2 (w+3)\right) e^{2 i \pi  \lambda 
}\ , \nn\\
\tilde{d}_3=&-2 \left(4 m^2 (w-1)^2+|q_3|^2 (w (w+2)+5)\right) e^{i \lambda  \left(2 \tan
^{-1}\frac{2| m|}{|q_3|}+\pi\right)}\ .
\end{align}

Another useful way to write the $J$ function is to use the following identities
\begin{align}\label{tanprop}
& \tan^{-1}\frac{2 m}{q}= \f{\pi }{2}-\tan^{-1}\f{q}{2m} \nn\\
& \tan^{-1}\f{q}{2 m}= \f{1}{2 i} \log\biggl(\f{1+\f{i q}{2 m}}{1-\f{i q}{2 m}}\biggr)
\end{align}
Using this relations, it is easy to write the $J$ functions in a factorized form as given
in \eqref{Jfunm}
\begin{align}\label{Jfunma}
 J_B(q,\la)= &\f{4 \pi q}{\ka} \f{N_1 N_2+ M_1}{D_1 D_2}\ ,\nn\\
 J_F(q,\la)= &\f{4 \pi q}{\ka} \f{N_1 N_2+ M_2}{D_1 D_2}\ ,
\end{align}
where
\begin{align}\label{Jparma}
N_1= &\left(\left(\frac{2 |m|+i q}{2 |m|-i q}\right)^{-\lambda }  (w-1) (2 m+i q)+(w-1) (2 m-i
q)\right)\ ,\nn\\
N_2= &\left(\left(\frac{2 |m|+i q}{2 |m|-i q}\right)^{-\lambda }  (q (w+3)+2 i m (w-1))+(q (w+3)-2 i
m (w-1))\right)\ ,\nn\\
M_1=&-8 m q ((w+3) (w-1)-4 w) \left(\frac{2 |m|+i q}{2 |m|-i q}\right)^{-\lambda }\ ,\nn\\
M_2= & -8 m q (1 + w)^2 \left(\frac{2 |m|+i q}{2 |m|-i q}\right)^{-\lambda }\ ,\nn\\
D_1= &\left(i \left(\frac{2 |m|+i q}{2 |m|-i q}\right)^{-\lambda } (w-1) (2 m+i q)-2 i m (w-1)+q
(w+3)\right)\ , \nn\\
D_2= & \left( \left(\frac{2 |m|+i q}{2 |m|-i q}\right)^{-\lambda }  (-q (w+3)-2 i m (w-1))+(w-1)
(q+2 i m)\right)\ .
\end{align}
Another useful property of the $J$ function is manifest in the above form is its reality under
complex conjugation
\beq\label{Jreal}
J_B(q,\la)=J_B^*(-q,\la)\ , \ J_F(q,\la)=J_F^*(-q,\la)\ .
\eeq

Yet another useful way to write the $J$ function is to note that the basic integral which appears in
the four point function of scalars in an ungauged theory has the form
\beq
H(q)=\int \f{d^3 r}{(2\pi)^3} \f{1}{(r^2+m^2)((r+q)^2+m^2)}= \f{1}{4\pi |q_3|}\tan^{-1}
\left(\left|\f{q_3}{2 m}\right|\right)
\eeq
for $q_\pm=0$. Thus we can also write
\begin{align}\label{JfunHq}
 J_B(|q|,\la)= &\f{4 \pi |q|}{\ka} \f{N_1 N_2+ M_1}{D_1 D_2}\ ,\nn\\
 J_F(|q|,\la)= &\f{4 \pi |q|}{\ka} \f{N_1 N_2+ M_2}{D_1 D_2}\ ,
\end{align}
where
\begin{align}\label{JparmHq}
N_1= &\left(e^{-8\pi i \la |q|H(q)}(w-1) (2 m+i |q|)+(w-1) (2 m-i
|q|)\right)\ ,\nn\\
N_2= &\left( e^{-8\pi i \la |q|H(q)}(|q| (w+3)+2 i m (w-1))+(|q|
(w+3)-2 i m (w-1))\right)\ ,\nn\\
M_1=&-8 m |q| ((w+3) (w-1)-4 w) e^{-8\pi i \la |q|H(q)}\ ,\nn\\
M_2= & -8 m |q| (1 + w)^2 e^{-8\pi i \la |q|H(q)}\ ,\nn\\
D_1= &\left(i e^{-8\pi i \la |q|H(q)} (w-1) (2 m+i |q|)-2 i m (w-1)+|q|
(w+3)\right)\ , \nn\\
D_2= & \left( e^{-8\pi i \la |q|H(q)}  (-|q| (w+3)-2 i m (w-1))+(w-1)
(|q|+2 i m)\right)\ .
\end{align}

\subsection{Limits of the \texorpdfstring{$J$}{J} function}
\subsubsection{\texorpdfstring{$\mN=2$}{N=2} point}\label{N2pt}

The $\mN=1$ theory studied in this paper enjoys an enhanced $\mN=2$ supersymmetry when $w=1$.
Naturally in this limit we expect the $J$ functions to have a simplification. In
particular we get
\begin{align}\label{N2Jfunctions}
J_B^{w=1}=& - \f{8 \pi m}{\ka}\ , \nn\\
J_F^{w=1}=&   \f{8 \pi m}{\ka}\ .
\end{align}

\subsubsection{Massless limit}
There exists a consistent massless limit for the $J$ functions 
\beq
J_B^{m=0}=J_F^{m=0}= \f{4  \pi |q_3|}{\ka }\frac{ (w-1) (w+3)\sin (\pi  \lambda )}{(w-1) (w+3) \cos
(\pi  \lambda )-w (w+2)-5}\ .
\eeq
This expression is self dual under the duality map \eqref{dtransf}. Note that when $w=1$ this
vanishes and is consistent with the $m\to0$ limit of \eqref{N2Jfunctions}.
\subsubsection{Non relativistic limit in the singlet channel}
The $J$ functions for the S channel are given in \eqref{SJfun}. The non-relativistic limit of the
$J$ functions is obtained by taking $\sqrt{s}\to 2m$ with all the other parameters held fixed. In
this
limit, remarkably we recover the $\mN=2$ result.
\begin{align}\label{NonrelJfunctions}
J_B^{\sqrt{s}\to 2m}=& - \f{8 \pi m}{\ka}\ , \nn\\
J_F^{\sqrt{s}\to 2m}=&   \f{8 \pi m}{\ka}\ .
\end{align}

\bibliography{massh.bib}
\end{document}